\documentclass[12pt,final]{amsart}
\usepackage{amsmath,amsthm,amssymb}
\usepackage{amsfonts}
\usepackage[varg]{txfonts}
\usepackage[mathscr]{eucal}
\usepackage{bxpapersize}
\usepackage{ascmac}
\usepackage{scalefnt}
\usepackage{epic, eepic}
\usepackage{graphicx}
\usepackage{lscape}
\usepackage{indentfirst}
\usepackage{cases}
\usepackage{color}
\usepackage{url}
\usepackage{bm}
\usepackage[all]{xy}
\usepackage[top=25truemm,bottom=25truemm,left=30truemm,right=30truemm]{geometry}
\usepackage[noadjust]{cite}
\usepackage{showkeys}
\usepackage{enumerate}
\usepackage{algorithm}
\usepackage{algorithmic}
\usepackage{hyperref}
\hypersetup{
setpagesize=false,
 bookmarksnumbered=true,%
 bookmarksopen=true,%
 colorlinks=true,%
 linkcolor=blue,
 citecolor=blue,
 urlcolor=blue,
}

\theoremstyle{plane}

\newtheorem{thm}{Theorem}[section]
\theoremstyle{definition}

\theoremstyle{remark}
\newtheorem{rem}[thm]{Remark}
 \newtheorem*{acknowledgements}{Acknowledgements}

\newcommand{\tr}{\operatorname{tr}}

\newcommand{\obs}{\operatorname{obs}}
\newcommand{\argmin}{\operatorname{argmin}}
\newcommand{\test}{\operatorname{test}}
\newcommand{\train}{\operatorname{train}}
\newcommand{\mis}{\operatorname{mis}}

\newcommand{\mse}{\operatorname{MSE}}
\newcommand{\lasso}{\operatorname{lasso}}
\newcommand{\smrm}{\operatorname{SMRM}}

\newcommand{\Y}{\bm{Y}}
\newcommand{\X}{\bm{X}}
\newcommand{\y}{\bm{y}}
\newcommand{\x}{\bm{x}}

\newcommand{\B}{\bm{B}}
\newcommand{\R}{\bm{R}}
\newcommand{\bE}{\mathbb{E}}
\newcommand{\bc}{\bm{c}}

\newcommand{\Sig}{\Sigma}
\renewcommand{\phi}{\varphi}
\newcommand{\eps}{\varepsilon}
\newcommand{\beps}{\bm{\varepsilon}}
\newcommand{\bmu}{\bm{\mu}}
\newcommand{\what}{\widehat}
\newcommand{\til}{\tilde}
\newcommand{\wtil}{\widetilde}

\numberwithin{equation}{section}

\title[Sparse multivariate regression]
{Sparse multivariate regression with missing values and its application to the prediction of material properties}

\author[K. Teramoto]{Keisuke Teramoto}
\address[K. Teramoto]{Institute of Mathematics for Industry, Kyushu University, 
Motooka 744, Fukuoka 819-0395, Japan}
\email{k-teramoto@imi.kyushu-u.ac.jp}

\author[K. Hirose]{Kei Hirose}
\address[K. Hirose]{Institute of Mathematics for Industry, Kyushu University, 
Motooka 744, Fukuoka 819-0395, Japan \&
RIKEN Center for Advanced Intelligence Project, Tokyo 103-0027, Japan}
\email{hirose@imi.kyushu-u.ac.jp}

\thanks{This work was supported by JST-Mirai Program Grant Number JPMJMI18A2, Japan.}

\date{\today}
\begin{document}
\maketitle

\begin{abstract}
In the field of materials science and engineering, 
statistical analysis and machine learning techniques have recently been used to predict multiple material properties from an experimental design.  
These material properties correspond to response variables in the multivariate regression model.  
This study conducts a penalized maximum likelihood procedure to estimate model parameters, 
including the regression coefficients and covariance matrix of response variables.  
In particular, we employ $l_1$-regularization to achieve a sparse estimation of 
regression coefficients and the inverse covariance matrix of response variables. 
In some cases, there may be a relatively large number of missing values in response variables, owing to the difficulty in collecting data on material properties.  
A method to improve prediction accuracy under the situation with missing values 
incorporates a correlation structure among the response variables into the statistical model.  
The expectation and maximization algorithm is constructed, 
which enables application to a data set with missing values in the responses.  
We apply our proposed procedure to real data consisting of 22 material properties.
\end{abstract}
\section{Introduction}
The importance of data analysis applications using statistics and machine learning 
for materials science and engineering has been steadily increasing
(cf. \cite{molecular,materialinfo,strategy,accel,machine-composite,machine-appl,applied,experiment,experiment-improve,parameter}). 
Owing to the recent development of machine learning methods, 
data-centric informatics applied to a sufficiently large amount of data is useful for identifying materials that have desirable properties, such as durability and flexibility. 
Desirable materials are often difficult to identify using only experiments 
and physical simulations. 

The data structure in that field often has two features that introduce new challenges.  
The first feature is that we often predict multiple properties of the materials; 
i.e., we must construct a regression model with multiple responses (multivariate regression).  
For example, in a study on adhesion structure, 
the yield strength, ultimate tensile strength, fracture, and Young's modulus must be predicted, 
among many other material properties that have not been described here. 
However, it would be difficult to find a material 
that satisfies multiple desired properties simultaneously 
because there are typically trade-offs among these properties (\cite{hclkl2020,jyw2019}).  
These trade-offs are expressed as a correlation matrix among the response variables. 
This study assumes the correlation structure among response variables and 
employs a likelihood procedure to estimate regression coefficients and a covariance matrix of response variables. 
The estimated covariance matrix assists engineers with interpreting the relationship between properties 
and improves the prediction accuracy (\cite{mrce,Wang2015}).

When the number of responses is relatively large, 
it may be difficult to estimate all correlation pairs because the number of parameters is proportional to the square of the number of variables.  
In such cases, regularization methods have typically been employed to achieve a stable estimation of 
the covariance matrix of response variables. 
In particular, the least absolute shrinkage and selection operator (LASSO)-type sparse estimation (cf. \cite{lasso1996}) conducts simultaneous variable selection and model estimation, 
which enables an interpretation of the relationship among material properties. 
A wide variety of regularization methods that induce a sparse structure have been proposed, 
such as elastic net (\cite{elasticnet}), group LASSO (\cite{sparse-lasso}), 
graphical LASSO (\cite{sparse-glasso}), generalized LASSO (\cite{generalized}), and overlapping group LASSO (\cite{overlapping}). 
Rothman et al. \cite{mrce} introduced the multivariate regression with covariance estimation ({\it MRCE}) method, 
in which sparse regression coefficients and a sparse inverse covariance matrix of the response variables are simultaneously estimated. 
This is a generalization of the LASSO regression to the sparse multivariate regression analysis. 
The model parameter is estimated using the penalized maximum likelihood procedure with the LASSO.
In this study, we perform data analysis based on the MRCE method.

The second feature that introduces challenges is that the data values in the response variables are often missing because it is difficult to observe all the properties of the materials, owing to large-scale experiments. 
When the ratio of missing data is small, we can exclude the corresponding observations and conduct a data analysis. 
This method is referred to as the complete-case analysis (\cite{missingdata}).  
However, responses tend to have many missing values. 
In fact, a data set used in this study comprised $50$\%  or more missing values (see Figure \ref{fig:ratio1} in Section \ref{sec:application}). 
If we conduct a complete-case analysis, the number of observations becomes 
extremely small, which results in low prediction accuracy. 
The {\it full information maximum likelihood} ({\it FIML}) approach (see \cite{FIML}) provides a means to handle a large number of missing values; 
Hirose et al. \cite{hirose2016FIML} showed that the FIML approach provides a good estimator 
when the ratio of missing data is $90$\%. 
The FIML produces a consistent estimator, even when 
the number of missing values is large under the missing at random (MAR) assumption (cf. \cite{missingdata}).
Moreover, the FIML approach enables missing value interpolation, 
which may assist with understanding some hidden structures/relations.
For multivariate data without response variables, 
St\"{a}dler and B\"{u}hlmann \cite{missglasso} proposed the  graphical LASSO with missing values {\it MissGLASSO} 
method for data with missing values. 
An $l_1$-regularized likelihood method is used to estimate the sparse inverse covariance matrix. 
Moreover, they proposed an efficient EM algorithm (see \cite{buck}) for optimization with provable 
numerical convergence properties. 
St\"{a}dler and B\"{u}hlmann \cite{missglasso} extended MissGLASSO to multiple (not multivariate) regression analysis. 
However, they only assumed the case where the exploratory variables have missing values; 
MissGLASSO cannot be directly applied to multivariate regression analysis when there are missing values in the response variables.

As mentioned above, the MRCE simultaneously estimates the regression coefficients and covariance matrix of the response variables. 
However, it is applicable only to complete multivariate data; 
thus, we cannot perform this method directly for data with missing values. 
Therefore, we need a suitable extension of the MRCE to apply to data with missing values. 
Notably, MissGLASSO can be applied to data with missing values. 
The aim of this study is to propose a multivariate regression model with missing values by combining these two methods.

In this study, we establish a new algorithm called the {\it sparse multivariate regression with missing data} 
({\it SMRM}) algorithm 
to estimate the inverse covariance matrix and interpolate the data with missing values (see Section \ref{sec:SMRM}). 
To estimate multivariate regression coefficients and the covariance structure, 
we need to solve a particular $l_1$-regularized likelihood type optimization problem 
with two regularization parameters; one is related to the correlation structure of the responses, 
and the other is related to the regression coefficients matrix.
Here, we note that multiple regularization parameters for regression coefficients 
are assumed because the error variances vary among the response variables. 
For this optimization problem, we employ the EM algorithm.  
As with the case of the MRCE method, the coordinate descent algorithm and graphical LASSO algorithm 
are conducted in the maximization (M) step of the EM algorithm.  
Using sparse estimation, 
the SMRM algorithm can conduct stable estimation, even for a dataset with a relatively large number of missing values. 
In addition, we can improve the prediction accuracy by using the correlation structure among the response variables. 
We estimate the sparse inverse covariance matrix 
to introduce our method instead of the covariance matrix itself 
because spurious correlations among responses may be excluded (\cite{high-lasso}). 
In the last section, we apply the SMRM algorithm to real data 
and investigate influences of regularization parameters. 
Furthermore, we compare the prediction accuracy obtained by our method to that of the LASSO.

\section{Preliminaries}\label{sec:prelim}
\subsection{Conditional distribution}
We briefly review some notions and facts from multivariate regression analysis. 
For a detailed explanation, refer to \cite{multivariate}. 

Let $\x_j=(x^1_{j},\ldots,x^n_{j})^T$ $(1\leq j\leq p)$ be the predictor variables, $\y_l=(y^1_{l},\ldots,y^n_{l})^T$ $(1\leq l\leq q)$ response variables. 
(We consider $\x_j$ and $\y_l$ as {\it column} vectors.) 
Then, we set matrices $\X$, $\wtil{\X}$, and $\Y$ as 
\begin{align}\label{eq:mat-XY}
\begin{aligned}
\X&=\left(\x_1 , \cdots , \x_p\right)
=\begin{pmatrix} 
x^1_{1} &  x^1_{2} & \cdots & x^1_{p} \\
x^2_{1} & x^2_{2} & \cdots & x^2_{p} \\
\vdots &\vdots  &\ddots & \vdots\\
x^n_{1} & x^n_{2} & \cdots & x^n_{p}
\end{pmatrix},\quad
\wtil{\X}=\left(\bm{1}_n, \x_1 , \cdots , \x_p\right)
=\begin{pmatrix}
1& x^1_{1} &  x^1_{2} & \cdots & x^1_{p} \\
1&x^2_{1} & x^2_{2} & \cdots & x^2_{p} \\
\vdots & \vdots &\vdots  &\ddots & \vdots\\
1& x^n_{1} & x^n_{2} & \cdots & x^n_{p}
\end{pmatrix},\\
\Y&=\left(\y_1,\y_2,\cdots,\y_q\right)
=\begin{pmatrix}
y^1_{1} &  y^1_{2} & \cdots & y^1_{q} \\
y^2_{1} & y^2_{2} & \cdots & y^2_{q} \\
\vdots &\vdots  &\ddots & \vdots\\
y^n_{1} & y^n_{2} & \cdots & y^n_{q}
\end{pmatrix}.
\end{aligned}
\end{align}
Let $\x^i=(x^i_{1},x^i_{2},\ldots,x^i_{p})$, $\til{\x}^i=(1,x^i_{1},x^i_{2},\ldots,x^i_{p})$, and 
$\y^i=(y^i_{1},y^i_{2},\ldots,y^i_{q})$ $(1\leq i\leq n)$ be the $i$-th row vectors of $\X$, $\wtil{\X}$, and $\Y$, as in \eqref{eq:mat-XY}, respectively. 
(We consider $\x^i$, $\til{\x}^i$, and $\y^i$ as {\it row} vectors.)
We then consider the multivariate linear regression model of the form 
\begin{equation}\label{eq:mvlr}
\Y=\wtil{\X}\wtil{\B}+\bm{E}=\left(\bm{1}_n,{\X}\right)
\begin{pmatrix}\bm{b}_0^T \\ \B\end{pmatrix}+\bm{E},
\end{equation}
where $\bm{b}_0=(b^0_{1},b^0	_{2},\ldots,b^0_{q})^T\in\R^q$ is a vector of the regression intercept. 
$\B$ is a regression coefficient matrix of the form
$$\B=\left(\bm{b}_1,\bm{b}_2,\cdots,\bm{b}_q\right)
=\begin{pmatrix}
b^1_{1} & b^1_{2} & \cdots & b^1_{q}\\
b^2_{1} & b^2_{2} & \cdots & b^2_{q}\\
\vdots & \vdots & \ddots & \vdots\\
b^p_{1} & b^p_{2} & \cdots & b^p_{q}
\end{pmatrix},\quad 
$$
and $\bm{E}$ is the error matrix given by 
$$\bm{E}=\left(\beps_1,\beps_2,\cdots,\beps_q\right)
=\begin{pmatrix}
\eps^1_{1} & \eps^1_{2} & \cdots & \eps^1_{q}\\
\eps^2_{1} & \eps^2_{2} & \cdots & \eps^2_{q}\\
\vdots & \vdots & \ddots & \vdots \\
\eps^n_{1} & \eps^n_{2} &\cdots & \eps^n_{q}
\end{pmatrix}.
$$
We denote the $i$-th row vector of $\bm{E}$ $(1\leq i\leq n)$ as $\beps^i=(\eps^i_{1},\eps^i_{2},\ldots,\eps^i_{q})$.
We assume that the $n$ subjects are independent. 
We then obtain the following:
\begin{itemize}
\item $\beps_{l}\sim N(\bm{0}_{n},\sigma^l_{l}\bm{I}_n)$ $(1\leq l \leq q)$,
\item $(\beps^i)^T\sim_{\mathrm{i.i.d}} N(\bm{0}_{q},\Sig)$ $(1\leq i\leq n)$, 
\end{itemize}
where $\bm{0}_q$ is the $q\times q$ zero vector, 
$\bm{I}_n$ is the $n\times n$ identity matrix, and $\Sig$ is the covariance matrix of the form 
$$\Sig=\begin{pmatrix}
\sigma_{1}^1 & \sigma_{2}^1 & \cdots & \sigma_{q}^1\\
\sigma_{1}^2 & \sigma_{2}^2 & \cdots & \sigma_{q}^2 \\
\vdots & \vdots & \ddots & \vdots\\
\sigma_{1}^q & \sigma_{2}^q & \cdots & \sigma_{q}^q
\end{pmatrix}$$,
where $\sigma_{l'}^l=\sigma_{l}^{l'}$ $(1\leq l,l' \leq q)$. 
We assume the independence condition in the following.
Under these assumptions, we note that $\y^i|\til{\x}^i$ follows $\y^{i}|\til{\x}^i\sim N(\bmu_i=\wtil{\B}^T(\til{\x}^i)^T,\Sig)$.

We now consider a partition $\y^i=(\y^{i,1},\y^{i,2})$ for each $i\in\{1,\ldots,n\}$. 
$(\y^{i,2})^T|(\y^{i,1})^T$ follows a linear regression on $(\y^{i,1})^T$ with a mean of $\bmu_{i,2}+\Sig_{i,21}\Sig_{i,11}^{-1}(Y_{i,1}^T-\bmu_{i,1})$ 
and covariance of $\Sig_{i,22}-\Sig_{i,21}\Sig_{i,11}^{-1}\Sig_{i,12}$ (\cite{multivariate,graphical-model}). 
Here, we divide $\bmu_i$ and $\Sig$ into 
$$\bmu_i=\begin{pmatrix} \bmu_{i,1} \\ \bmu_{i,2}\end{pmatrix},\quad 
\Sig=\begin{pmatrix} \Sig_{i,11} & \Sig_{i,12} \\ \Sig_{i,21} & \Sig_{i,22} \end{pmatrix}$$
for each $i$. 
(For example, if we divide $\y^i$ into $\y^{i,1}=(y^i_{1},\ldots,y^i_{l})$ and $\y^{i,2}=(y^i_{l+1},\ldots,y^i_{q})$, 
then $\bmu_{i,1}$, $\bmu_{i,2}$, $\Sig_{i,11}$, $\Sig_{i,12}$, $\Sig_{i,21}$, and $\Sig_{i,22}$ are a
$l\times1$ matrix, $(q-l)\times 1$ matrix, $l\times l$ matrix, $l\times (q-l)$ matrix, $(q-l)\times l$ matrix, 
and a $(q-l)\times(q-l)$ matrix, respectively.)
Thus, it can be observed that 
\begin{equation}\label{eq:seigen}
(\y^{i,2})^T|(\y^{i,1})^T\sim N(\bmu_{i,2}+\Sig_{i,21}\Sig_{i,11}^{-1}((\y^{i,1})^T-\bmu_{i,1}),\Sig_{i,22}-\Sig_{i,21}\Sig_{i,11}^{-1}\Sig_{i,12}).
\end{equation}

Let $K$ be a $q\times q$ matrix that satisfies $K\Sig=\bm{I}_q$. 
We call $K$ the {\it precision matrix}. 
For $i$, if we divide $(\y^i)^T$ into $(\y^{i,1},\y^{i,2})^T$, then it holds that  
\begin{equation}\label{eq:prec-covar}
\begin{pmatrix} K_{i,11} & K_{i,12} \\ K_{i,21} & K_{i,22} \end{pmatrix}
\begin{pmatrix} \Sig_{i,11} & \Sig_{i,12} \\ \Sig_{i,21} & \Sig_{i,22} \end{pmatrix}=
\begin{pmatrix} I & 0 \\ 0 & I \end{pmatrix}. 
\end{equation}
By \eqref{eq:seigen} and \eqref{eq:prec-covar}, we obtain 
\begin{equation}\label{eq:seigen-prec}
(\y^{i,2})^T|(\y^{i,1})^T\sim N(\bmu_{i,2}-K_{i,22}^{-1}K_{i,21}((\y^{i,1})^T-\bmu_{i,1}),K_{i,22}^{-1}).
\end{equation}
This relation is the key part of our algorithm.

\subsection{The LASSO}
We briefly review the LASSO. 
(For further details, refer to \cite{lasso1996}.)
This method will be used in Section \ref{sec:application} to evaluate 
the prediction accuracy obtained by our proposed method according to real data. 
Let $\x_j=(x_{j}^1,\ldots,x_{j}^n)^T$ be predictor variables $(1\leq j\leq p)$ and $\y=(y^1,\ldots,y^n)^T$ be the response variables. 
We set a matrix $\wtil{\X}$ using $\x_j$, as in \eqref{eq:mat-XY}. 
We then consider the linear regression model 
$$\y=\beta_0\bm{1}_n+\X\bm{\beta}+\beps,\quad \beps\sim N(\bm{0},\sigma^2\bm{I}_n),$$
where $\beta_0\in\R$ and $\bm{\beta}=(\beta_1,\ldots,\beta_p)^{T}\in\R^{p}$ are parameters of the regression.  
In this case, the LASSO optimizes the following form
\begin{equation}\label{eq:lasso}
\min_{\bm{\beta}\in\R^p}\left\{\dfrac{1}{n}||\y-\beta_0\bm{1}_n-\X\bm{\beta}||^2_2
+\lambda||\bm{\beta}||_1\right\}\quad \left(~||\bm{\beta}||_1=\sum_{j=1}^p|\beta_j|~\right),
\end{equation}
where $\lambda>0$. 
By solving the optimization problem, as in \eqref{eq:lasso}, we obtain 
estimators of $\bm{\beta}$ and $\beta_0$. 
Generally, the regularization parameter $\lambda$ of the LASSO 
is chosen to minimize predicted errors of each response.
Such a regularization parameter is typically called the `best' regularization parameter.

\section{Interpolation for data with missing values}\label{sec:SMRM}
\subsection{Responses with missing values}
Let $\x_j\in\R^n$ $(1\leq j\leq p)$ be the predictor varieties and $\y_l\in\R^n$ the response varieties $(1\leq l\leq q)$. 
We assume that the relation \eqref{eq:mvlr} holds. 
We consider the case in which the matrix of responses $\Y$, as in \eqref{eq:mat-XY}, has missing values. 
Then, we divide the $i$-th row vector $\y^i$ of $\Y$ into 
$$\y^i=(\y^{i,\obs},\y^{i,\mis}),$$
where $\y^{i,\obs}$ is a vector that consists of the observed values, and $\y^{i,\mis}$ consists of missing values. 
By \eqref{eq:seigen-prec}, it follows that 
\begin{equation}\label{eq:seigen-mis}
(\y^{i,\mis})^T|(\y^{i,\obs})^T\sim 
N\left(\bmu_{i,\mis}-K_{i,\mis,\mis}^{-1}K_{i,\mis,\obs}((\y^{i,\obs})^T-\bmu_{i,\obs}),K_{i,\mis,\mis}^{-1}\right),
\end{equation}
where we divide the mean vector $\bmu_i=\wtil{\B}^T(\til{\x}^i)^T$ and the precision matrix $K$ into 
$$\bmu_i=\begin{pmatrix} \bmu_{i,\obs} \\ \bmu_{i,\mis} \end{pmatrix}, \quad 
K=\begin{pmatrix} K_{i,\obs,\obs} & K_{i,\obs,\mis} \\ K_{i,\mis,\obs} & K_{i,\mis,\mis} \end{pmatrix}$$
for each $i$. 
For remainder of this paper, we assume that for the matrix $\Y$ given by 
\begin{equation}\label{eq:mat-Y}
\Y=
\left(\y_{1},\cdots,\y_{q}\right)=
\begin{pmatrix}
\y^{1}\\ \vdots \\ \y^{n}
\end{pmatrix}=
\begin{pmatrix} 
y^1_{1} & y^1_{2} & \cdots & y^1_{q}\\
\vdots & \vdots & \ddots & \vdots\\
y^n_{1} & y^n_{2} & \cdots & y^n_{q}
\end{pmatrix},
\end{equation}
there are no columns with entries that are missing values.

\subsection{Algorithm to interpolate the data with missing values}
We derive an algorithm that performs multivariate regression and interpolates data with missing values. 
We assume the same conditions as in the previous subsection. 
For each $i$, it follows that $(\y^i)^T|(\til{\x}^i)^T\sim N(\bmu_i=\wtil{\B}^T(\til{\x}^i)^T,\Sig)$; 
hence, the likelihood function $L((\y^i)^T|(\til{\x}^i)^T)$ is 
$$L((\y^i)^T|(\til{\x}^i)^T)=(2\pi)^{\frac{p}{2}}|\Sig|^{-\frac{1}{2}}\exp\left(-\dfrac{1}{2}((\y^i)^T-\bmu_i)^T\Sig^{-1}((\y^i)^T-\bmu_i)\right),$$
where $|\Sig|$ is the determinant of $\Sig$.
Thus, the log-likelihood function can be expressed as 
\begin{align*}
\sum_{i=1}^n \log L((\y^i)^T|(\til{\x}^i)^T)&=-\dfrac{np}{2}\log(2\pi)-\dfrac{n}{2}\log|\Sig|
-\dfrac{1}{2}\sum_{i=1}^n((\y^i)^T-\bmu_i)^{T}\Sig^{-1}((\y^i)^T-\bmu_i)\\
&=-\dfrac{np}{2}\log(2\pi)+\dfrac{n}{2}\log|K|
-\dfrac{1}{2}\sum_{i=1}^n((\y^i)^T-\bmu_i)^{T}K((\y^i)^T-\bmu_i).
\end{align*}
Using this function, we set 
\begin{equation}\label{eq:function-l}
l(\wtil{\B},K;\Y)=\dfrac{n}{2}\log|K|-\dfrac{n}{2}\sum_{i=1}^n\bmu_i^TK\bmu_i+\dfrac{1}{2}\sum_{i=1}^n\bmu_i^T K(\y^i)^T
-\dfrac{1}{2}\tr\left(K\Y^T\Y\right),
\end{equation}
where $\Y$ is a matrix given by \eqref{eq:mat-Y}. 

We set the following $l_1$-regularization of the function $l$:
\begin{align}\label{eq:penalty}
\begin{aligned}
&-l(\wtil{\B},K;\Y)+\lambda_1\sum_{l\neq l'}|k_{l'}^l|+2\sum_{j=1}^p\sum_{l=1}^q\lambda_{2,l}^j|b_{l}^j|\\
&=\dfrac{n}{2}\log|K|-\dfrac{n}{2}\sum_{i=1}^n\bmu_i^TK\bmu_i+\dfrac{1}{2}\sum_{i=1}^n\bmu_i^T K(\y^i)^T
-\dfrac{1}{2}\tr\left(K\Y^T\Y\right)+\lambda_1\sum_{l\neq l'}|k_{l'}^l|+2\sum_{j=1}^p\sum_{l=1}^q\lambda_{2,l}^j|b_{l}^j|,
\end{aligned}
\end{align}
where $\lambda_1\geq0$ and $\lambda_{2,l}^j\geq0$ are regularization parameters. 
We consider the following conditional mean of 
$-l(\wtil{\B},K;\Y)+\lambda_1\sum_{l\neq l'}|k_{l'}^l|+2\sum_{j=1}^p\sum_{l=1}^q\lambda_{2,l}^j|b_{l}^j|$ 
as in \eqref{eq:penalty}: 
\begin{equation}\label{eq:function-q}
Q(\wtil{\B},K|\wtil{\B}',K')=-\bE[l(\wtil{\B},K;\Y)|\Y_{\obs},\wtil{\B}',K']+\lambda_1\sum_{l\neq l'}|k_{l'}^l|+2\sum_{j=1}^p\sum_{l=1}^q\lambda_{2,l}^j|b_{l}^j|.
\end{equation}
In this case, we derive an algorithm to impute the data $\y^{i,\mis}$ by solving the optimization problem for $Q$ 
by applying the EM-algorithm. 
We call this procedure the {\it sparse multivariate regression for responses with missing data} ({\it SMRM}) algorithm. 
First, we provide initial values $\wtil{\B}^{(0)}$ and $K^{(0)}$. 
Then, we compute the E-steps and M-steps as follows (cf. \cite{missglasso}).

\noindent
{\bf E-step:} For each $i$, we denote the mean vector and precision matrix in the $m$-step $(m=0,1,2,\ldots)$ as 
$\bmu_i^{(m)}=\wtil{\B}^{(m)T}(\til{\x}^i)^T$ and $K^{(m)}$, respectively. 
We set 
\begin{equation}\label{eq:ci}
\bc^{i,(m)}=\bmu_{i,\mis}^{(m)}-\left(K_{i,\mis,\mis}^{(m)}\right)^{-1}K_{i,\mis,\obs}^{(m)}((\y^{i,\obs})^T-\bmu^{(m)}_{i,\obs})
\end{equation}
for each $i$. 
We consider $\bc^{i,(m)}$ as a column vector and use vectors $\bc^{i,(m)}$ to impute the missing values $(\y^{i,\mis, (m)})^T$ in the $m$-step for each $i$. 
Then, the conditional means $\bE[y^i_{l}|(\y^{i,\obs})^T,\bmu_i^{(m)},K^{(m)}]$ and $\bE[y^i_{l}y^i_{l'}|(\y^{i,\obs})^T,\bmu_i^{(m)},K^{(m)}]$ 
can be calculated as 
\begin{equation}\label{eq:E-yij}
\bE[y^i_{l}|(\y^{i,\obs})^T,\bmu_i^{(m)},K^{(m)}]=\begin{cases} y^i_{l} & \text{if $y^i_{l}$ is observed}\\ 
c^{i,(m)}_{l} & \text{if $y^i_{l}$ is missing}\end{cases},
\end{equation}
\begin{equation}\label{eq:E-yijyij}
\bE[y^i_{l}y^i_{l'}|(\y^{i,\obs})^T,\bmu_i^{(m)},K^{(m)}]=\begin{cases}
y^i_{l}y^i_{l'}& \text{if both $y^i_{l}$ and $y^i_{l'}$ are observed}\\
y^i_{l}c^{i,(m)}_{l'} & \text{if $y^i_{l}$ is observed and $y^i_{l'}$ is missing}\\
\left(K^{(m)}_{i,\mis,\mis}\right)^{-1}_{ll'}+c^{i,(m)}_{l}c^{i,(m)}_{l'} & \text{if both $y^i_{l}$ and $y^i_{l'}$ are missing}
\end{cases}.
\end{equation}
Using this method, we compute the function $Q(\wtil{\B},K|\wtil{\B}^{(m)},K^{(m)})$, as in \eqref{eq:function-q}. 

\noindent
{\bf M-step:} We compute the updates $(\wtil{\B}^{(m+1)},K^{(m+1)})$ as the minimizer of $Q(\wtil{\B},K|\wtil{\B}^{(m)},K^{(m)})$. 
To do this, we define the following function: 
$$g(\wtil{\B},K)=\tr\left[\dfrac{1}{n}(\Y-\wtil{\X}\wtil{\B})^T(\Y-\wtil{\X}\wtil{\B})K\right]-\log|K|.$$
Then, our aim corresponds to solving the following optimization problem: 
\begin{equation}\label{eq:opt}
(\what{\wtil{\B}},\hat{K})=\argmin_{\wtil{\B},K}\left\{g(\wtil{\B},K)
+\lambda_1\sum_{l\neq l'}|k_{l'}^l|+2\sum_{j=1}^p\sum_{l=1}^q\lambda_{2,l}^j|b_{l}^j|\right\}.
\end{equation}
The algorithms used to numerically solve the above problem are called the {\it MRCE algorithm} (\cite{mrce}) and {\it MissGLASSO} (\cite{missglasso}). 
By applying the MRCE-type and MissGLASSO-type algorithms, we solve the above problem and provide updates. 

To summarize, the algorithm works as follows: \\
\noindent
{\bf Algorithm} (SMRM): For fixed $\lambda_1$ and $\lambda_{2,jl}$, 
initialize $\wtil{\B}^{(0)}$ and $K^{(0)}$. 
\begin{itemize}
\item[Step 1:] Impute $\y^{i,{(m)}}$ using $\bc^{i,{(m)}}$ given by \eqref{eq:ci} for each $i=1,\ldots,n$. 
\item[Step 2:] Compute $(\what{\wtil{\B}}^{(m+1)},\hat{K}^{(m+1)})(=(\what{\wtil{\B}(K^{(m)})},\what{K(\wtil{\B}^{(m)})}))$ 
by solving \eqref{eq:opt} using MRCE and MissGLASSO (\cite{mrce,missglasso}). 
(See also Remark \ref{rem:compute} below.) 
\item[Step 3:] If $\sum_{j,l}|b_{l}^{j,(m+1)}-b_{l}^{j,(m)}|<\eps$ for given a sufficiently small $\eps>0$, then stop. 
Otherwise, go to Step 1.
\end{itemize}

\begin{rem}\label{rem:compute}
To compute the mean of $g$, we remark the following process: 
\begin{enumerate}
\item The case of optimizing $\wtil{\B}$ for fixed $K$: to compute the mean of $g$,  we use $\Y$ as 
a matrix with a row vector that consists of the complement vector $(\hat{\y}^{i,(m)})^{T}=(\y^{i,\obs},(\bc^{i,(m)})^T)^T$ 
given by the rule $\bE[y^i_{l}|(\y^{i,\obs})^T,\bmu_i^{(m)},K^{(m)}]$, as in \eqref{eq:E-yij}.
\item The case of optimizing $K$ for fixed $\wtil{\B}$: to compute the mean of $\Y^T\Y$, 
we use the rule of computation for $\bE[y^i_{l}y^i_{l'}|(\y^{i,\obs})^T,\bmu_i^{(m)},K^{(m)}]$, as in \eqref{eq:E-yijyij}.  
\end{enumerate}
\end{rem}

\begin{rem}
For complete data, the SMRM algorithm performs similarly to the LASSO for large $\lambda_1$. 
Thus, we may consider the SMRM algorithm as a generalization of the LASSO 
for multivariate regression analysis. 
\end{rem}

\section{Applying the algorithm to real data}\label{sec:application}
\begin{figure}[t]
\begin{center}
\centerline{\includegraphics[width=150mm]{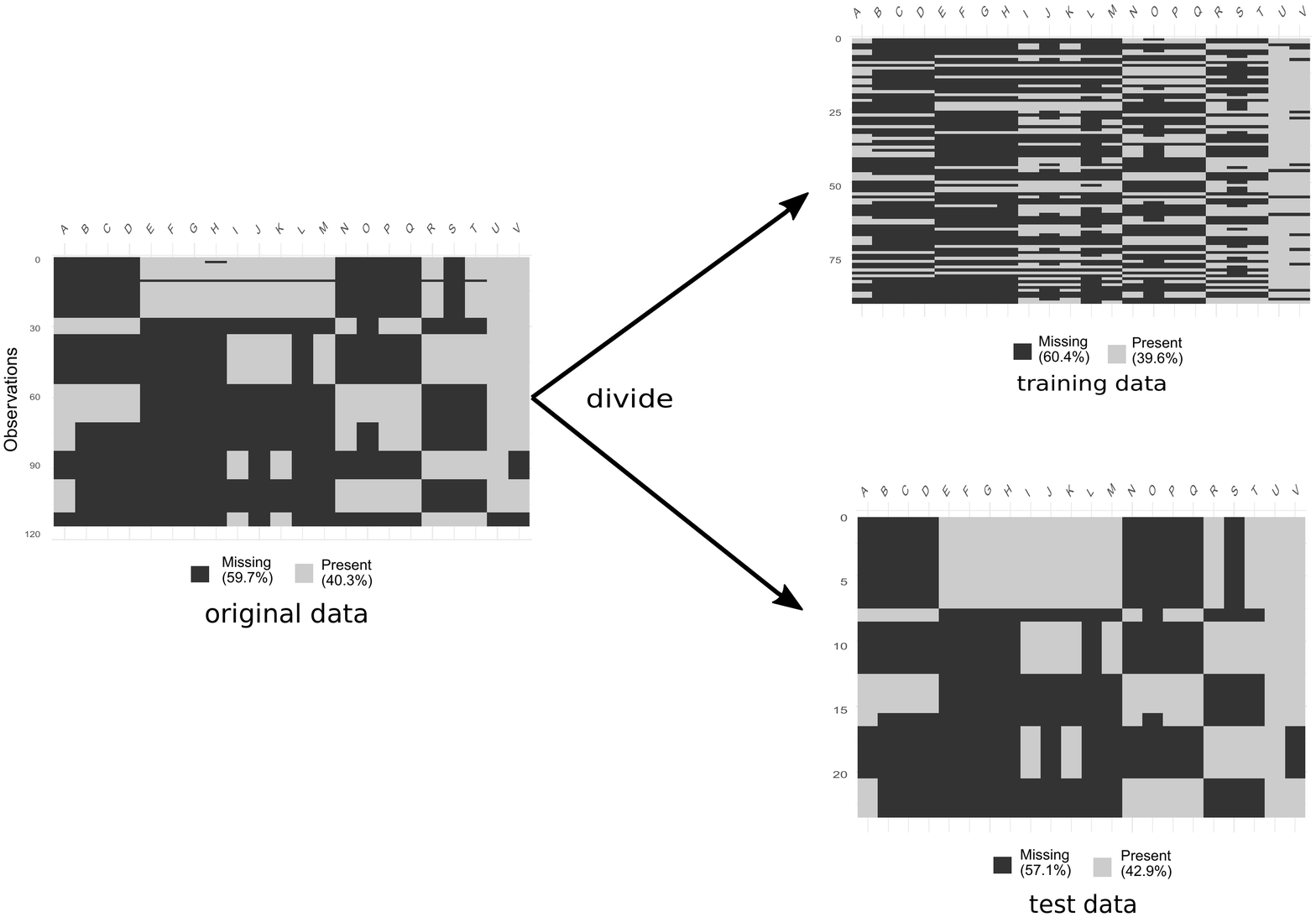}}
\caption{Left: Ratios of missing values for the original data. 
We use symbols ``A'' to ``V'' to represent particular mechanical characteristics of the polymer compounds.
Right: Ratios of missing values for training and test data, which is divided from the original data.}
\label{fig:ratio1}
\end{center}
\end{figure}

In this section, we apply the SMRM algorithm to real data provided by Toray Industries, Inc.
This data consists of physical/mechanical properties of particular polymer compounds. 
To maintain confidentiality, we cannot display the full dataset; however, 
we describe the size and components of the data. 
The sample size of the data is $n=114$, and the number of predictors and responses are 
$p=26$ and $q=22$, respectively.
Predictor variables consist of compounding ratios of the source materials. 
Response variables consist of mechanical characteristics 
created by the source materials, 
such as Young's modulus, tensile strength, elongation at break, flexural modulus, flexural strength 
and the Charpy impact strength of polymer compounds. 
Responses have missing values, of which the rates range from $5$\%  to $80$ \% for each observation. 
In particular, the total ratio of missing values in the responses is $59.7$\% 
which is typical in materials science field due to the development process of focusing on the specific properties
(see Figure \ref{fig:ratio1}).

During data analysis, we apply the SMRM algorithm to the data and  
compare the prediction accuracy of our proposed method with that of the {LASSO}.   
In this study, we use R version 4.0.2. 

\subsection{The procedure}
Let $\x_j=(x^1_{j},\ldots,x^n_{j})^T$ $(1\leq j\leq p)$ be predictor varieties, and let $\y_l=(y^1_{l},\ldots,y^n_{l})^T$ $(1\leq l\leq q)$ be response varieties. 
Then, {we divide the original data into $\text{training data}:\text{test data}=8:2$; 
that is, we partition $\x_j$ and $\y_l$ into $\x_j^T=(\x_{j,\train}^T,\x_{j,\test}^T)$ and $\y_l^T=(\y_{l,\train}^T,\y_{l,\test}^T)$, respectively,
with a partition rate of $8:2$ for each $j$ and $l$ (see Figure \ref{fig:ratio1})}. 
Although $\y_{l,\train}$ and $\y_{l,\test}$ may have missing values, 
we assume that $\x_j$, $\x_{j,\train}$, and $\x_{j,\test}$ are complete data. 
We perform the following analysis to compare the SMRM algorithm and the LASSO: 

\begin{itemize}
\item[Step 1:] 
For each $l$, we set $\y_{l,\train,\obs}$ and $\X_{l,\train,\obs}$, where $\y_{l,\train,\obs}$ is a vector, 
of which the elements are observed values in $\y_{l,\train}$, and $\X_{l,\train,\obs}$ is the matrix corresponding to $\y_{l,\train,\obs}$. 
Then, we apply the LASSO to the data set $(\y_{l,\train,\obs},\wtil{\X}_{\train,\obs})$ for each $l$. 
Regularization parameters, such as $\lambda_{l,\train}$ $(1\leq l \leq q)$, are chosen via cross-validation. 
The prediction values, $\hat{\y}_{l,\test,\obs}$, are then computed. 
For each $l$, we calculate the mean squared errors for the LASSO estimation, $\mse_l^{\lasso}$, for each $l$ 
using 
\begin{equation}\label{eq:l-mse}
\mse_l^{\lasso}=\dfrac{||\y_{l,\test,\obs}-\hat{\y}_{l,\test,\obs}^{\mathrm{lasso}}||^2}{\mathrm{length}(\y_{l,\test,\obs})},
\end{equation}
where $||\bm{w}||^2=\bm{w}^T\bm{w}$ for $\bm{w}\in\R^{d}$. 
\item[Step 2:]
We assume the multivariate linear regression model $\Y_{\text{train}}=\wtil{\X}_{\text{train}}\wtil{\B}_{\train}+\bm{E}_{\train}$ 
for the training data, where $\bm{\varepsilon}_{l,\train}\sim N(\bm{0},\Sig_{\train})$ ($1\leq l\leq q$). 
Then, we apply the SMRM method to the training data $(\wtil{\X}_{\train},\Y_{\train})$ 
for the appropriate pair $(\lambda_1,\bm{\lambda}_2)$, where $\bm{\lambda}_{2}=(\lambda_{2,l}^{j})_{1\leq j\leq p,1\leq l\leq q}$ 
is a matrix with elements that are defined based on the regularization parameters $\lambda_{l,\train}$. 
Then, we obtain the estimator $\what{\wtil{\B}}_{\text{train}}$ of parameter $\wtil{\B}_{\text{train}}$. 
Using $\what{\wtil{\B}}_{\text{train}}$, we can compute the matrix of the prediction value of $\Y_{\test}$, $\hat{\Y}_{\test}^{\smrm}$. 
We remark that $\hat{\Y}_{\test}^{\smrm}$ is {\it complete} data, whereas $\Y_{\test}$ is data that has missing values. 
We calculate $\mse^{\smrm}_l$ using 
\begin{equation}\label{eq:s-mse}
\mse_l^{\smrm}=\dfrac{||\y_{l,\test,\obs}-\hat{\y}_{l,\test,\obs}^{\smrm}||^2}{\mathrm{length}(\y_{l,\test,\obs})}
\end{equation}
for each $l$, similarly to the LASSO. 
\item[Step3:]  
We set $\mse^{\lasso}$ and $\mse^{\smrm}$ using 
$$\mse^{\lasso}=\sum_{l=1}^q\mse_l^{\lasso},\quad 
\mse^{\smrm}=\sum_{l=1}^q\mse_l^{\smrm}.$$ 
The above MSEs are primarily affected by response variables with large variances. 
Thus, we define the MSEs that are not affected by the variance of the response variables as follows:
\begin{equation}\label{eq:mod-mse}
\wtil{\mse}^{\lasso}=\sum_{l=1}^q(\mse_l^{\lasso})^{-1}\mse^{\lasso}_l(=q),\quad 
\wtil{\mse}^{\smrm}=\sum_{l=1}^q(\mse_l^{\lasso})^{-1}\mse_l^{\smrm}.
\end{equation}
Then, we compare $\wtil{\mse}^{\lasso}$ and $\wtil{\mse}^{\smrm}$. 
\end{itemize}

\begin{rem}\label{rem:log}
Because the SMRM algorithm is based on the multivariate normal distribution, 
the predicted $\hat{\Y}_{\test}^{\mathrm{SMRM}}$ contains negative values. 
However, the physical property $\y_l$ cannot assume negative values in a real-world situation. 
To avoid this, we first set $\log(\Y)$ and consider it as the response matrix. 
Then, applying $\exp(\log(\hat{\Y}))$ to the predicted matrix $\log(\hat{\Y})$, we have $\hat{\Y}$. 
\end{rem}

In Step 2, we use the $\bm{\lambda}_2$ matrix for the SMRM algorithm. 
If the responses are standardized, we define the $\bm{\lambda}_{2}$ matrix as $\bm{\lambda}_2=r\bm{\lambda}$, 
where $r\in\R\setminus\{0\}$ and 
\begin{equation}\label{eq:mat-lambda2}
\bm{\lambda}=\left.\begin{pmatrix}
\lambda_{1,\train} & \cdots & \lambda_{q,\train}\\
\vdots & \ddots & \vdots\\
\lambda_{1,\train} & \cdots & \lambda_{q,\train}
\end{pmatrix}\right\}p.
\end{equation}

However, when the responses are not standardized, 
the variance of the response variable affects the regularization parameter; 
$\bm{\lambda}_2$ must be different among the response variables. 
Thus, we conduct the following procedure to reduce the effect of the variances: 
\begin{itemize}
\item[Step 1:] For each $l\in\{1,\ldots,q\}$, we estimate $\y_{l,\train,\obs}$ using the LASSO 
with the regularization parameter $\lambda_{l,\train}$, which is chosen via cross-validation. 
\item[Step 2:] For each $l$, we calculate the MSEs, 
$t_l=||{\y}_{l,\train,\obs}-\hat{\y}_{l,\train,\obs}^{\mathrm{lasso}}||^2/\mathrm{length}(\y_{l,\train,\obs})$, for the training data, 
where $\hat{\y}_{l,\train,\obs}^{\mathrm{lasso}}$ is the estimator for $\y_{l,\train,\obs}$ by the LASSO in Step 1. 
\item[Step 3:] Using $t_l$, which was obtained in Step 2, we define a vector $\bm{a}$ as 
\begin{equation}\label{eq:vec-a}
\bm{a}=\begin{pmatrix} a_1 \\ \vdots \\ a_{q} \end{pmatrix}=\begin{pmatrix} t_1^{-1} \\ \vdots \\ t_{q}^{-1} \end{pmatrix}.
\end{equation}
\item[Step 4:] We define a matrix as
\begin{equation}\label{eq:mat-lambda}
\bm{\lambda}=
\left.\begin{pmatrix}
\lambda_{1,\train}a_1 & \lambda_{2,\train}a_2 & \cdots & \lambda_{q,\train}a_q\\
\vdots & \vdots & \ddots & \vdots\\
\lambda_{1,\train}a_1& \lambda_{2,\train}a_2 & \cdots & \lambda_{q,\train}a_q
\end{pmatrix}\right\}p.
\end{equation}
\item[Step 5:] We set $\bm{\lambda}_{2}=r\bm{\lambda}$ $(r\in\R\setminus\{0\})$ and apply the SMRM algorithm using this matrix.
\end{itemize}

\subsection{Comparison between the SMRM algorithm and the LASSO}
Following the procedure that we explained in the previous subsection, 
we compare our method (the SMRM algorithm) to the LASSO using the data provided by Toray Industries, Inc. 
Regularization parameters $\lambda_{l,\train}$ $(1\leq l\leq 22)$ for the training data and 
MSEs ($\mse_l^{\lasso}$) for the test data by the LASSO are summarized in the first and second rows of Table \ref{tab:lasso1}. 
\begingroup
\scalefont{0.6}
\begin{table}[t]
\centering
\caption{List of $\lambda_{l,\train}$, $\mse^{\lasso}_l$ elements of $\bm{a}$ and elements of $\bm{\lambda}$ for each mechanical characteristic.}
\begin{tabular}{|r|rrrrrrrrrrr|}
  \hline
Mech. Char. & A & B & C & D & E & F & G & H & I  & J & K \\ 
  \hline
$\lambda_{l,\train}$ & 1.3055 & 0.1899 & 3.3045 & 9.2906 & 2.5037 & 1.1784 & 23.8220 & 310.7017 & 0.9794 & 0.0976 & 22.6453 \\ 
  $\mse^{\lasso}_l$ & 445.3167 & 0.6529 & 0.4304 & 88.7628 & 4.6077 & 0.1835 & 5873.6704 & 40911.2809 & 4.0712 & 0.0368 & 374.3443 \\   
 elements of $\bm{a}$& 159.71 & 541.05 & 18.02 & 8.02 & 2114.25 & 12.76 & 21.94 & 23.55 & 2575.31 & 24.63 & 8.00\\
  elements of $\bm{\lambda}$  & 0.22 & 17.34 & 17.20 & 14.24 & 80.39 & 0.37 & 5.54 & 20.93 & 32.77 & 1.82 & 4.13  \\
  \hline
\end{tabular}
\begin{tabular}{|r|rrrrrrrrrrr|}
  \hline
Mech. Char.  & L & M & N & O & P & Q & R & S & T & U & V \\ 
  \hline
$\lambda_{l,\train}$ & 2.2927 & 53.0732 & 4.6505 & 0.3983 & 0.0208 & 0.5416 & 1.6233 & 0.0035 & 0.0365 & 0.1239 & 2.6903 \\ 
 $\mse^{\lasso}_l$ & 1490.9948 & 866.9672 & 1799.3309 & 0.4578 & 0.1281 & 4.2093 & 9.8900 & 0.0149 & 0.0046 & 3.6632 & 1358.2532 \\ 
  elements of $\bm{a} $& 15.25 & 6.57 & 169.18 & 18.09 & 1816.92 & 63.96 & 3385.57 & 6670.35 & 1950.88 & 39.45 & 9.43 \\
  elements of $\bm{\lambda}$ & 0.33 & 6.09 & 1.27 & 20.02 & 2.66 & 2.13 & 46.68 & 4.00 & 28.08 & 0.30 & 0.16 \\ 
       \hline
\end{tabular}
\label{tab:lasso1}
\end{table}
\endgroup
The second row of Table \ref{tab:lasso1} shows that the mechanical characteristics 
A, G, H, K, L, M, N, and V have large MSE values. 
These values significantly act on $\mse^{\lasso}$ and also act on $\mse^{\smrm}$. 
Therefore, it is better to use $\wtil{\mse}$, as in \eqref{eq:mod-mse}, to compare the prediction accuracy 
between the LASSO and the SMRM algorithm by avoiding the dependence of the variance of responses. 

We subsequently apply the SMRM algorithm. 
Because the responses of the data are not standardized, 
we calculate the vector $\bm{a}$, as in \eqref{eq:vec-a} (this is a $22\times1$ matrix), to obtain the $\bm{\lambda}_2$ matrix. 
The vector $\bm{a}$ is listed in the third row of Table \ref{tab:lasso1}. 
Using the regularization parameters $\lambda_{l,\train}$ $(1\leq l\leq 22)$, as in the first row of Table \ref{tab:lasso1} and $\bm{a}$, 
we obtain the $\bm{\lambda}$ matrix, as given in \eqref{eq:mat-lambda}.
The row vector of $\bm{\lambda}$ is shown in the fourth row of Table \ref{tab:lasso1}.

The values in the third row of Table \ref{tab:lasso1} correspond to the reciprocals of the MSEs for each mechanical characteristic of the training data obtained by the LASSO. 
These MSE values vary significantly. 
The values in the fourth row of Table \ref{tab:lasso1} can be considered as modified regularization parameters obtained by the LASSO in Step 1. 
A comparison of the first and fourth rows of Table \ref{tab:lasso1} shows that the values in the fourth row vary only slightly.
Thus, one may make a stable estimation using the SMRM algorithm with $\bm{\lambda}_2$ constructed by \eqref{eq:mat-lambda} 
instead of the matrix given by \eqref{eq:mat-lambda2}. 

We consider the following cases of the pair $(\lambda_1,\bm{\lambda}_2)$ for the SMRM algorithm: 
\begin{itemize}
\item For $\lambda_1$, we consider $6.5\times10^{-3}\leq\lambda_1\leq 1$ divided into $200$ points of equal length under the log scale.
\item We consider $\bm{\lambda}_2$ as $\bm{\lambda}_2=r\bm{\lambda}$ for $r=3,2,1,0.75,0.5,0.225,0.2,0.175,0.1$. 
\end{itemize}
As we apply the SMRM algorithm, we use the {\it warm start} method, which is outlined as follows. 
Let $\{\lambda_1^{(s)}\}_{s=1}^{200}$ be a sequence of $\lambda_1$,
of which the initial value is $\lambda_1^{(1)}=1$, and the end is $\lambda_1^{(200)}=6.5\times 10^{-3}$. 
For a fixed $\bm{\lambda}_2$, we start with $(\lambda_1^{(1)},\bm{\lambda}_2)=(1,\bm{\lambda}_2)$. 
Then, we obtain $\what{\wtil{\B}}^{(1)}$ and $\what{K}^{(1)}$ via the SMRM algorithm. 
Next, we apply the SMRM algorithm to the pair of regularization parameters $(\lambda_1^{(2)},\bm{\lambda}_2)$ 
with the initial values $\wtil{\B}=\what{\wtil{\B}}^{(1)}$ and $K=\what{K}^{(1)}$, which were obtained in the previous step. 
Inductively, we practice a similar analysis until $\lambda_1^{(200)}=6.5\times 10^{-3}$.

In our observation, the values of $\wtil{\mse}^{\smrm}$ improve gradually whenever $\lambda_1^{(s)}$ is updated 
for a fixed $r\leq1$ (see Figure \ref{fig:mse} and Figure \ref{fig:tilde_mse} in Appendix \ref{append}). 
However, when $\lambda_1$ becomes smaller than a particular number, the SMRM algorithm is not stable, and 
$\wtil{\mse}^{\smrm}$ deteriorates. 
The following reasons can be considered: 
\begin{itemize}
\item When the regularization parameter $\lambda_1$ for the SMRM algorithm is large, 
the correlation structure ise not considered. 
In this case, imputation of missing values may not be effective 
because missing completion is achieved by taking advantage of the correlation structure of the responses.
Hence, the prediction accuracy may be worse than that of the LASSO.
\item When $\lambda_1$ is appropriately small, missing values are well imputed using the correlation structure among the responses (or among the residuals). 
The prediction accuracy improves, owing to the contribution of the correlation structure. 
\item When $\lambda_1$ is too small, the estimated model overfits the data. 
Hence, the prediction error increases again.
\end{itemize} 
In the case of $r>1$, $\wtil{\mse}^{\smrm}$ deteriorates independent of $\lambda_1$. 
When $r>1$, elements in $\bm{\lambda}_2$ assume large values. 
For these data, the elements of the row vectors of the estimator $\what{\wtil{\B}}$ tend to be zeros. 
This is the reason why $\wtil{\mse}^{\smrm}$ is worse than the case of $r\leq1$. 

We observe the influence of $r$ on the prediction accuracy. 
As $r$ gradually decreases, 
the prediction accuracy for the SMRM algorithm improves. 
In particular, we observe that $r=0.2$, that is, $\bm{\lambda}_2=0.2\bm{\lambda}$, 
with $\log(\lambda_1)=-4.96$ provide the best prediction accuracy (see Figure \ref{fig:mse}). 
When $r=0.1$, the prediction accuracy decreases compared with the case where $r=0.2$ 
(see Figure \ref{fig:tilde_mse} in Appendix \ref{append}). 

Furthermore, comparing $\wtil{\mse}^{\smrm}$ with $\bm{\lambda}_2=0.2\bm{\lambda}$ and $\wtil{\mse}^{\lasso}$, 
it can be observed that $\wtil{\mse}^{\smrm}<\wtil{\mse}^{\lasso}$ holds; that is, the SMRM algorithm is superior to the LASSO for a suitably small $\lambda_1$. 
Because $\lambda_1$ affects the correlation structure among responses, 
the prediction accuracy may be improved 
by estimating responses multivariately with the appropriate correlation structure, instead of individually. 

\begin{figure}[t]
\begin{center}
\centerline{\includegraphics[width=100mm]{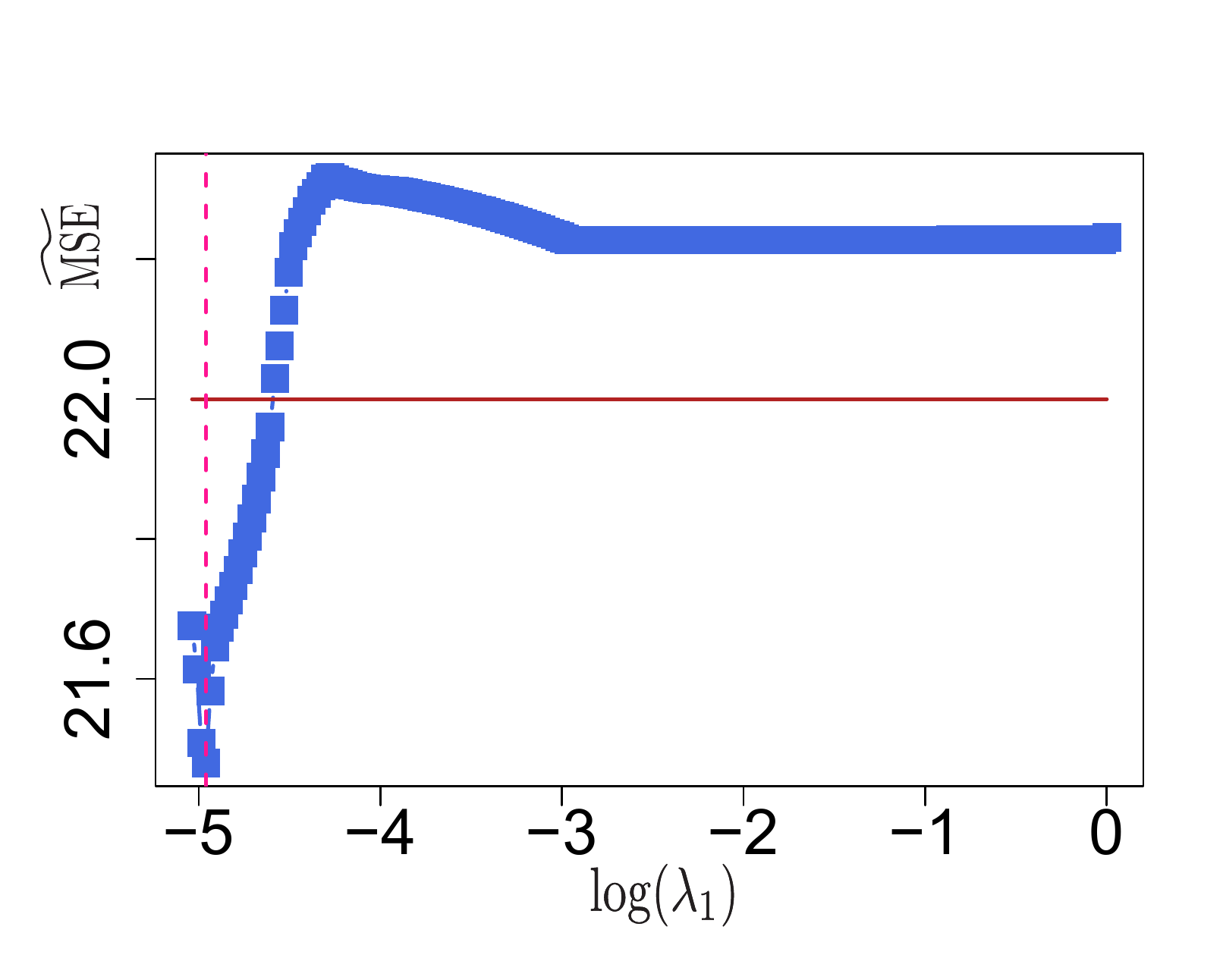}}
\caption{$\wtil{\mse}^{\smrm}$ (blue) and $\wtil{\mse}^{\lasso}$ (red). 
The $x$-axis represents $\log(\lambda_1)$ and the $y$-axis $\wtil{\mse}$. 
The dotted vertical line (pink) indicates $\log(\lambda_1)=-4.96$, which provides the best case.}
\label{fig:mse}
\end{center}
\end{figure}

\begin{figure}[t]
\begin{center}
    \begin{tabular}{c}

      \begin{minipage}{0.5\hsize}
        \begin{center}
          \centerline{\includegraphics[clip, width=6.5cm]{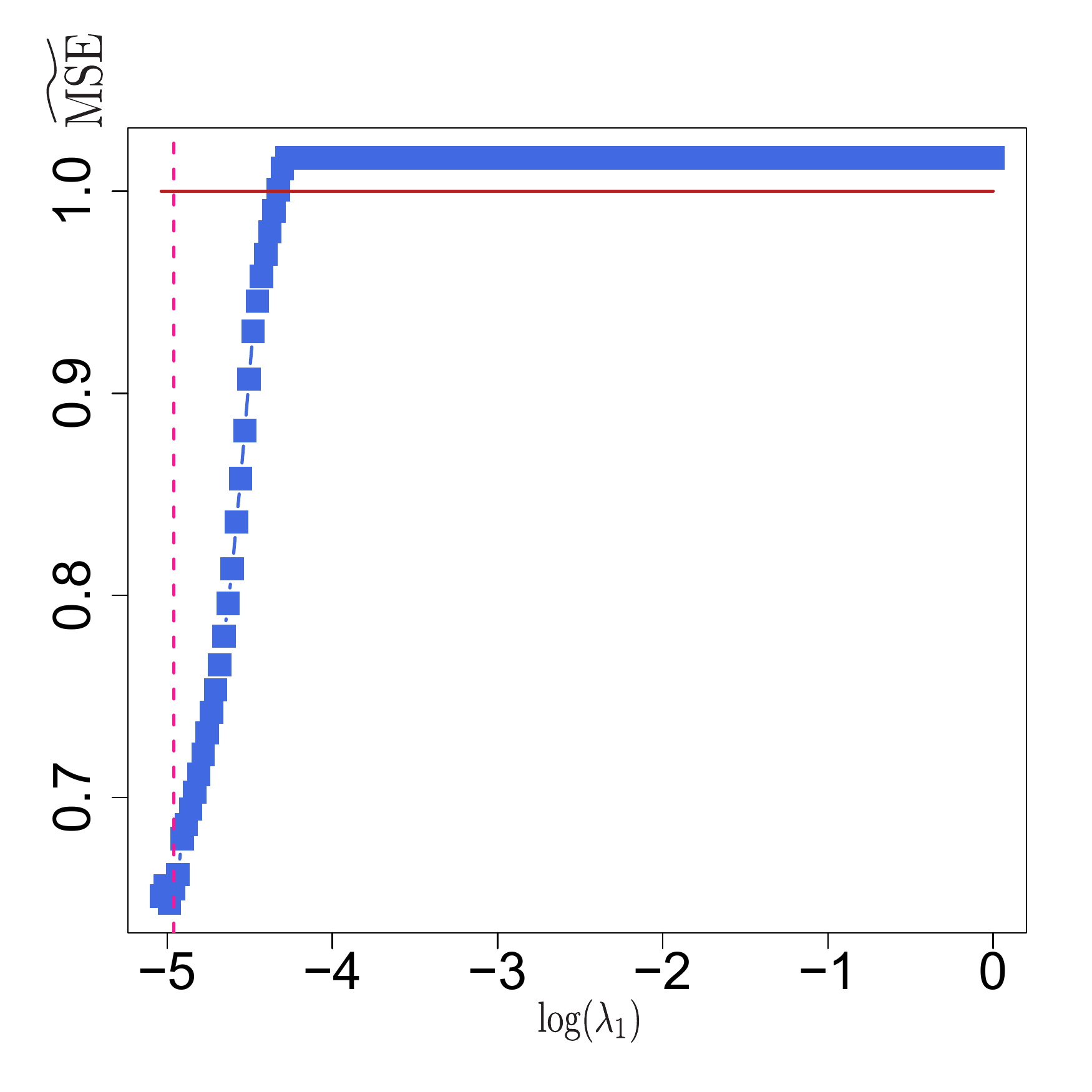}}
          \hspace{1cm} C
        \end{center}
      \end{minipage}

\begin{minipage}{0.5\hsize}
        \begin{center}
          \centerline{\includegraphics[clip, width=6.5cm]{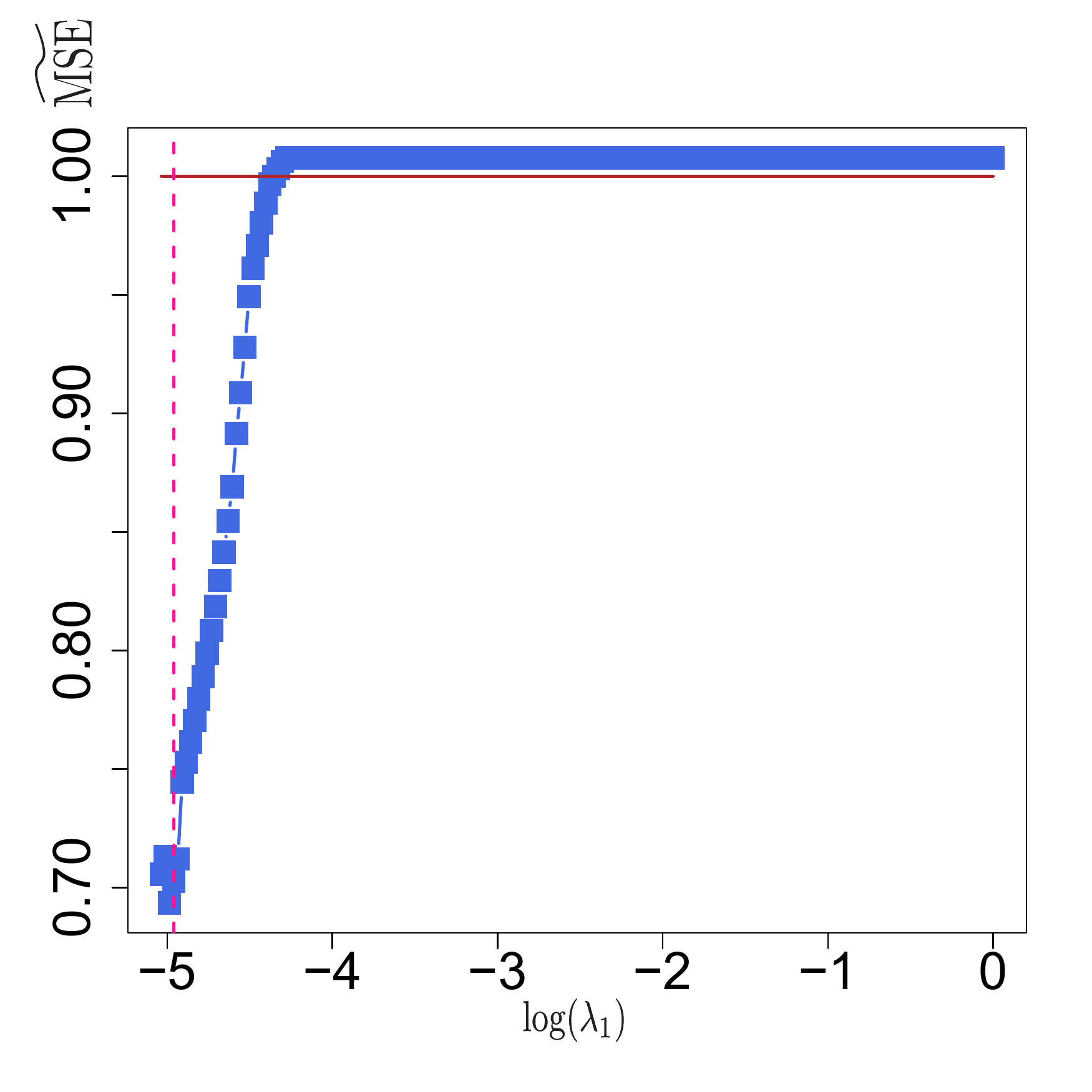}}
          \hspace{1cm} D
        \end{center}
      \end{minipage}
      \end{tabular}
      \caption{$\wtil{\mse}^{\smrm}$ (blue) and $\wtil{\mse}^{\lasso}$ (red) of mechanical characteristics C (left) and D (right). 
The $x$-axis represents $\log(\lambda_1)$.
The dotted vertical line (pink) represents $\log(\lambda_1)=-4.96$, which indicates the best case for $\wtil{\mse}^{\smrm}$.}
\label{fig:char_C}
\end{center}
\end{figure}

Next, we consider the MSEs in the case of $\bm{\lambda}_2=0.2\bm{\lambda}$ for mechanical characteristics individually 
(see Figure \ref{fig:char_mse}). 
In the individual analysis, we consider the mechanical characteristics C and D, i.e., the third and fourth mechanical characteristics (see Figure \ref{fig:char_C}).
For $-4.35\leq\log(\lambda_1)\leq 0$, $\wtil{\mse}^{\smrm}_l>\wtil{\mse}^{\lasso}_l(=1)$ holds, where $l=3,4$. 
This implies that the prediction accuracy obtained by the SMRM algorithm is lower than that of the LASSO. 
However, for $\log(\lambda_1)\leq-4.38$, we find that $\wtil{\mse}^{\smrm}_l<\wtil{\mse}^{\lasso}_l$ ($l=3,4$) holds. 
Thus, it can be observed that the prediction accuracy is improved by using the correlation structure among the responses. 
Although $\log(\lambda_1)=-4.96$ exhibits the best $\wtil{\mse}^{\smrm}$ prediction accuracy, as mentioned above, 
$\wtil{\mse}^{\smrm}_3$ and $\wtil{\mse}^{\smrm}_4$ provide the best values
for $\log(\lambda_1)=-4.99$ (one after `the best' with respect to $\wtil{\mse}$). 

\begin{figure}[t]
  \begin{center}
    \begin{tabular}{c}

\begin{minipage}{0.5\hsize}
        \begin{center}
          \includegraphics[clip, width=6.5cm]{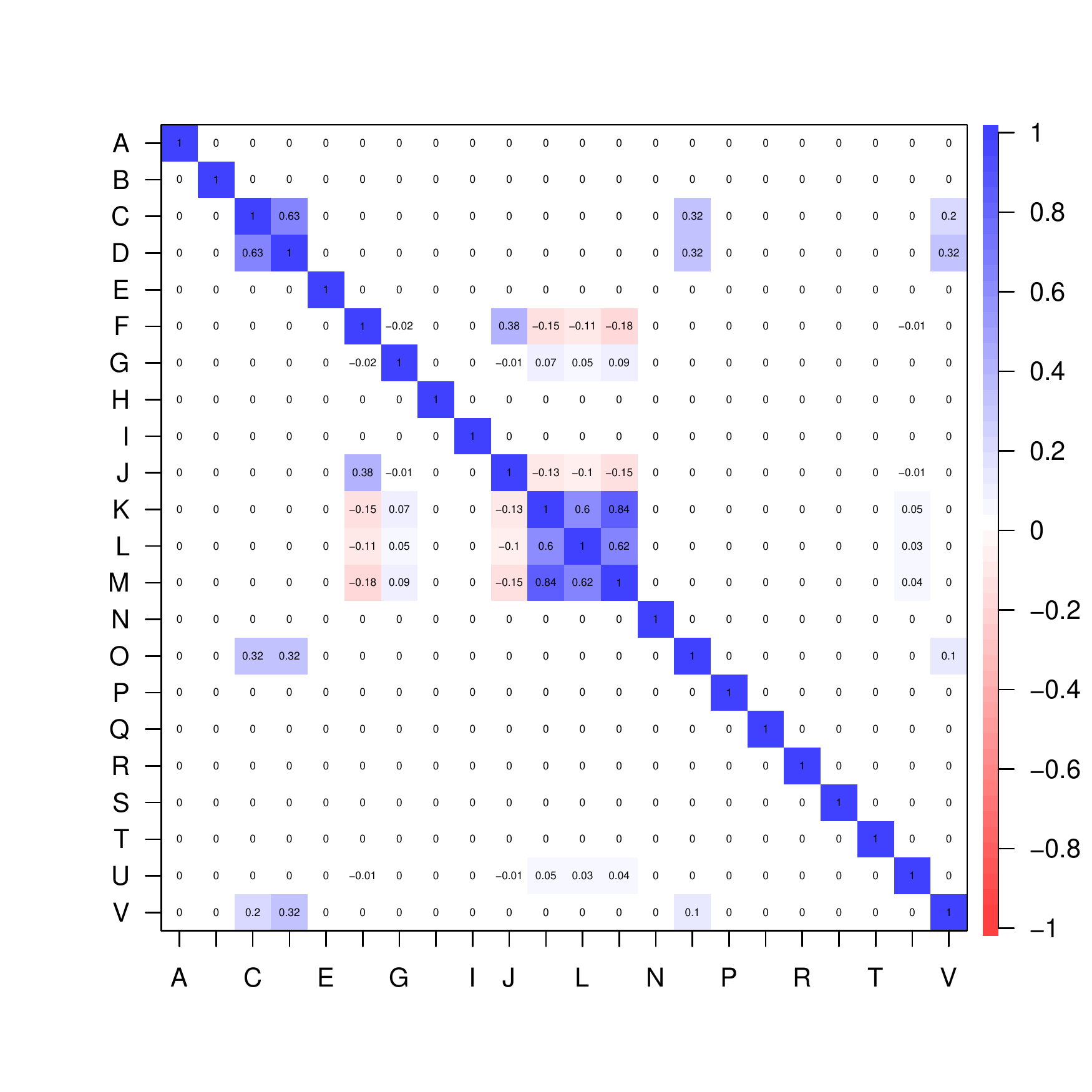}
          \hspace{1cm} [i] better1 (one before the best)
        \end{center}
      \end{minipage}
      \begin{minipage}{0.5\hsize}
        \begin{center}
          \includegraphics[clip, width=6.5cm]{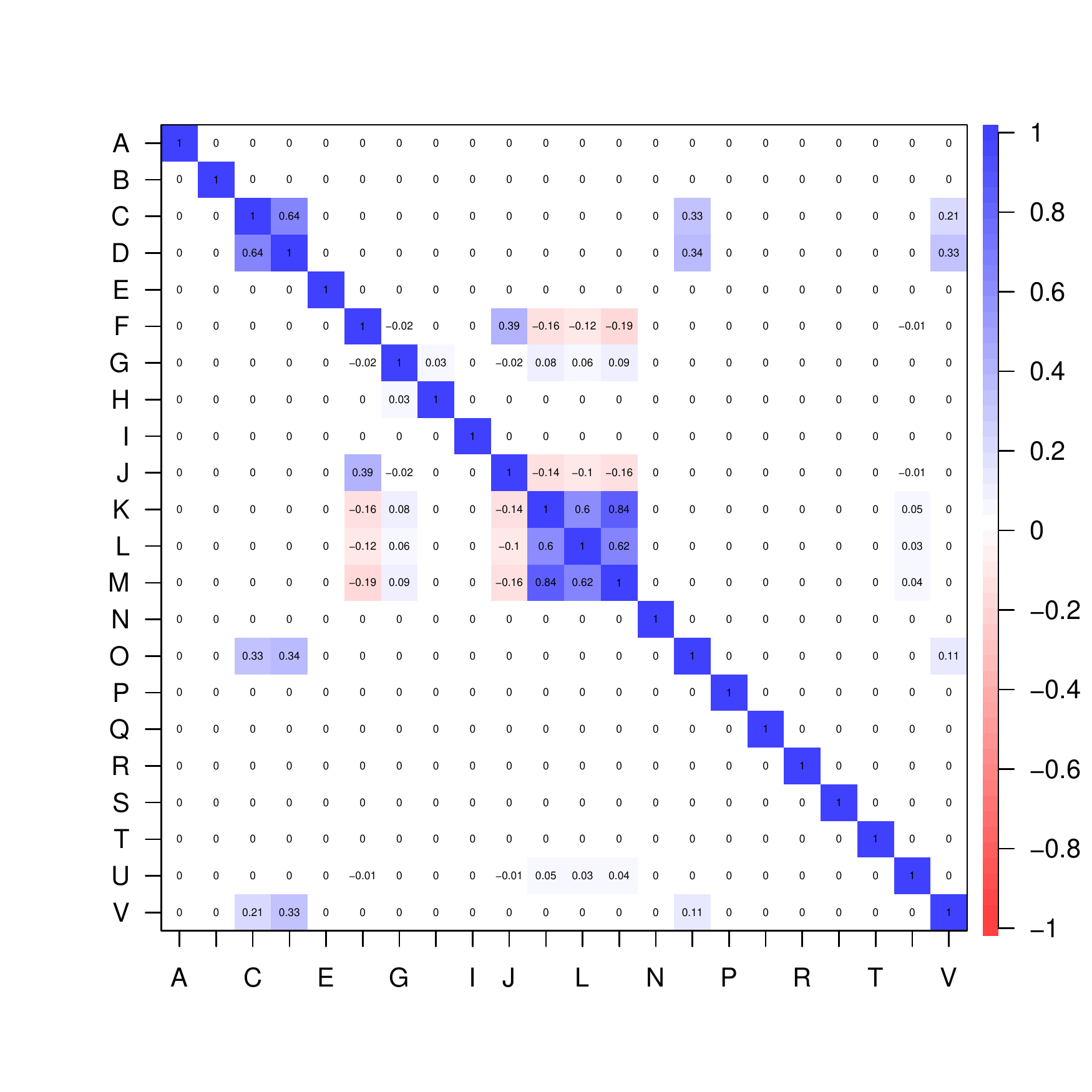}
          \hspace{1cm}[ii] the best
        \end{center}
      \end{minipage}\\

		      \begin{minipage}{0.5\hsize}
        \begin{center}
          \includegraphics[clip, width=6.5cm]{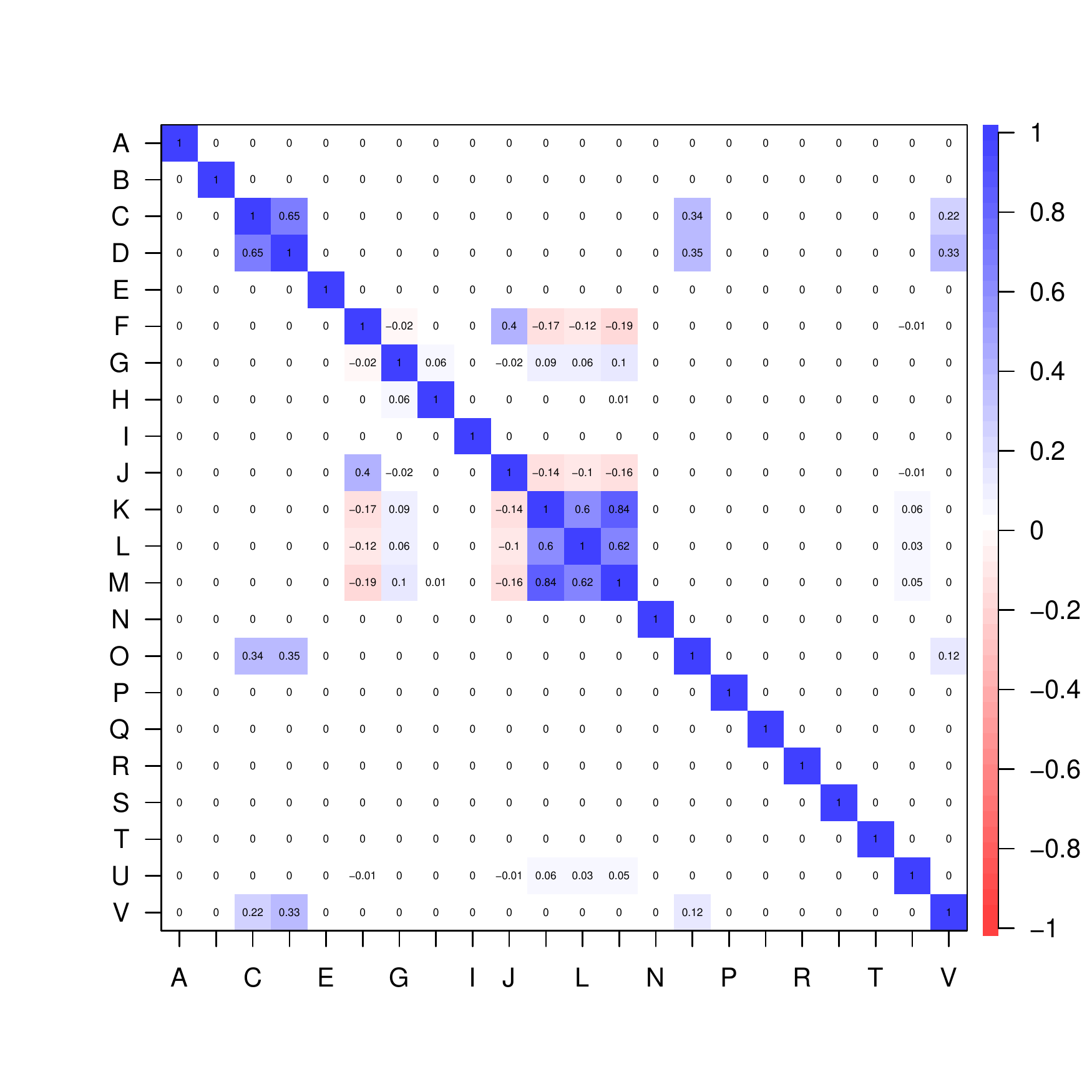}
          \hspace{1cm} [iii] better2 (one after the best)
        \end{center}
      \end{minipage}

      \begin{minipage}{0.5\hsize}
        \begin{center}
          \includegraphics[clip, width=6.5cm]{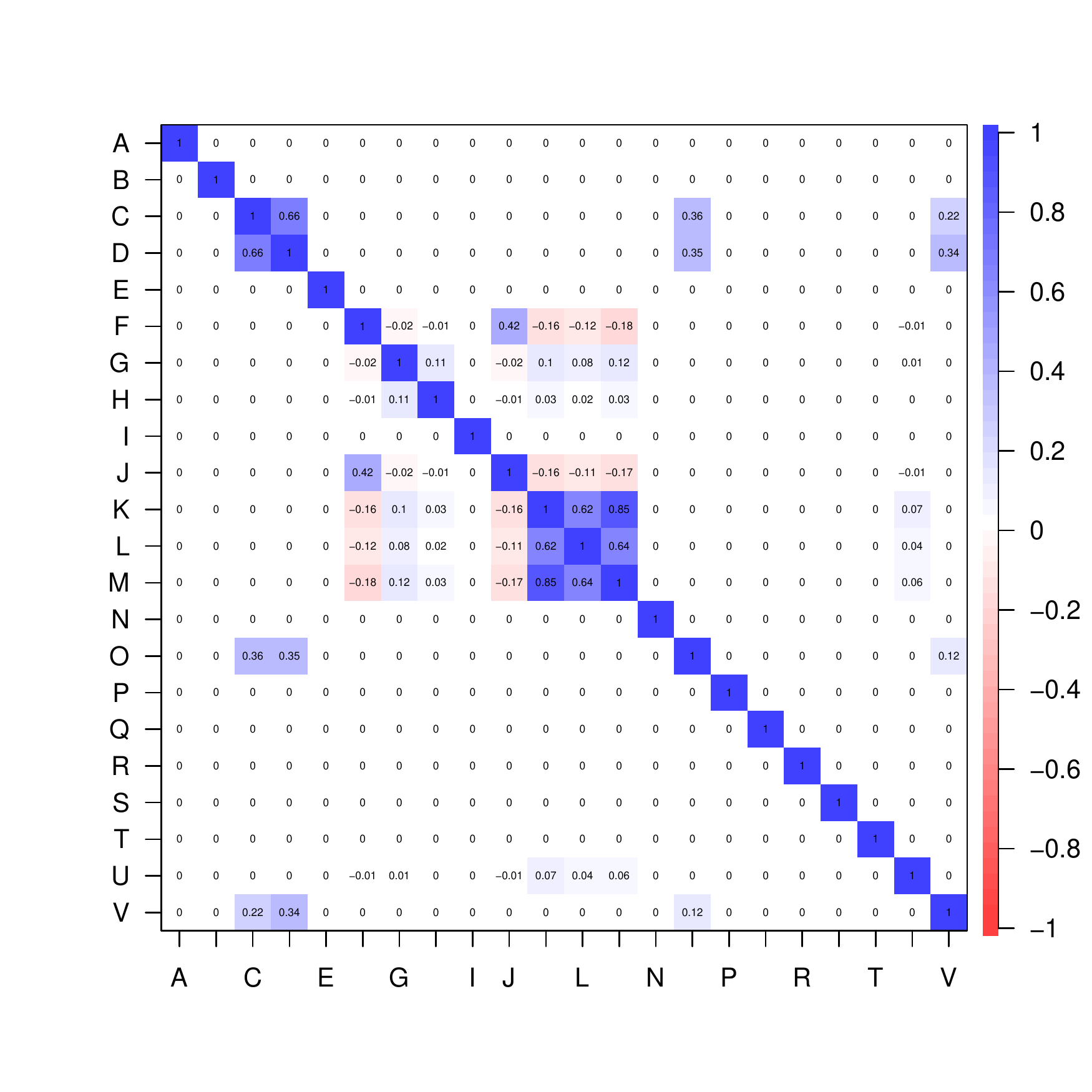}
          \hspace{1cm}[iv] last
        \end{center}
      \end{minipage}

    \end{tabular}
    \caption{Heat maps of correlation structures among responses for $\bm{\lambda}_2=0.2\bm{\lambda}$. 
    From top left to bottom right, we present heat maps for $\log(\lambda_1)=-4.93$, $-4.96$, 
    $-4.99$, $-5.04$.}
    \label{fig:heat}
  \end{center}
\end{figure}

To consider the reason for these differences, we use heat maps of correlation structures (see Figure \ref{fig:heat}). 
Here, `the best,' `better1,' and `better2' are named with respect to $\wtil{\mse}^{\smrm}$. 
(When we observe $\wtil{\mse}_l$ individually, note that `better2' represents the case for the best prediction accuracy of C and D.) 
We first note that C and D have a particular physical/mechanical relationship with each other 
\footnote{This is indicated by K. Nomura, S. Kobayashi, and K. Koyanagi, who provided this data.}. 
It is noticeable that C and D have a strong positive correlation. 
Therefore, it appears that if a positive correlation between C and D gradually increases, 
the prediction accuracy improves. 

\begin{figure}[htbp]
  \begin{center}
  \includegraphics[clip,width=6.5cm]{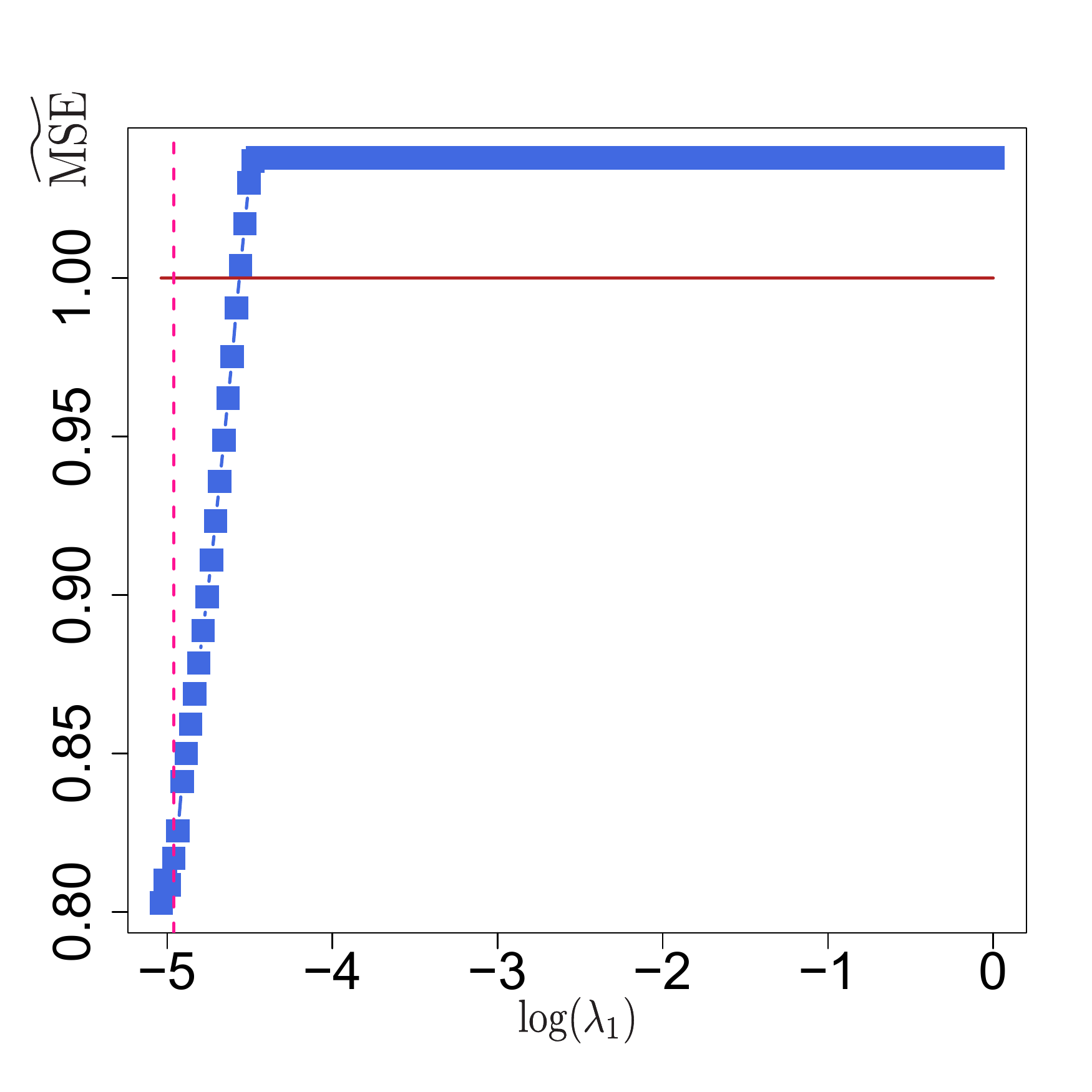}
    \caption{$\wtil{\mse}^{\smrm}$ (blue) and $\wtil{\mse}^{\lasso}$ (red) of the mechanical characteristic O. 
The $x$-axis represents $\log(\lambda_1)$. The dotted vertical line (pink) represents $\log(\lambda_1)=-4.96$.}
    \label{fig:char_O}
  \end{center}
\end{figure}

On the other hand, the SMRM algorithm estimates 
the correlation structure among C, D, and O (fifteenth mechanical characteristic).
Since C (resp. D) and O represent different mechanical properties, 
it is difficult to emphasize their correlation structure by experiments. 
Therefore this may be considered as a hidden relation among mechanical properties, 
and it seems that the sparse multivariate regression method using a precision matrix 
contributes to identifying such relations. 
If a positive correlation among C, D, and O is suitably set, 
the prediction accuracy of C and D improve. 
For the mechanical characteristic O, 
we notice that the prediction accuracy is better when the positive correlation 
of O with C and D increases (see Figure \ref{fig:char_O}). 

By the above observations, a suitable positive correlation structure among C, D, and O for this dataset
affects the prediction accuracy of these mechanical characteristics. 
Furthermore, we can identify unexpected relations among responses, 
similar to the above characteristics, using our method.

For other mechanical characteristics, see Appendix \ref{append}. 

\section{Conclusion}\label{sec:conclusion}
In this study, we proposed a novel method called the sparse multivariate regression with missing values (SMRM) algorithm 
with the intention to apply it to materials science. 
Since the data structure of materials science often contains two features; 
(1) material properties are multivariate, and  
(2) they have often missing values, 
it seems that the MRCE and the MissGLASSO work effectively. 
Unfortunately, these methods cannot apply directly to our setting. 
However, by modifying these and establishing suitable framework, 
we constructed the proposed algorithm (Section \ref{sec:SMRM}). 
Owing to the regularization for the correlation structure, 
we can improve the prediction accuracy and may find the unexpected relation among the response variables.
Actually, in the real data analysis, we found the unexpected relation 
among response variables of data (Section \ref{sec:application}). 
Further, we verified that our proposed method was superior to the LASSO for the data.  
For these reasons, we expect that our proposed method has a possibility to contribute 
to the progress of materials science and related area. 

Our proposed procedure performed worse than the lasso for some material properties.  
The poor performance is probably due to a large number of missing values.  
As the ratio of missing values increases, the prediction accuracy of our proposed method becomes poor.  
As future work, it would be interesting to investigate the inﬂuence of the ratio of missing values on our proposed method.

\begin{acknowledgements}
The authors are grateful to Keiichiro Nomura, Sadayuki Kobayashi, and Kohei Koyanagi for fruitful discussions and valuable advise. 
They also express their gratitude to Professors Keiji Tanaka, Satoru Yamamoto, Shigeru Kuchii, and Shigeru Taniguchi for their valuable comments. 
\end{acknowledgements}

\bibliographystyle{plain}
\bibliography{references}

\begin{thebibliography}{10}

\bibitem{experiment-improve}
H.~Blockeel and J.~Vanschoren.
\newblock Experiment databases: Towards an improved experimental methodology in
  machine learning.
\newblock In {\em European Conference on Principles of Data Mining and
  Knowledge Discovery}, pages 6--17. Springer, 2007.

\bibitem{applied}
R.~J. Brook and G.~C. Arnold.
\newblock {\em Applied regression analysis and experimental design}.
\newblock CRC Press, 1985.

\bibitem{buck}
S.~F. Buck.
\newblock A method of estimation of missing values in multivariate data
  suitable for use with an electronic computer.
\newblock {\em J. Roy Statist. Soc. Ser. B.}, 22:302--306, 1960.

\bibitem{molecular}
K.~T. Butler, D.~W. Davies, H.~Cartwright, O.~Isayev, and A.~Walsh.
\newblock Machine learning for molecular and materials science.
\newblock {\em Nature}, 559(7715):547--555, 2018.

\bibitem{multivariate}
C.~Chatfield and A.~Collins.
\newblock {\em Introduction to multivariate analysis}.
\newblock Chapman and Hall, 1980.

\bibitem{machine-composite}
C.-T. Chen and G.~X. Gu.
\newblock Machine learning for composite materials.
\newblock {\em MRS Communications}, 9(2):556--566, 2019.

\bibitem{parameter}
S.~Da~Ros, M.~Schwaab, and J.~C. Pinto.
\newblock Parameter estimation and statistical methods.
\newblock Elsevier, 2017.

\bibitem{FIML}
C.~K. Enders and D.~L. Bandalos.
\newblock The relative performance of full information maximum likelihood
  estimation for missing data in structural equation models.
\newblock {\em Struct. Equ. Model.}, 8(3):430--457, 2001.

\bibitem{sparse-glasso}
J.~Friedman, T.~Hastie, and R.~Tibshirani.
\newblock Sparse inverse covariance estimation with the graphical lasso.
\newblock {\em Biostatistics}, 9:432--441, 2008.

\bibitem{missingdata}
J.~W. Graham.
\newblock {\em Missing data}.
\newblock Statistics for Social and Behavioral Sciences. Springer, New York,
  2012.
\newblock Analysis and design.

\bibitem{hclkl2020}
S.~Z. Han, E.-A. Choi, S.~H. Lim, S.~Kim, and J.~Lee.
\newblock Alloy design strategies to increase strength and its trade-offs
  together.
\newblock {\em Prog. Mater. Sci.}, page 100720, 2020.

\bibitem{hirose2016FIML}
K.~Hirose, S.~Kim, Y.~Kano, M.~Imada, M.~Yoshida, and M.~Matsuo.
\newblock Full information maximum likelihood estimation in factor analysis
  with a large number of missing values.
\newblock {\em J. Stat. Comput. Simul.}, 86(1):91--104, 2016.

\bibitem{jyw2019}
Z.~Jia, Y.~Yu, and L.~Wang.
\newblock Learning from nature: Use material architecture to break the
  performance tradeoffs.
\newblock {\em Materials \& Design}, 168:107650, 2019.

\bibitem{graphical-model}
S.~L. Lauritzen.
\newblock {\em Graphical models}, volume~17 of {\em Oxford Statistical Science
  Series}.
\newblock The Clarendon Press, Oxford University Press, New York, 1996.
\newblock Oxford Science Publications.

\bibitem{high-lasso}
N.~Meinshausen and P.~B\"{u}hlmann.
\newblock High-dimensional graphs and variable selection with the lasso.
\newblock {\em The Annals of Statistics}, 34(3):1436--1462, 2006.

\bibitem{accel}
G.~Pilania, C.~Wang, X.~Jiang, S.~Rajasekaran, and R.~Ramprasad.
\newblock Accelerating materials property predictions using machine learning.
\newblock {\em Scientific reports}, 3(1):1--6, 2013.

\bibitem{materialinfo}
R.~Ramprasad, R.~Batra, G.~Pilania, A.~Mannodi-Kanakkithodi, and C.~Kim.
\newblock Machine learning in materials informatics: recent applications and
  prospects.
\newblock {\em Npj Comput. Mater.}, 3(1):1--13, 2017.

\bibitem{generalized}
V.~Roth.
\newblock The generalized {LASSO}.
\newblock {\em IEEE transactions on neural networks}, 15(1):16--28, 2004.

\bibitem{mrce}
A.~J. Rothman, E.~Levina, and J.~Zhu.
\newblock Sparse multivariate regression with covariance estimation.
\newblock {\em J. Comput. Graph. Statist.}, 19(4):947--962, 2010.
\newblock Supplementary materials available online.

\bibitem{sparse-lasso}
N.~Simon, J.~Friedman, T.~Hastie, and R.~Tibshirani.
\newblock A sparse-group lasso.
\newblock {\em J. Comput. Graph. Statist.}, 22(2):231--245, 2013.

\bibitem{missglasso}
N.~St\"{a}dler and P.~B\"{u}hlmann.
\newblock Missing values: sparse inverse covariance estimation and an extension
  to sparse regression.
\newblock {\em Stat. Comput.}, 22(1):219--235, 2012.

\bibitem{lasso1996}
R.~Tibshirani.
\newblock Regression shrinkage and selection via the lasso.
\newblock {\em J. Roy. Statist. Soc. Ser. B.}, 58(1):267--288, 1996.

\bibitem{experiment}
J.~Vanschoren, H.~Blockeel, B.~Pfahringer, and G.~Holmes.
\newblock Experiment databases.
\newblock {\em Machine Learning}, 87(2):127--158, 2012.

\bibitem{Wang2015}
J.~Wang.
\newblock Joint estimation of sparse multivariate regression and conditional
  graphical models.
\newblock {\em Statistica Sinica}, 25(3):831--851, 2015.

\bibitem{machine-appl}
J.~Wei, X.~Chu, X.-Y. Sun, K.~Xu, H.-X. Deng, J.~Chen, Z.~Wei, and M.~Lei.
\newblock Machine learning in materials science.
\newblock {\em InfoMat}, 1(3):338--358, 2019.

\bibitem{overlapping}
L.~Yuan, J.~Liu, and J.~Ye.
\newblock Efficient methods for overlapping group lasso.
\newblock {\em IEEE Trans. Pattern Anal. Machine Intell.}, 35(9):2104--2116,
  2013.

\bibitem{strategy}
Y.~Zhang and C.~Ling.
\newblock A strategy to apply machine learning to small datasets in materials
  science.
\newblock {\em Npj Comput. Mater.}, 4(1):1--8, 2018.

\bibitem{elasticnet}
H.~Zou and T.~Hastie.
\newblock Regularization and variable selection via the elastic net.
\newblock {\em J. Roy. Statist. Soc. Ser. B.}, 67(2):301--320, 2005.

\end{thebibliography}

\appendix
\renewcommand{\thefigure}{\Alph{section}.\arabic{figure}}
\setcounter{figure}{0}
\section{Figures of the prediction accuracy of real data}\label{append}
We show figures obtained by real data analysis (see Section \ref{sec:application}). 
We first show figures that $\wtil{\mse}^{\smrm}$ for $r=3,2,1,0.75,0.5,0.225,0.2,0.175,0.1$,where $\bm{\lambda}_2=r\bm{\lambda}$, 
and $\wtil{\mse}^{\lasso}$ in Figure \ref{fig:tilde_mse}. 
When $r>1$, $\wtil{\mse}^{\smrm}$ is quite inferior to $\wtil{\mse}^{\lasso}$. 
One can recognize that taking $r$ smaller and smaller, $\wtil{\mse}^{\smrm}$ is improved step by step 
and $\wtil{\mse}^{\smrm}$ reaches the best at $r=0.2$. 
After that, $\wtil{\mse}^{\smrm}$ gets worse again. 

\begingroup
\scalefont{0.7}
\begin{figure}[t]
  \begin{center}
    \begin{tabular}{c}

      \begin{minipage}{0.33\hsize}
        \begin{center}
          \includegraphics[clip, width=4cm]{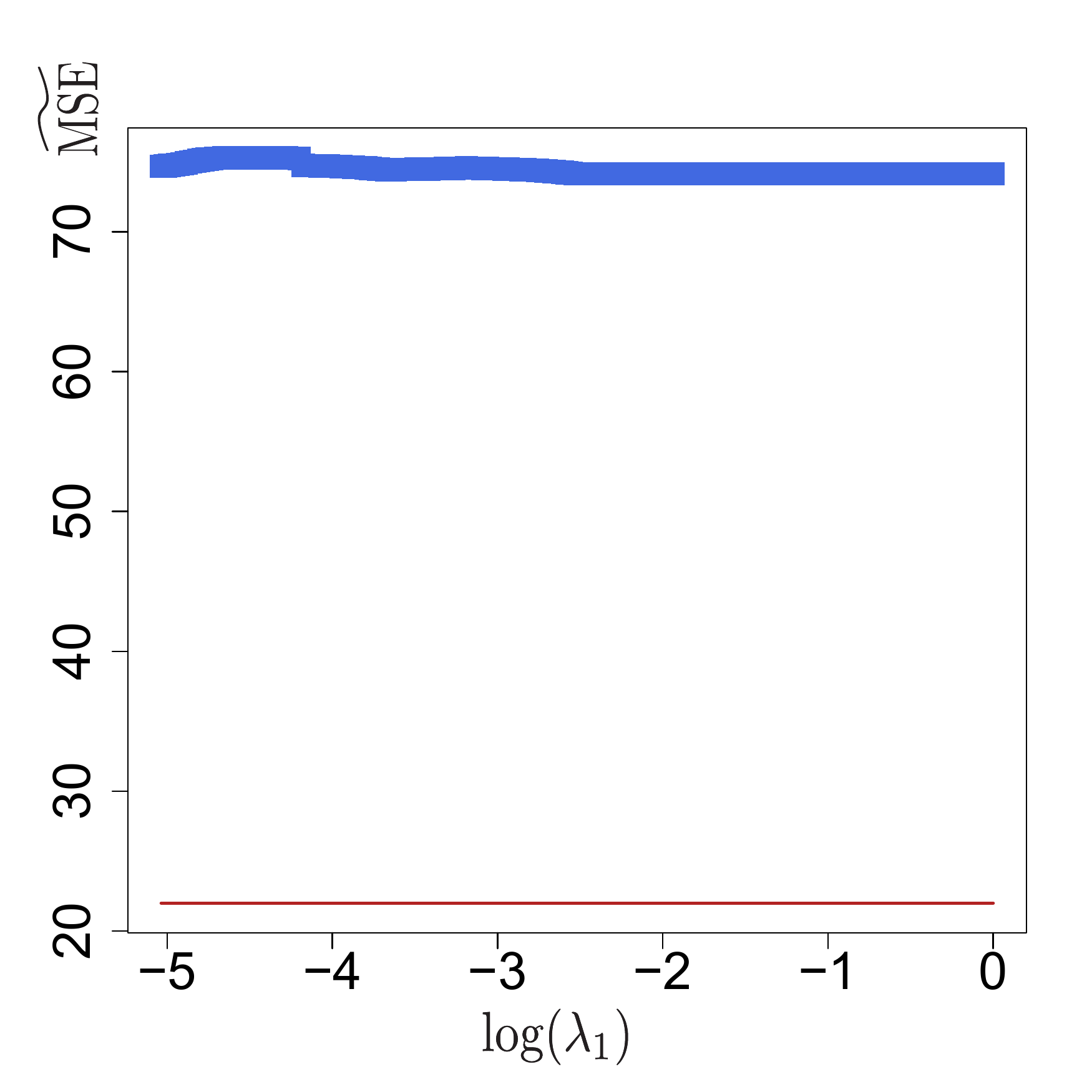}
         \hspace{1.6cm} \tiny{$r=3$}
        \end{center}
      \end{minipage}

      \begin{minipage}{0.33\hsize}
        \begin{center}
          \includegraphics[clip, width=4cm]{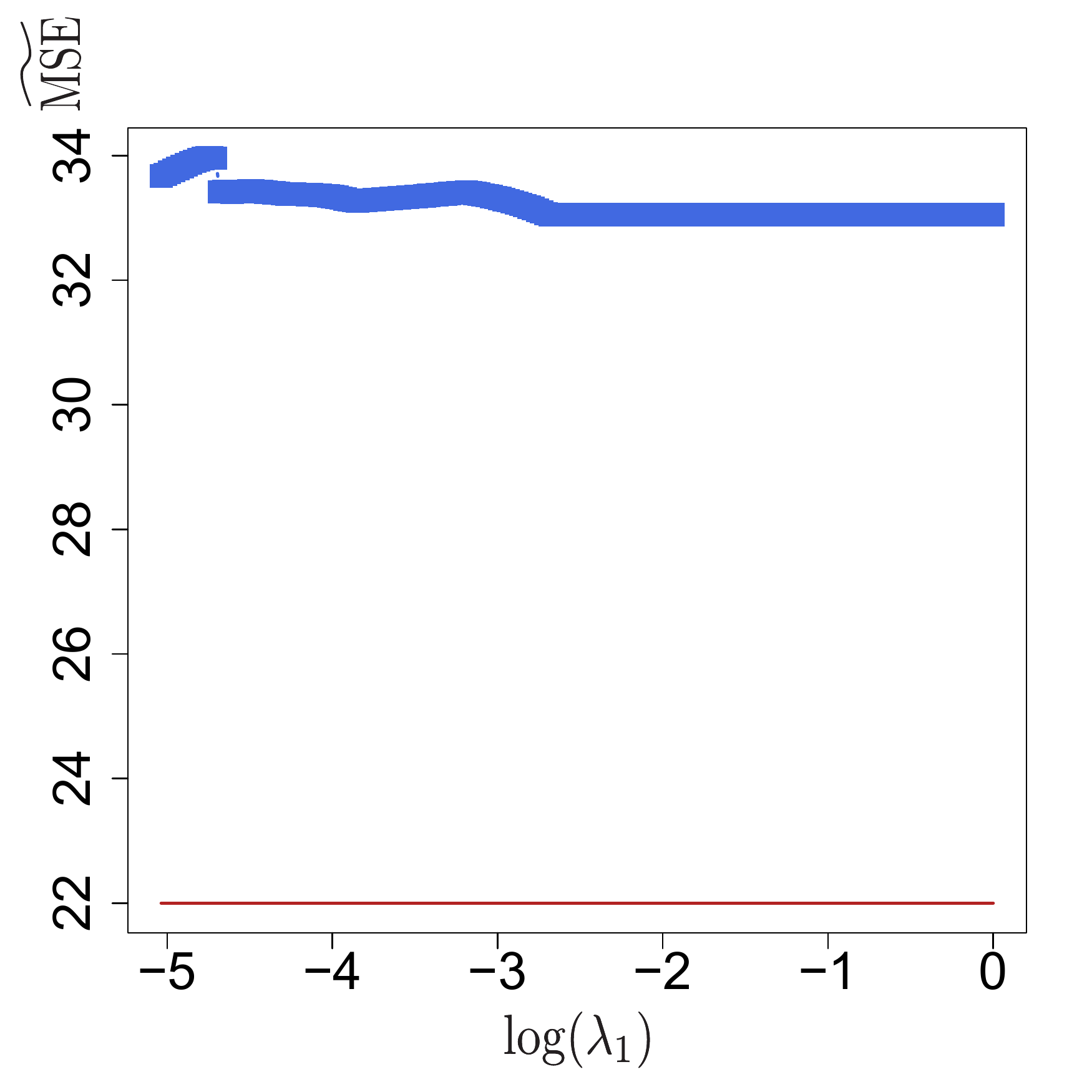}
          \hspace{1.6cm} \tiny{$r=2$}
        \end{center}
      \end{minipage}

      \begin{minipage}{0.33\hsize}
        \begin{center}
          \includegraphics[clip, width=4cm]{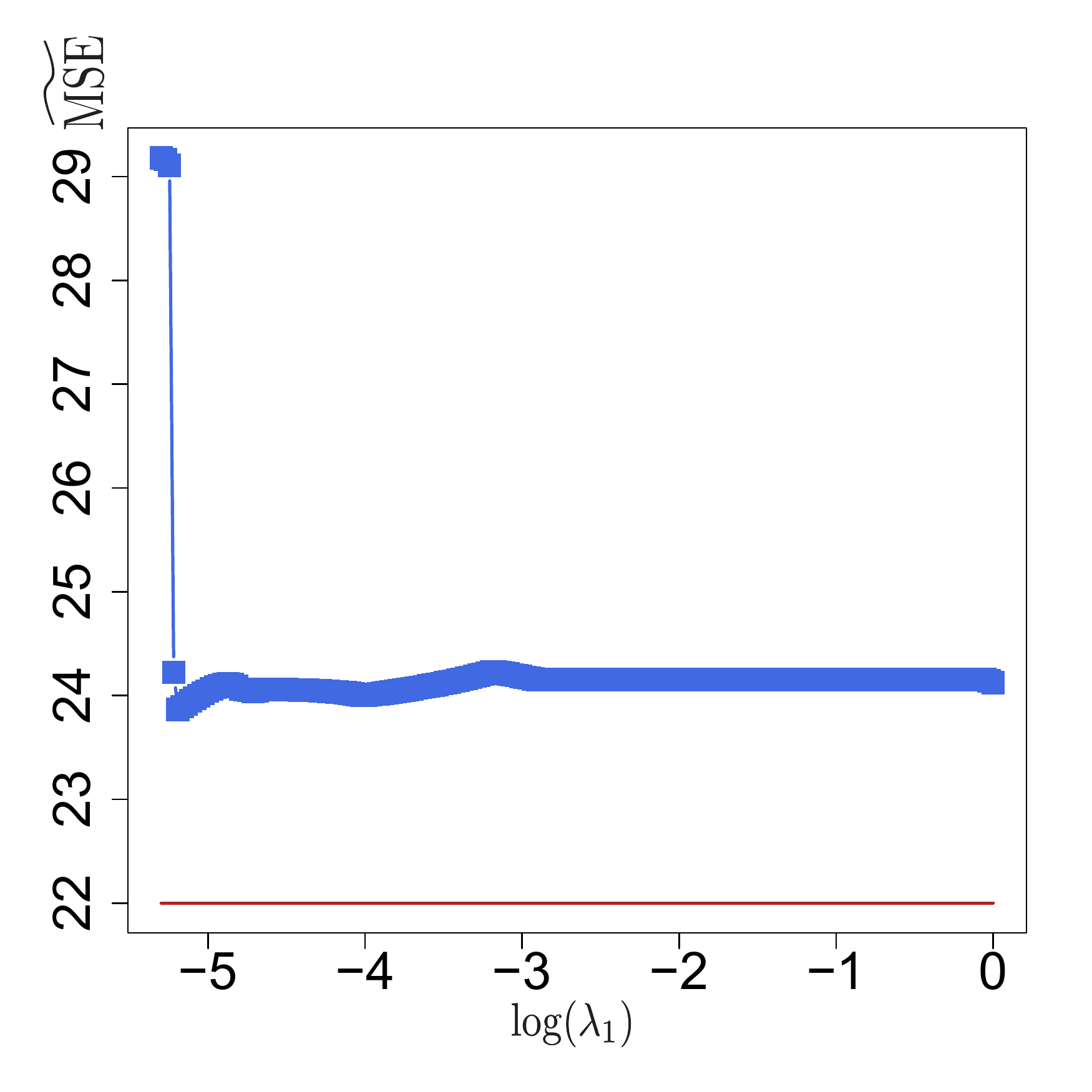}
          \hspace{1.6cm} \tiny{$r=1$}
        \end{center}
      \end{minipage} \\
      
      \begin{minipage}{0.33\hsize}
        \begin{center}
          \includegraphics[clip, width=4cm]{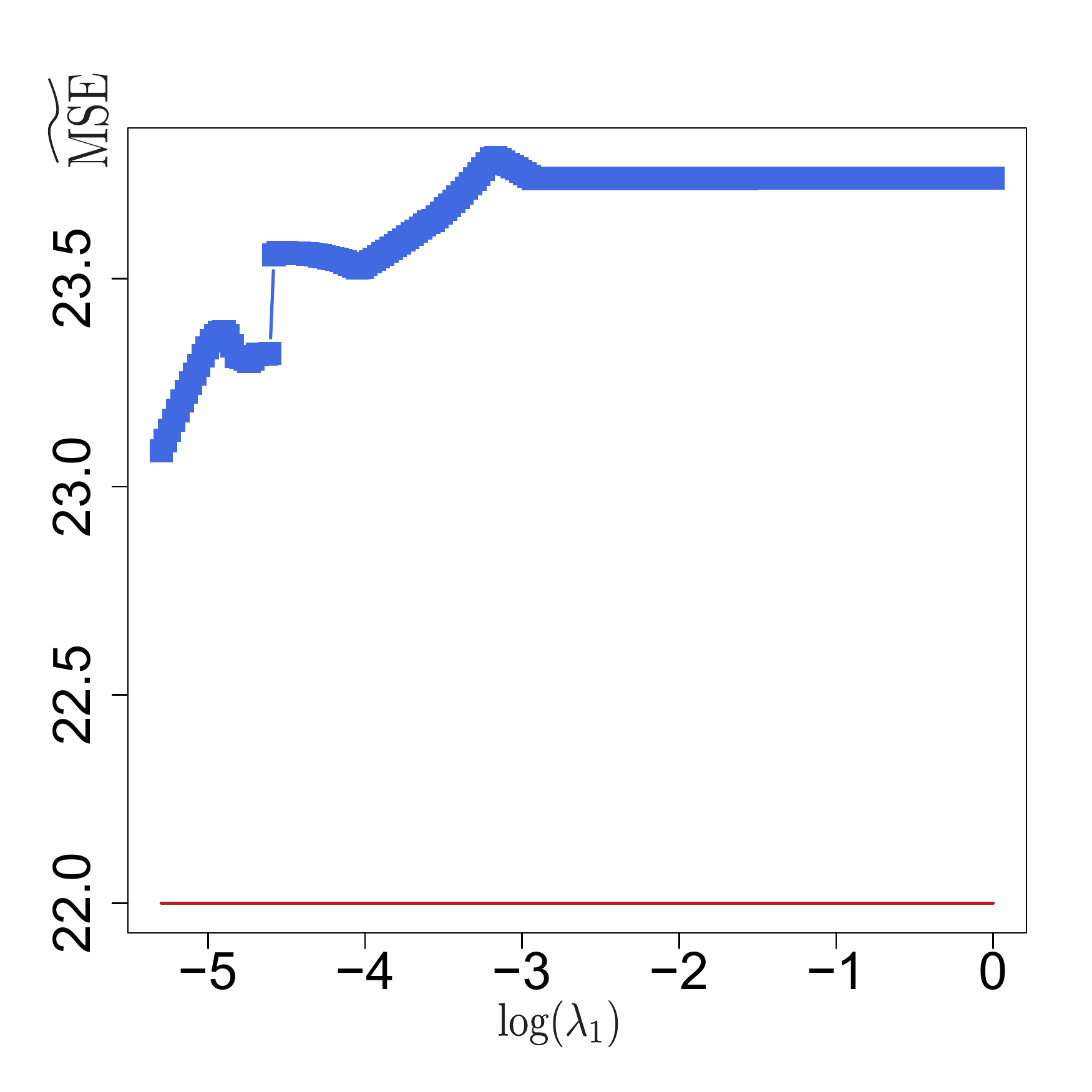}
          \hspace{1.6cm} \tiny{$r=0.75$}
        \end{center}
      \end{minipage}
      \begin{minipage}{0.33\hsize}
        \begin{center}
          \includegraphics[clip, width=4cm]{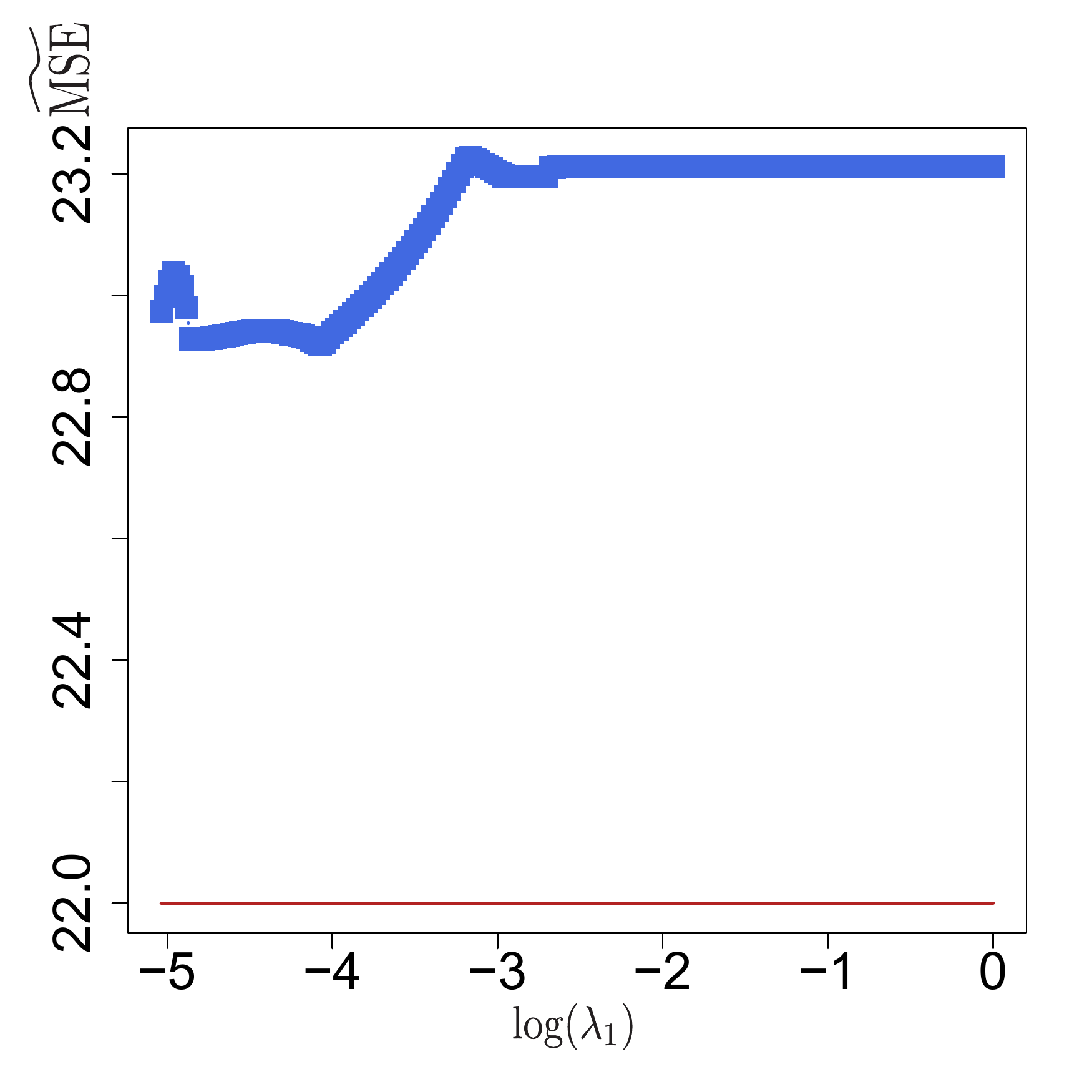}
         \hspace{1.6cm} \tiny{$r=0.5$}
        \end{center}
      \end{minipage}

      \begin{minipage}{0.33\hsize}
        \begin{center}
          \includegraphics[clip, width=4cm]{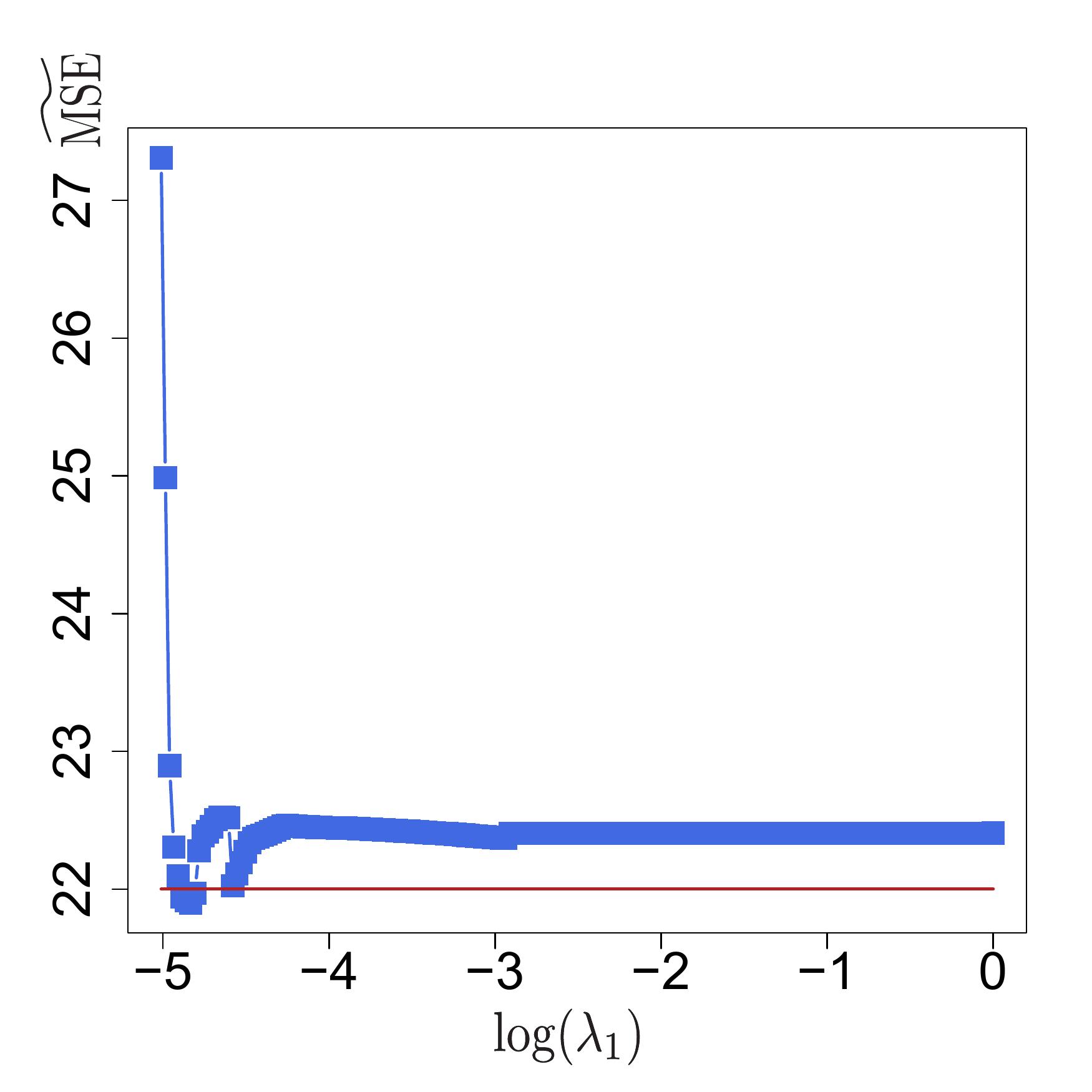}
         \hspace{1.6cm} \tiny{$r=0.225$}
        \end{center}
      \end{minipage}\\

      \begin{minipage}{0.33\hsize}
        \begin{center}
          \includegraphics[clip, width=4cm]{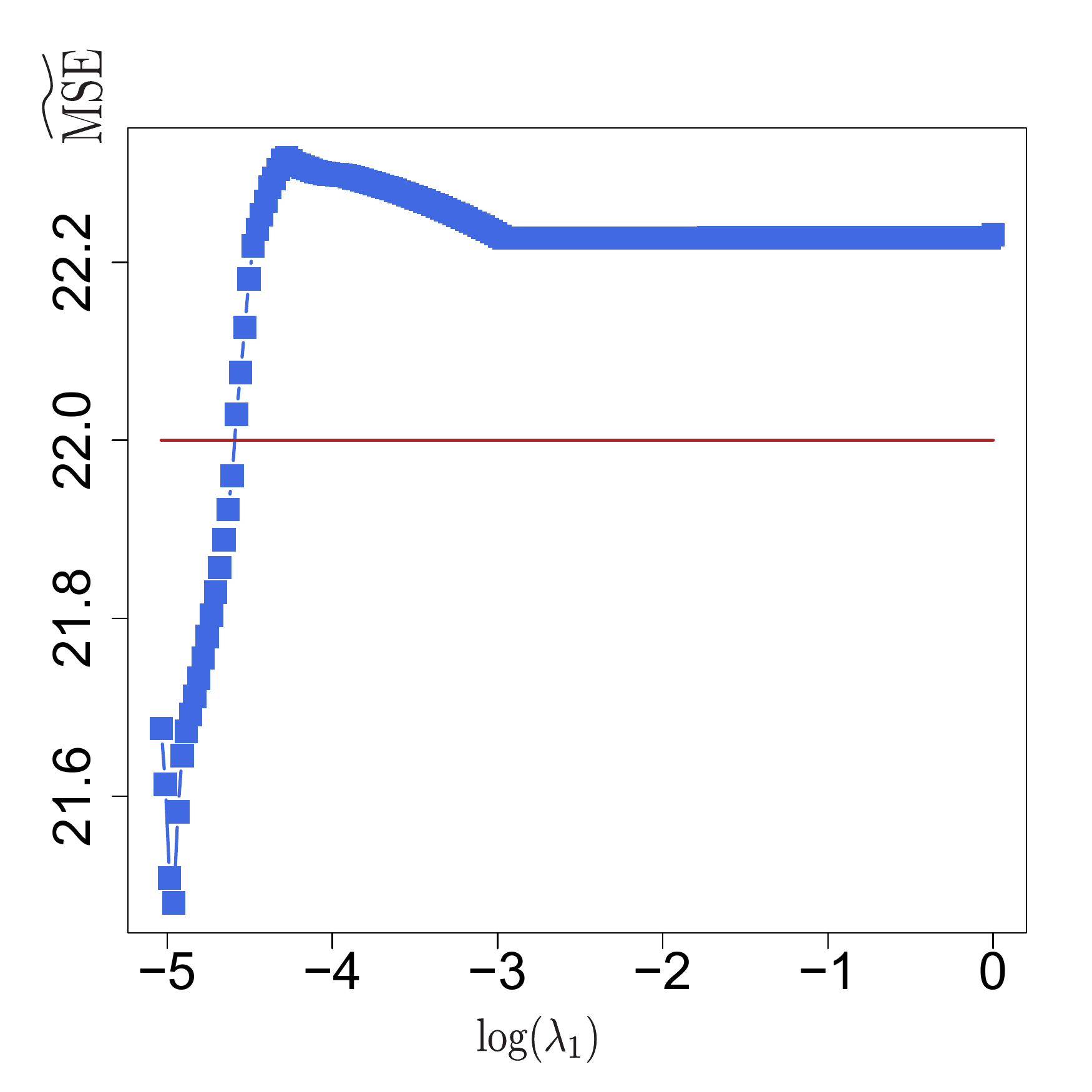}
          \hspace{1.6cm} \tiny{$r=0.2$}
        \end{center}
      \end{minipage} 
      
      \begin{minipage}{0.33\hsize}
        \begin{center}
          \includegraphics[clip, width=4cm]{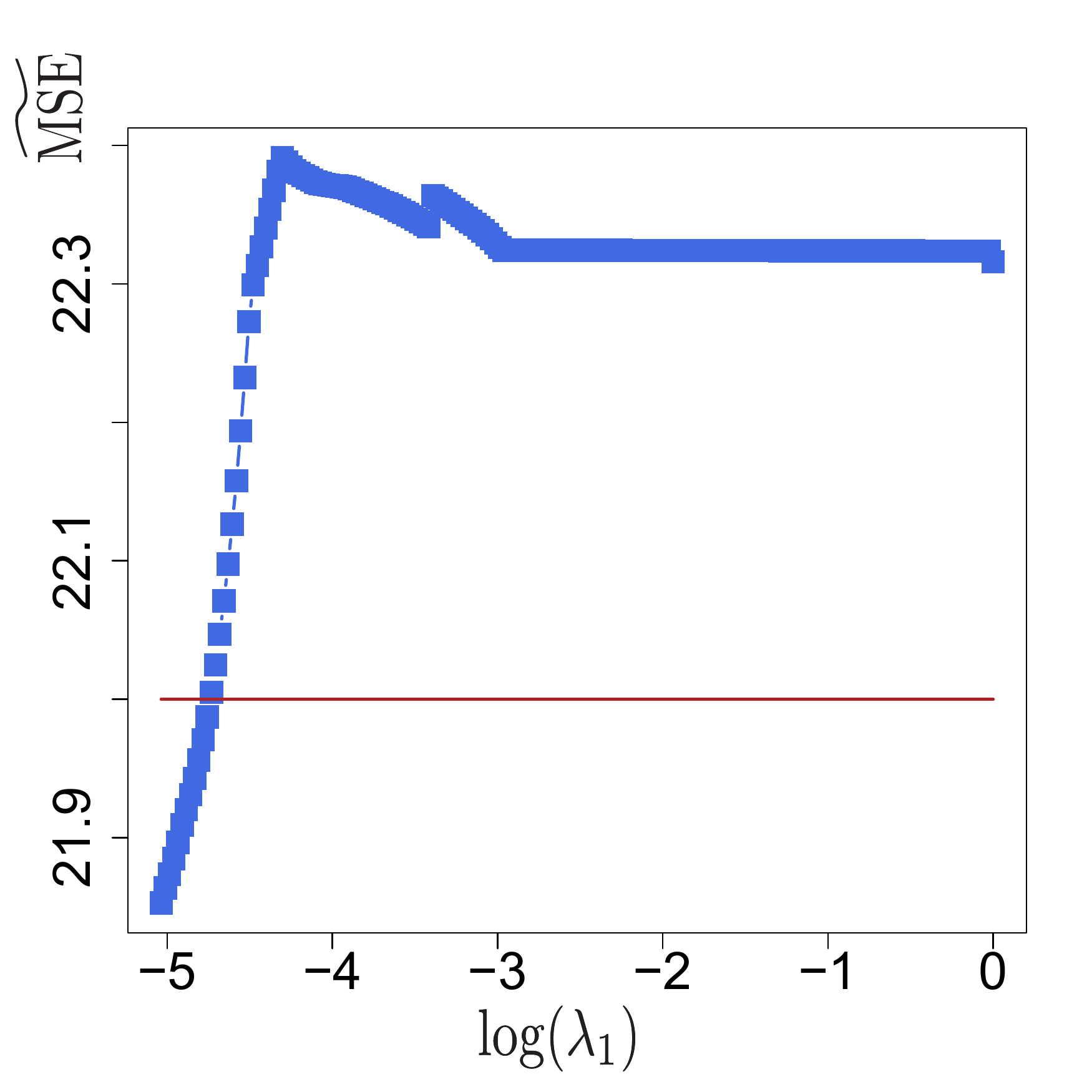}
          \hspace{1.6cm} \tiny{$r=0.175$}
        \end{center}
      \end{minipage}
      
      \begin{minipage}{0.33\hsize}
        \begin{center}
          \includegraphics[clip, width=4cm]{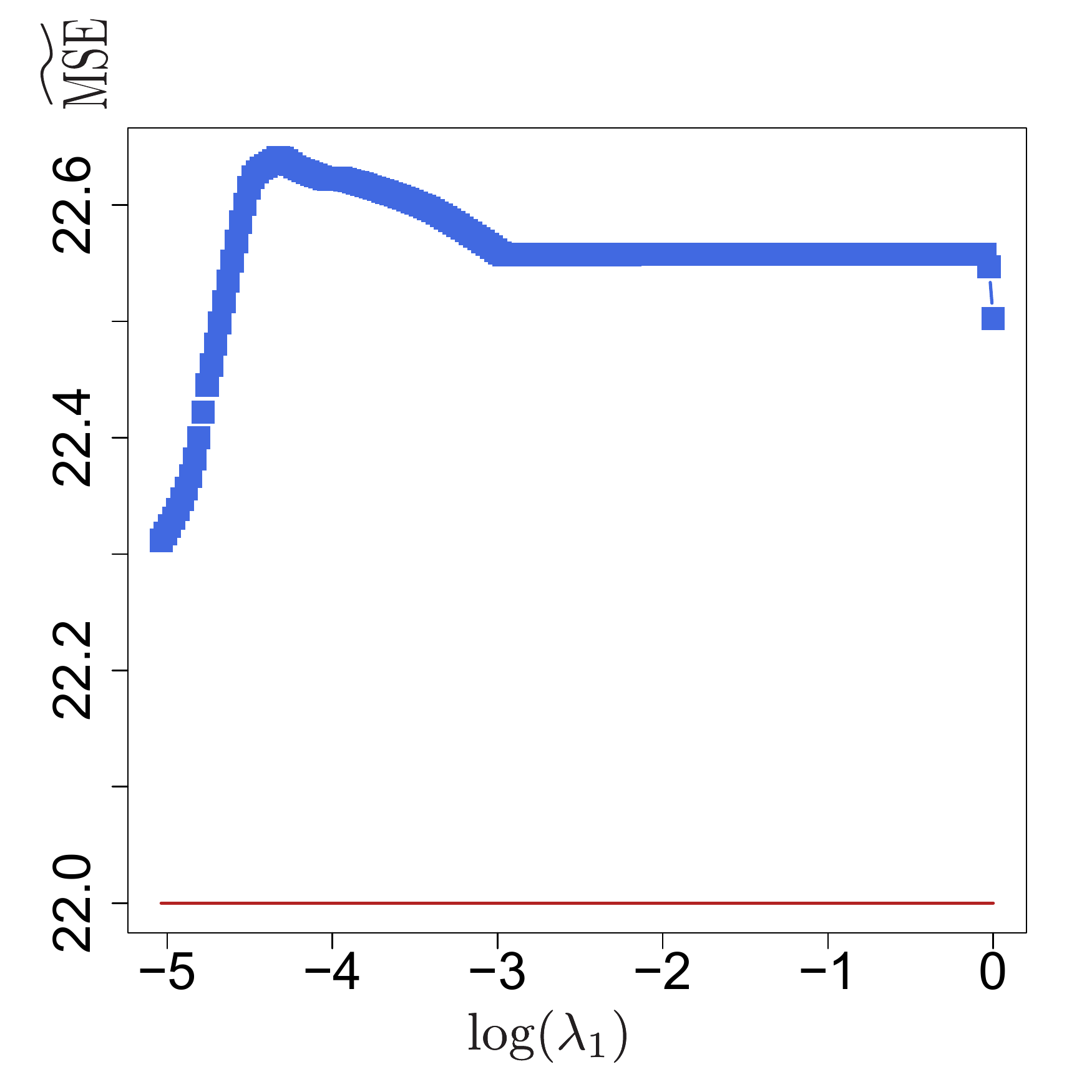}
         \hspace{1.6cm} \tiny{$r=0.1$}
        \end{center}
      \end{minipage}
\end{tabular}
    \caption{$\wtil{\mse}^{\smrm}$ (blue) and $\wtil{\mse}^{\lasso}$ (red). }
    \label{fig:tilde_mse}
  \end{center}
\end{figure}
\endgroup

We next list $\wtil{\mse}^{\smrm}$ with $\bm{\lambda}_2=0.2\bm{\lambda}$ 
and $\wtil{\mse}^{\lasso}$ for each mechanical characteristics in Figure \ref{fig:char_mse}.
By the modification, $\wtil{\mse}^{\lasso}_l$ takes $1$ for each mechanical characteristics. 
One can observe that $\wtil{\mse}^{\smrm}_l$ for C, D, N, O, Q and T (i.e., $l=3,4,15,17,20$) affect the total $\wtil{\mse}^{\smrm}$. 
\begingroup
\scalefont{0.7}
\begin{figure}[t]
  \begin{center}
    \begin{tabular}{c}

      \begin{minipage}{0.22\hsize}
        \begin{center}
          \includegraphics[clip, width=3cm]{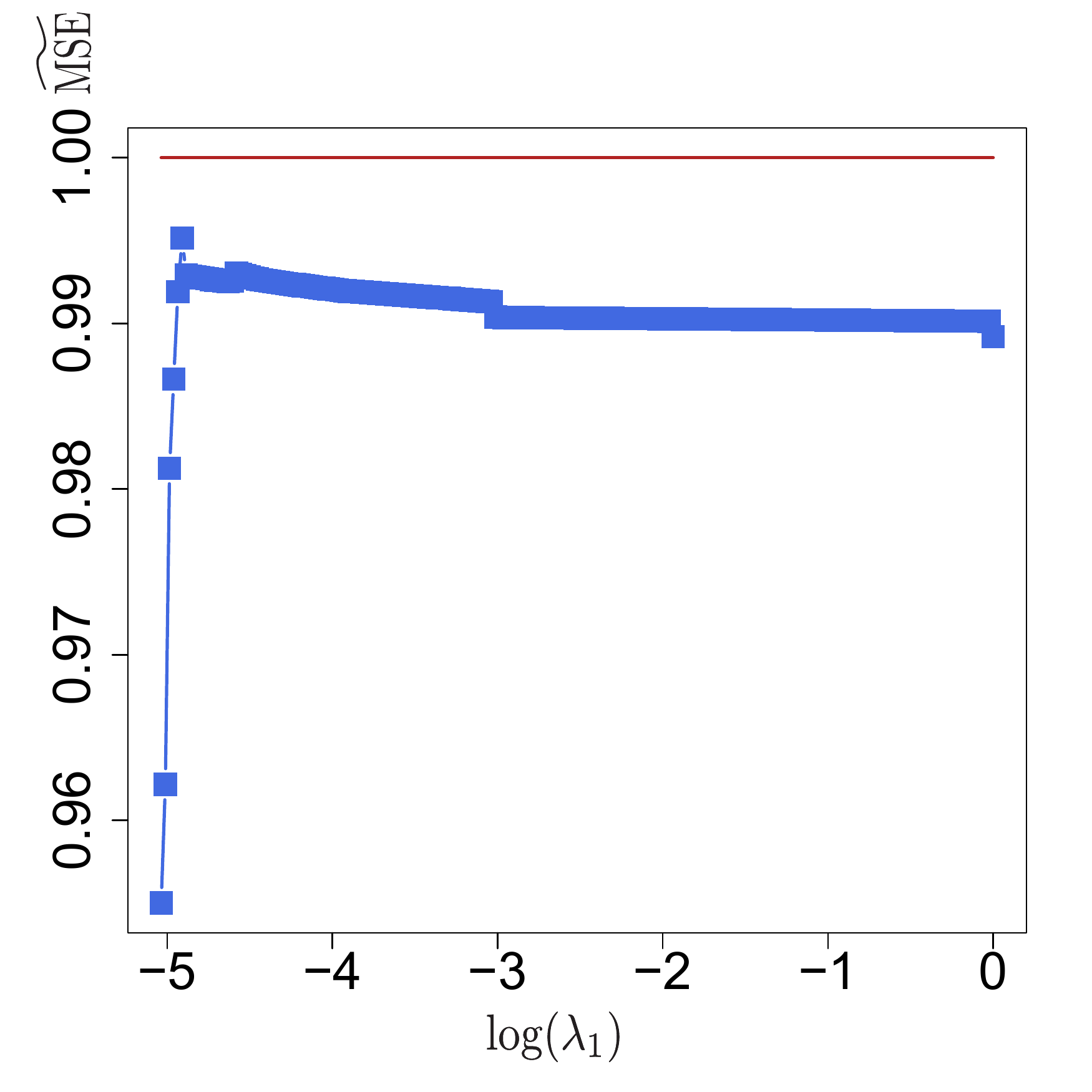}
         \hspace{1.6cm} \tiny{A}
        \end{center}
      \end{minipage}

      \begin{minipage}{0.22\hsize}
        \begin{center}
          \includegraphics[clip, width=3cm]{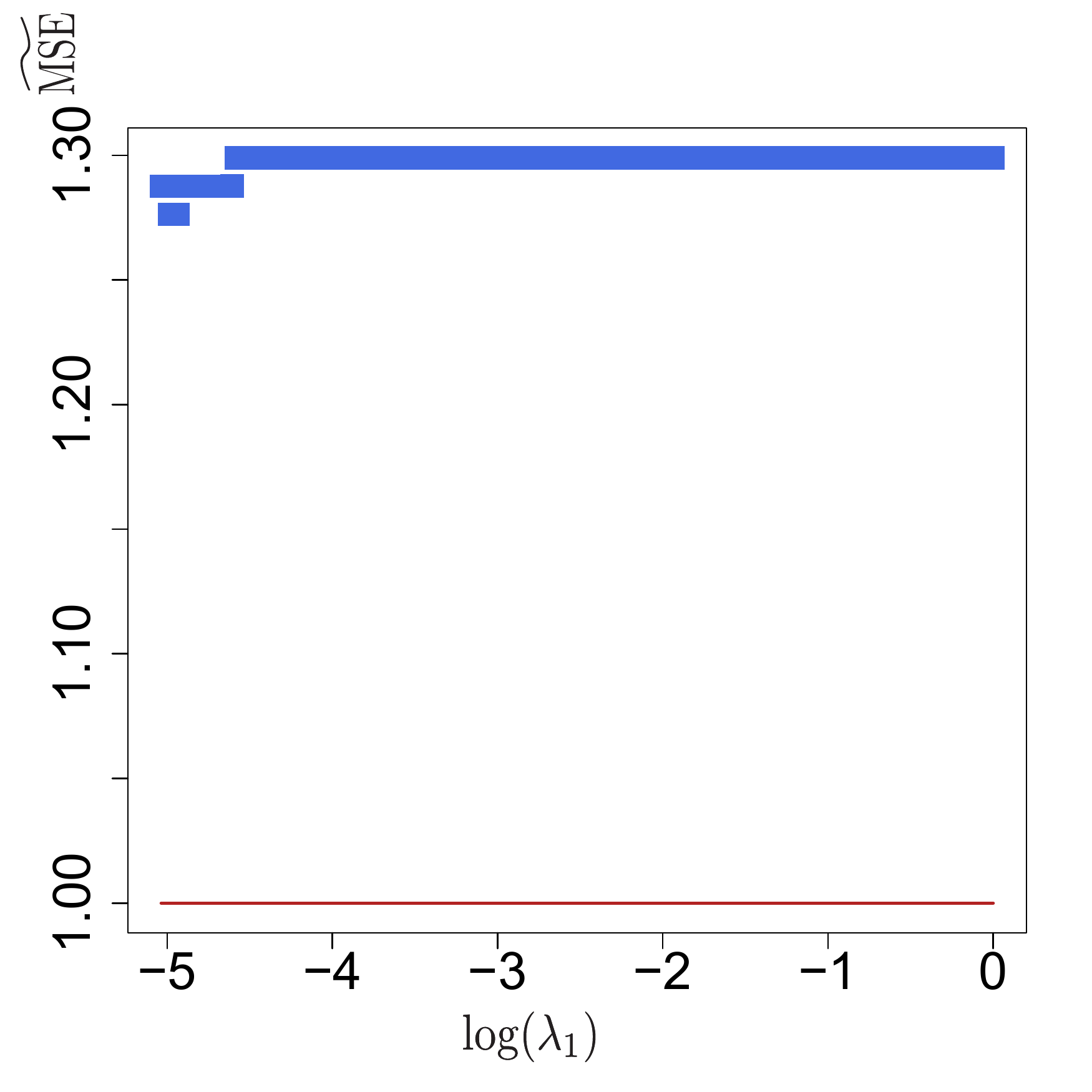}
          \hspace{1.6cm} \tiny{B}
        \end{center}
      \end{minipage}

      \begin{minipage}{0.22\hsize}
        \begin{center}
          \includegraphics[clip, width=3cm]{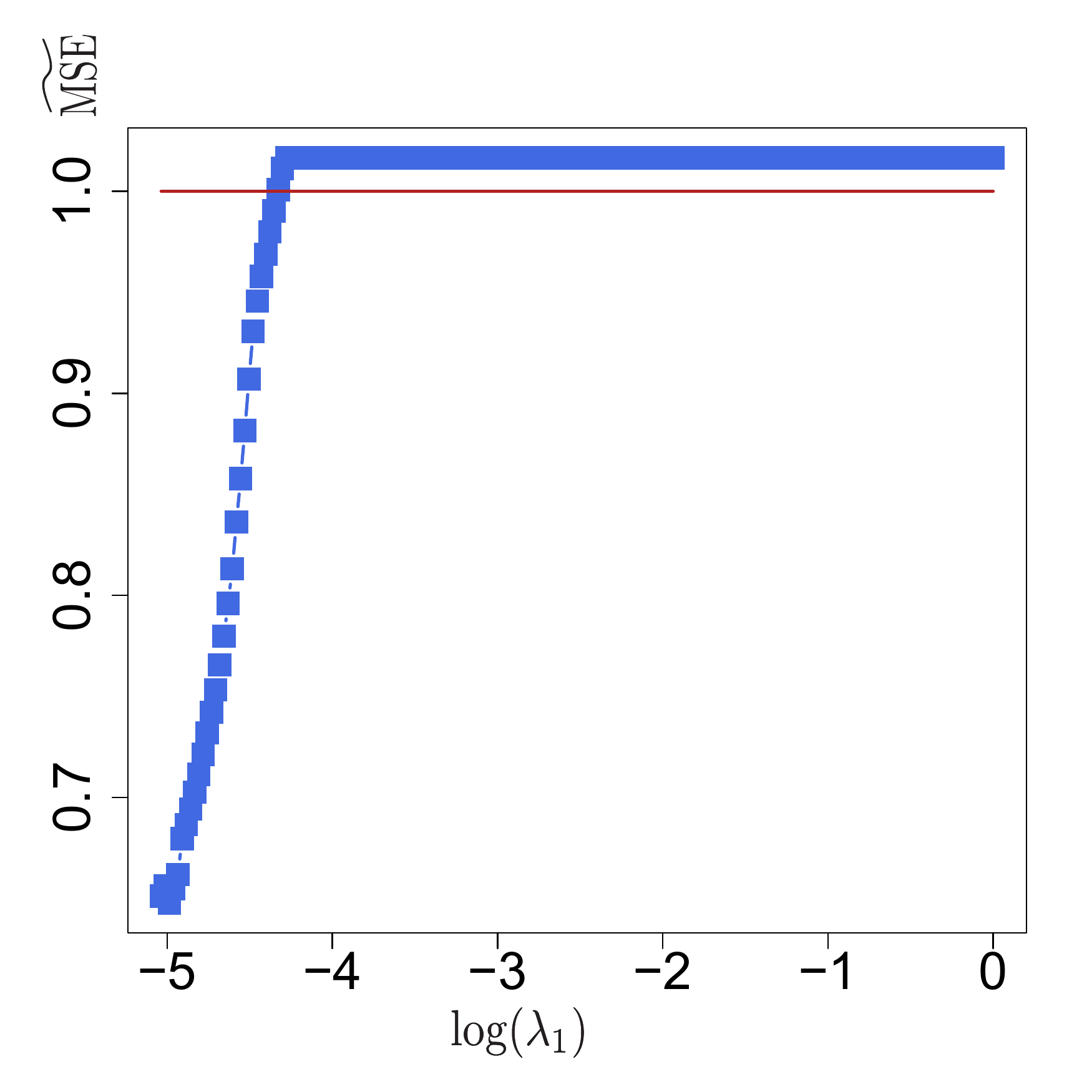}
          \hspace{1.6cm} \tiny{C}
        \end{center}
      \end{minipage} 
      
      \begin{minipage}{0.22\hsize}
        \begin{center}
          \includegraphics[clip, width=3cm]{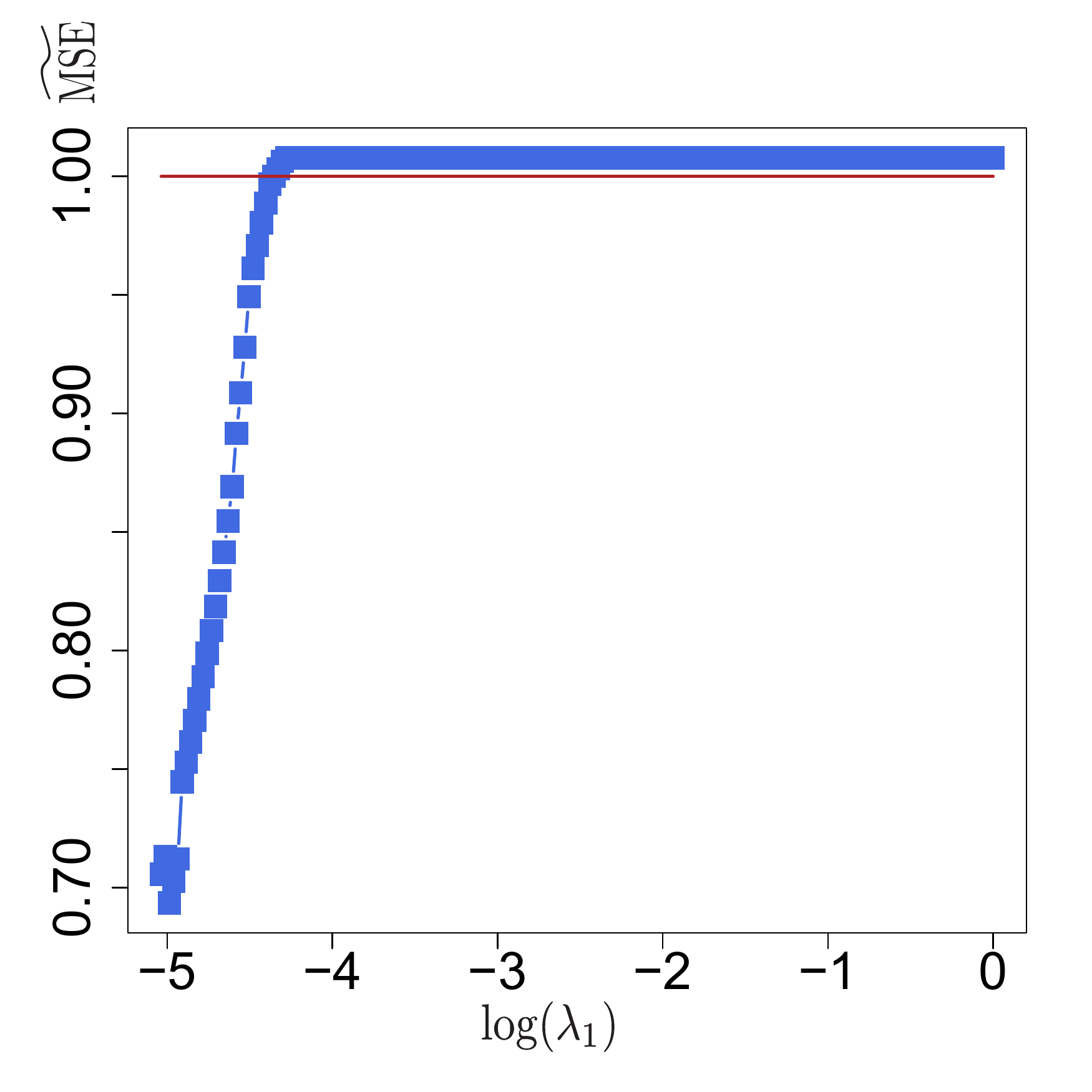}
          \hspace{1.6cm} \tiny{D}
        \end{center}
      \end{minipage}\\
      \begin{minipage}{0.22\hsize}
        \begin{center}
          \includegraphics[clip, width=3cm]{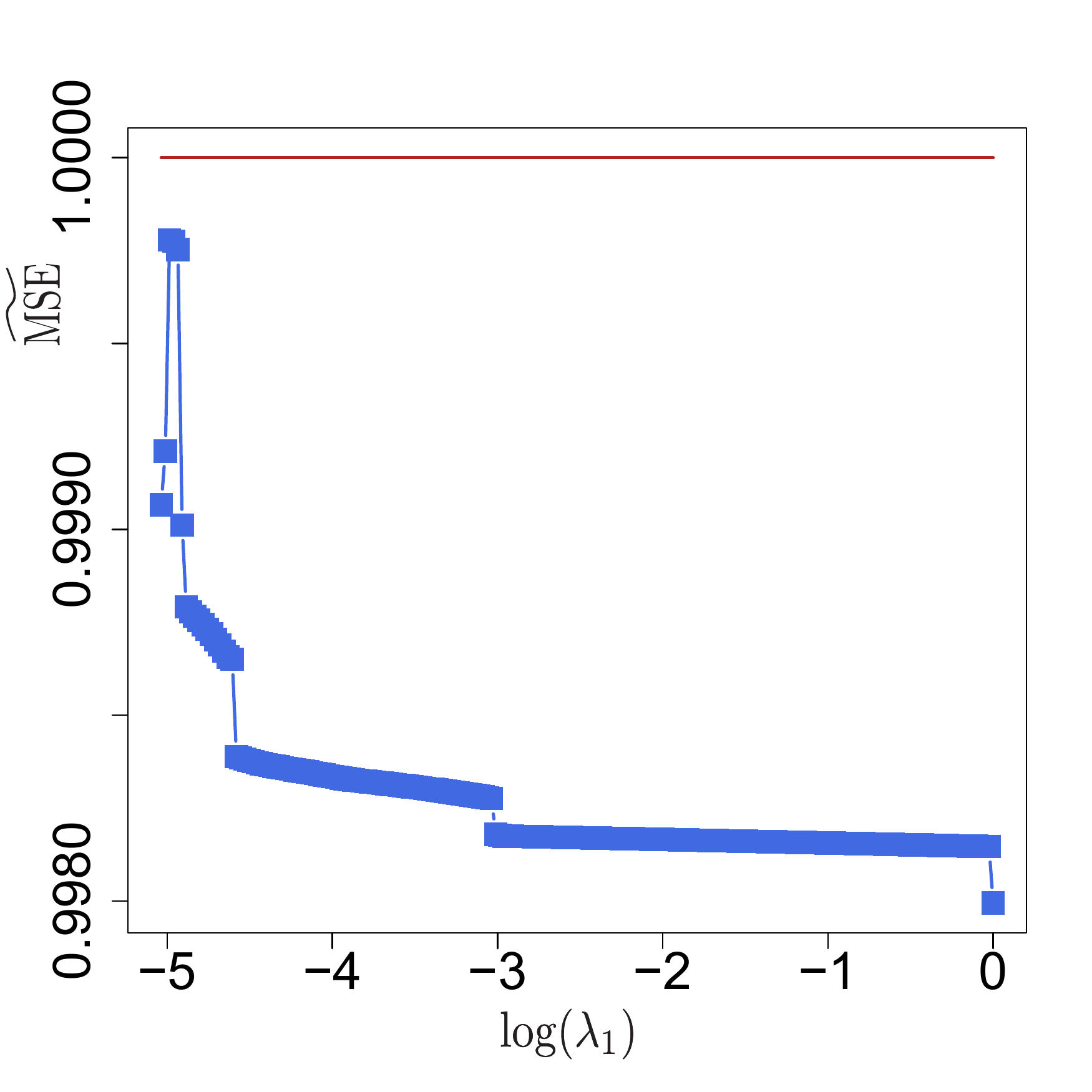}
         \hspace{1.6cm} \tiny{E}
        \end{center}
      \end{minipage}

      \begin{minipage}{0.22\hsize}
        \begin{center}
          \includegraphics[clip, width=3cm]{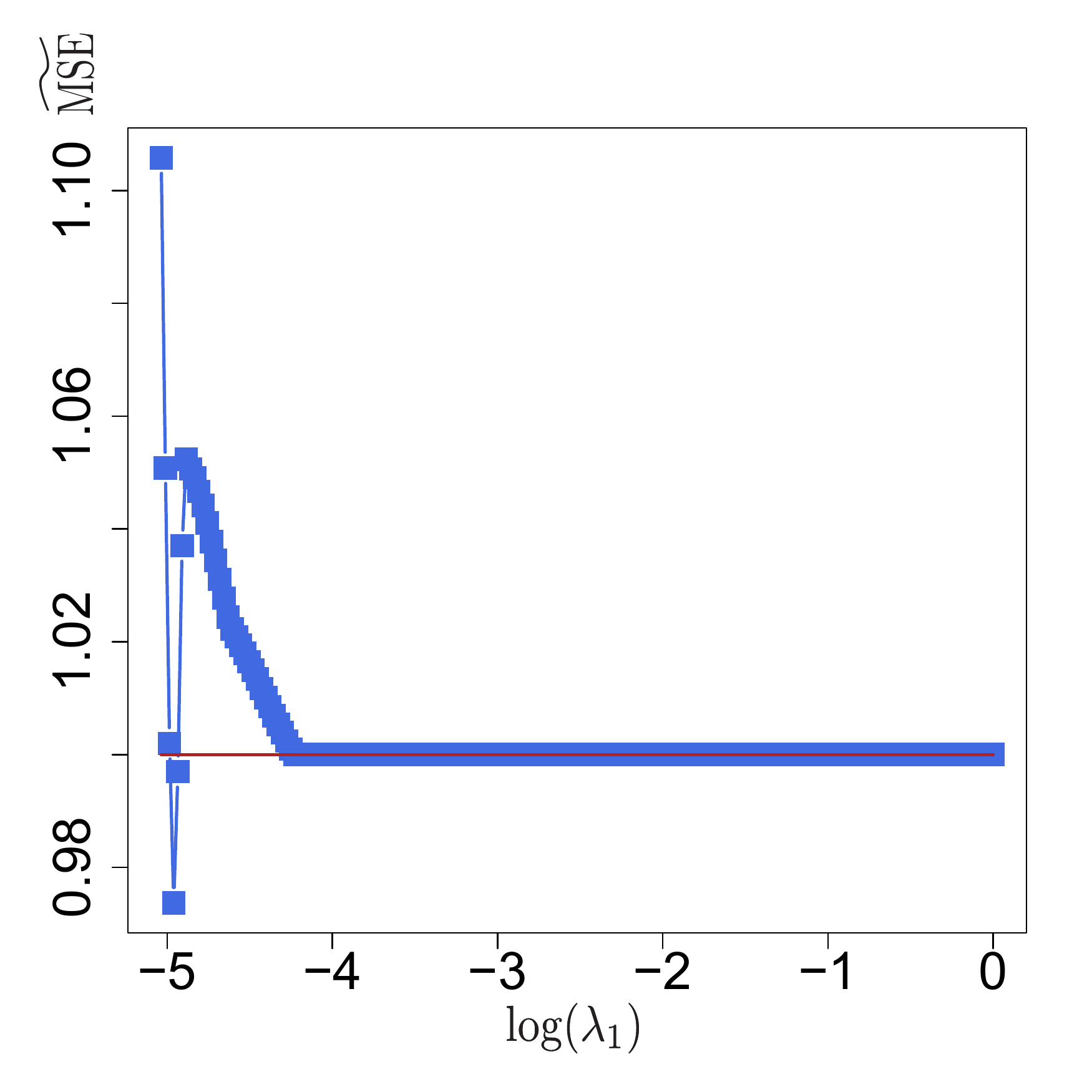}
         \hspace{1.6cm} \tiny{F}
        \end{center}
      \end{minipage}

      \begin{minipage}{0.22\hsize}
        \begin{center}
          \includegraphics[clip, width=3cm]{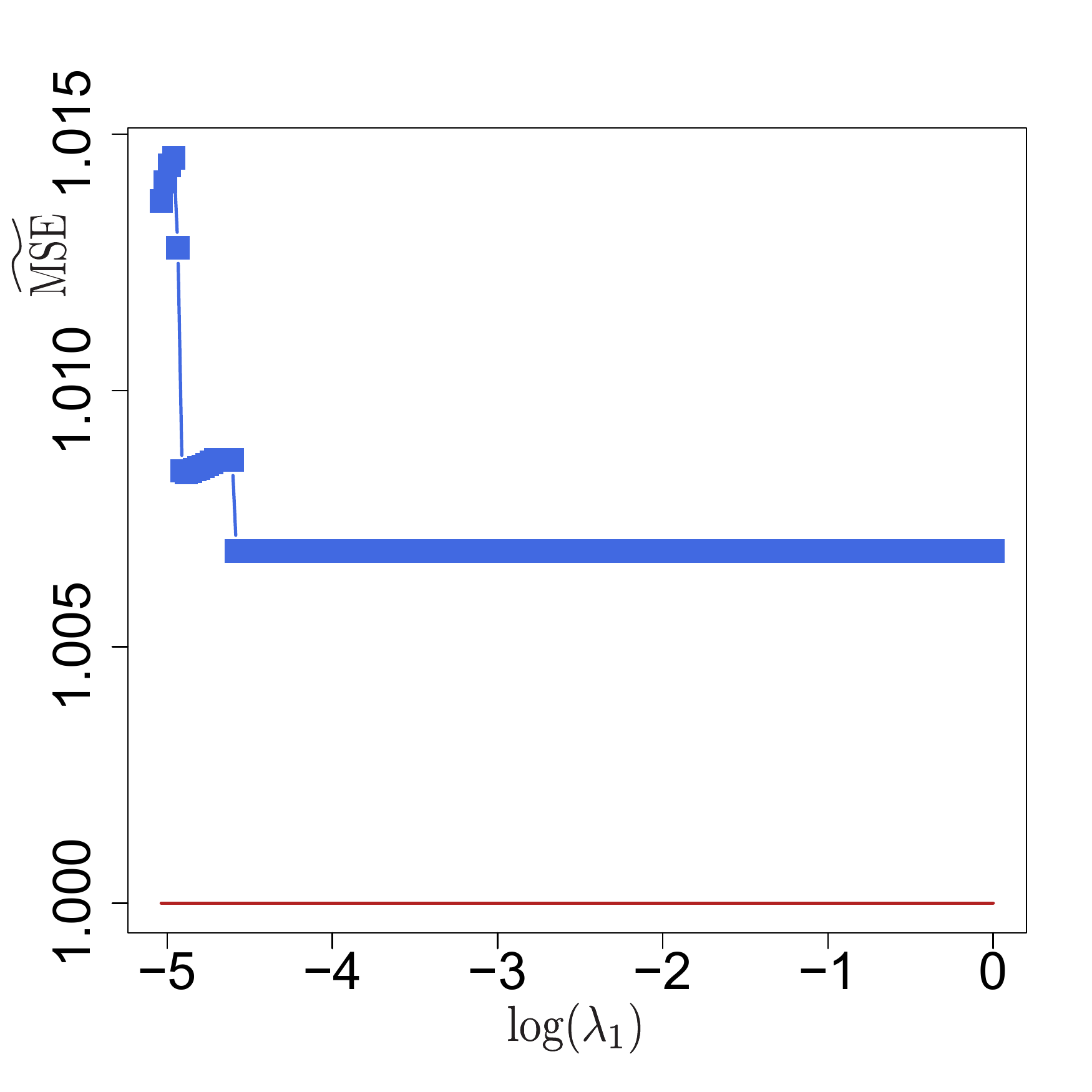}
          \hspace{1.6cm} \tiny{G}
        \end{center}
      \end{minipage} 
      
      \begin{minipage}{0.22\hsize}
        \begin{center}
          \includegraphics[clip, width=3cm]{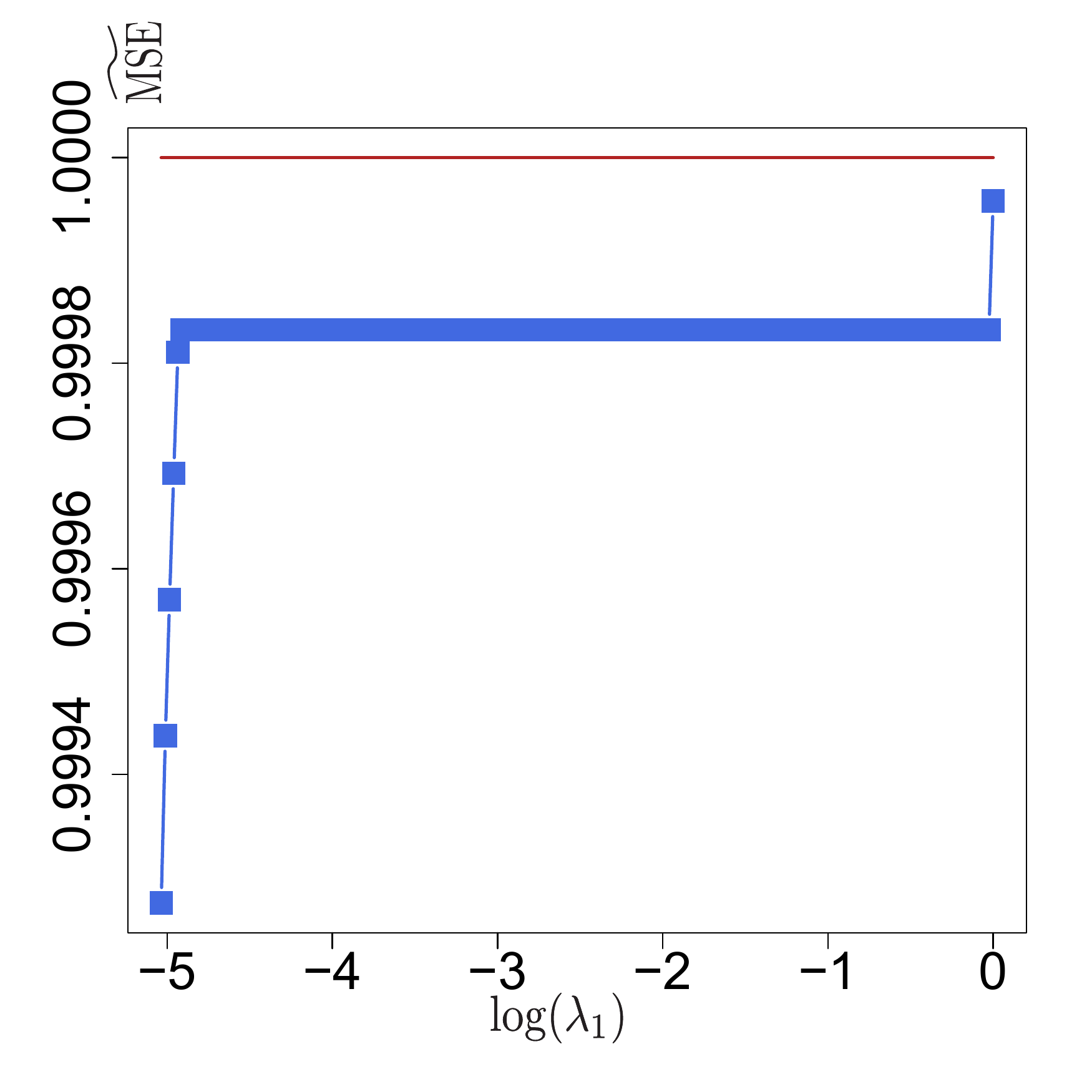}
          \hspace{1.6cm} \tiny{H}
        \end{center}
      \end{minipage}\\ 
      
      \begin{minipage}{0.22\hsize}
        \begin{center}
          \includegraphics[clip, width=3cm]{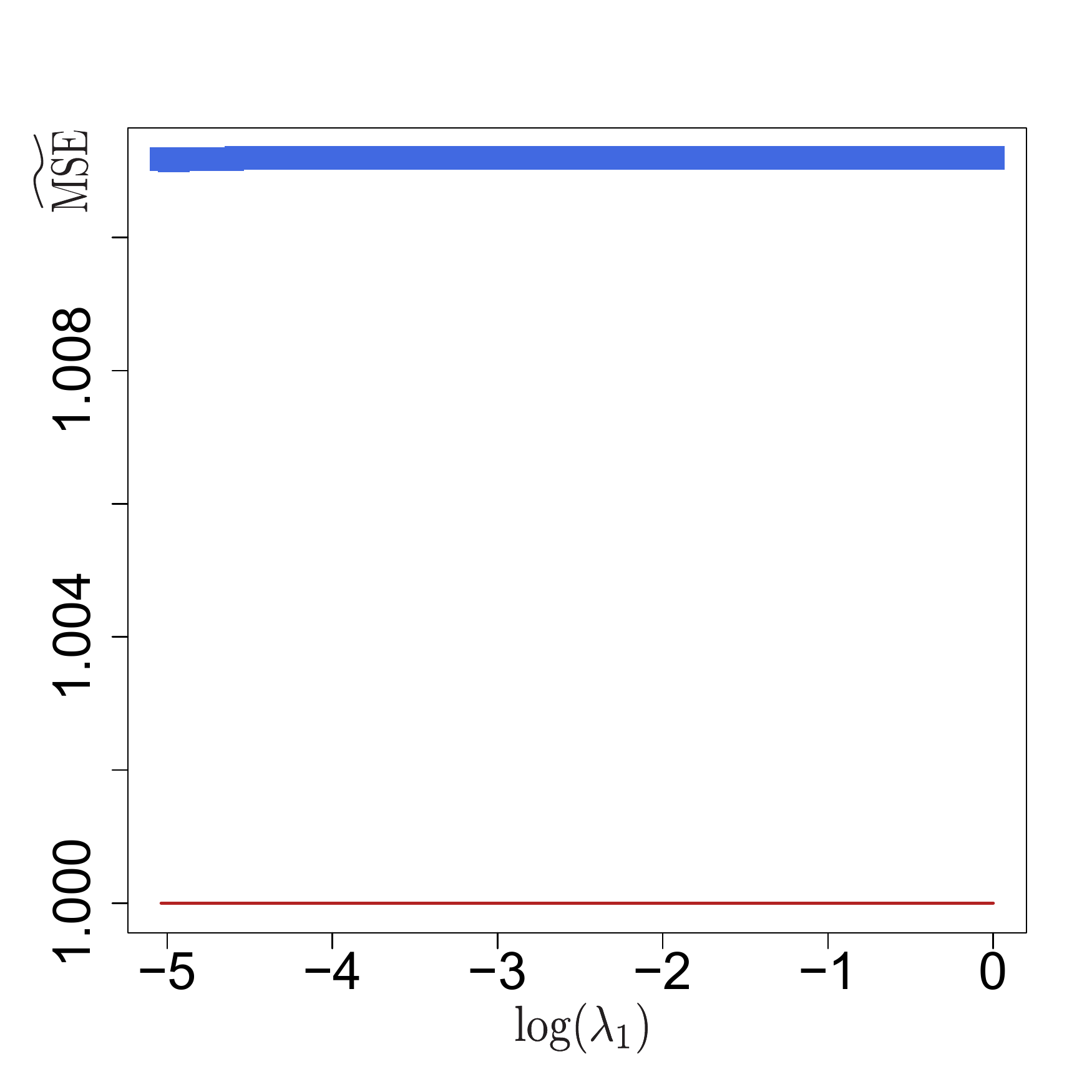}
         \hspace{1.6cm} \tiny{I}
        \end{center}
      \end{minipage}

      \begin{minipage}{0.22\hsize}
        \begin{center}
          \includegraphics[clip, width=3cm]{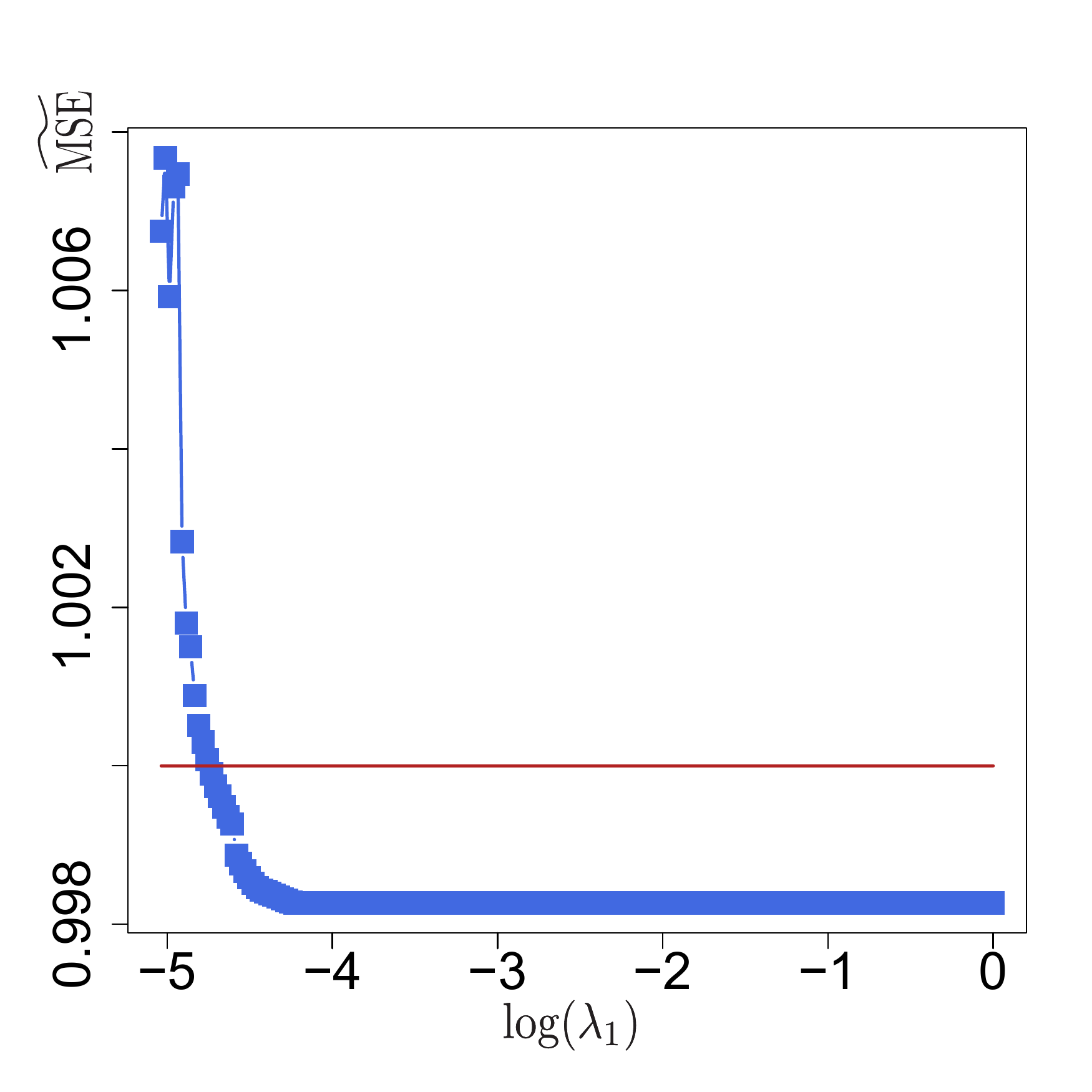}
          \hspace{1.6cm} \tiny{J}
        \end{center}
      \end{minipage}

      \begin{minipage}{0.22\hsize}
        \begin{center}
          \includegraphics[clip, width=3cm]{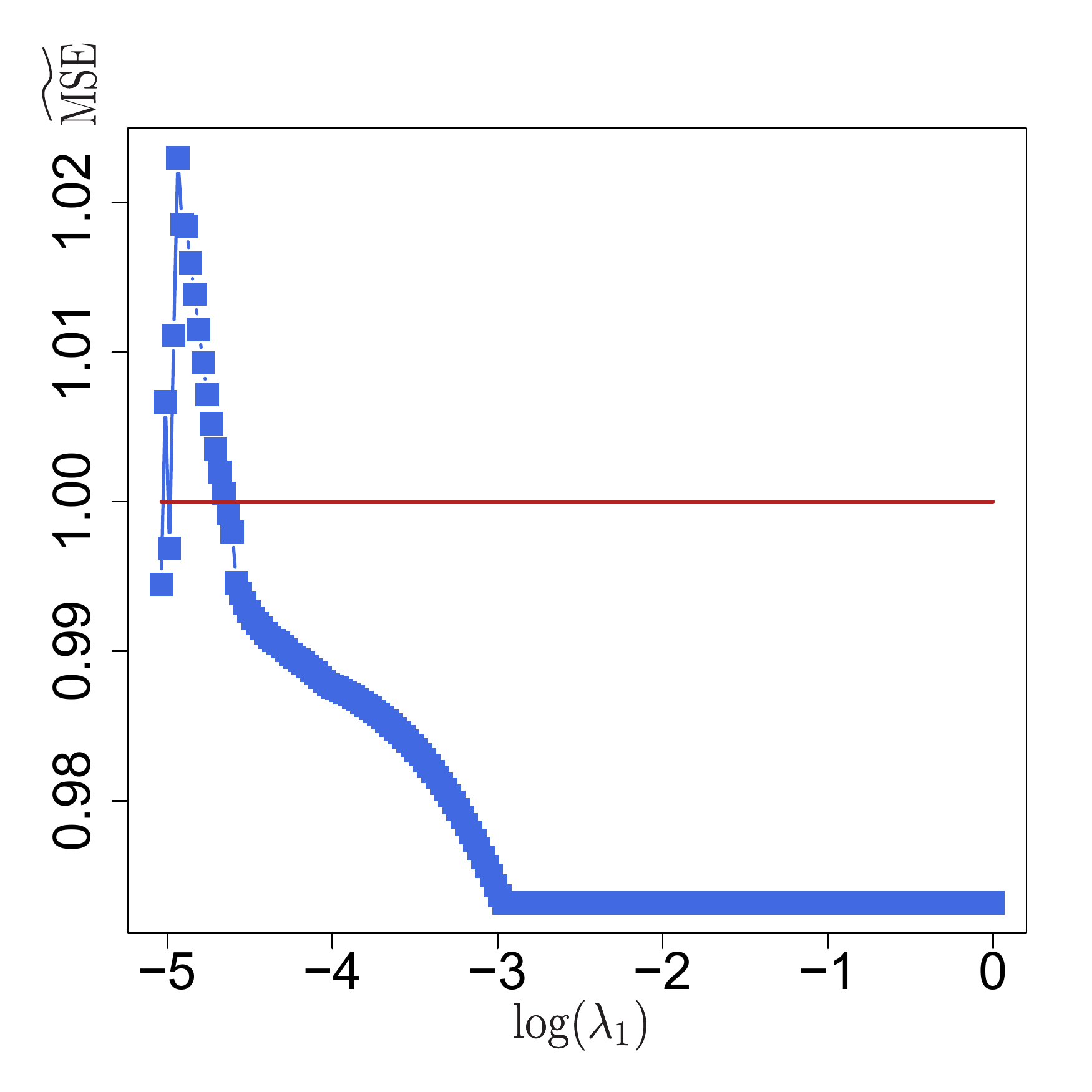}
         \hspace{1.6cm} \tiny{K}
        \end{center}
      \end{minipage} 
      
      \begin{minipage}{0.22\hsize}
        \begin{center}
          \includegraphics[clip, width=3cm]{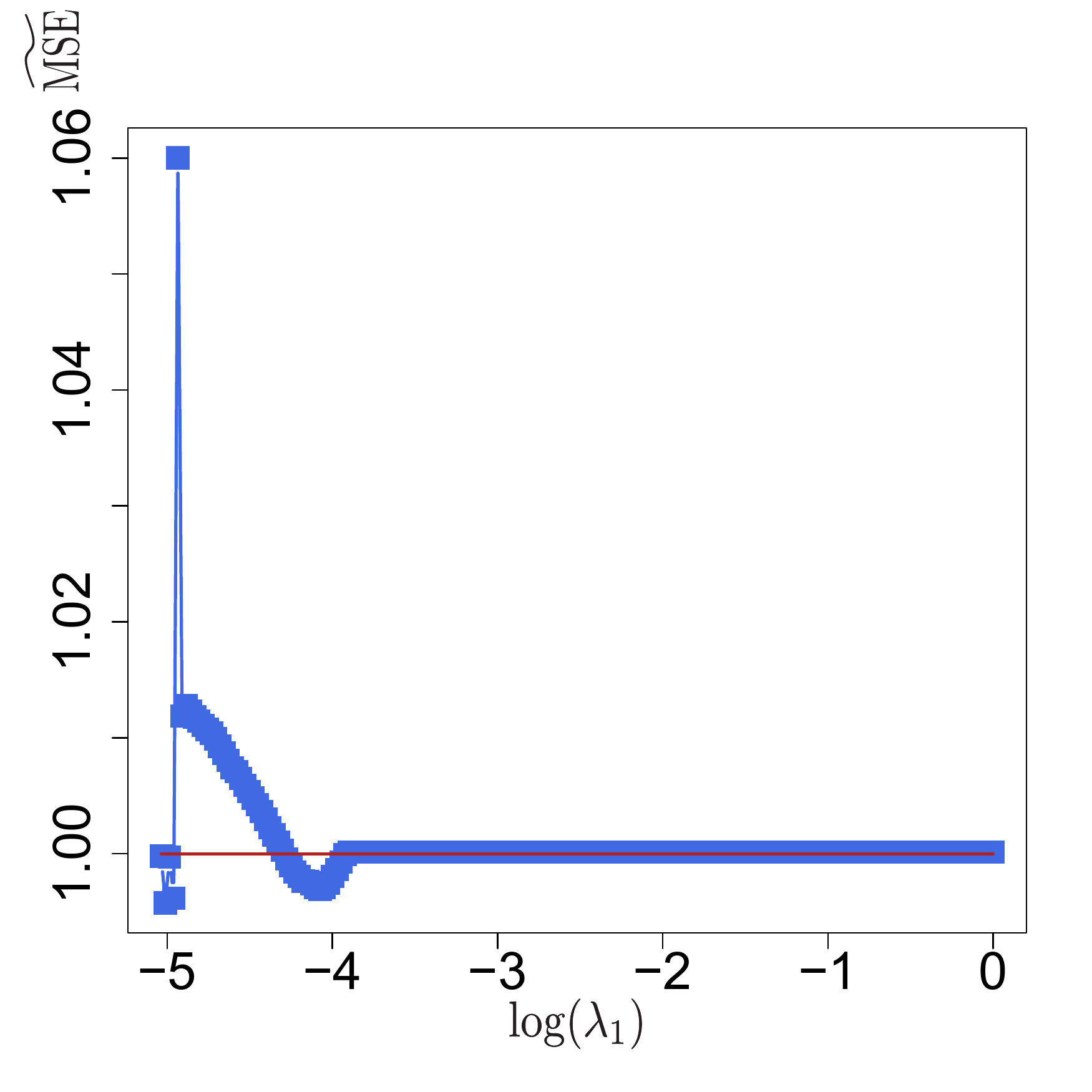}
        \hspace{1.6cm} \tiny{L}
        \end{center}
      \end{minipage}\\
      
      \begin{minipage}{0.22\hsize}
        \begin{center}
          \includegraphics[clip, width=3cm]{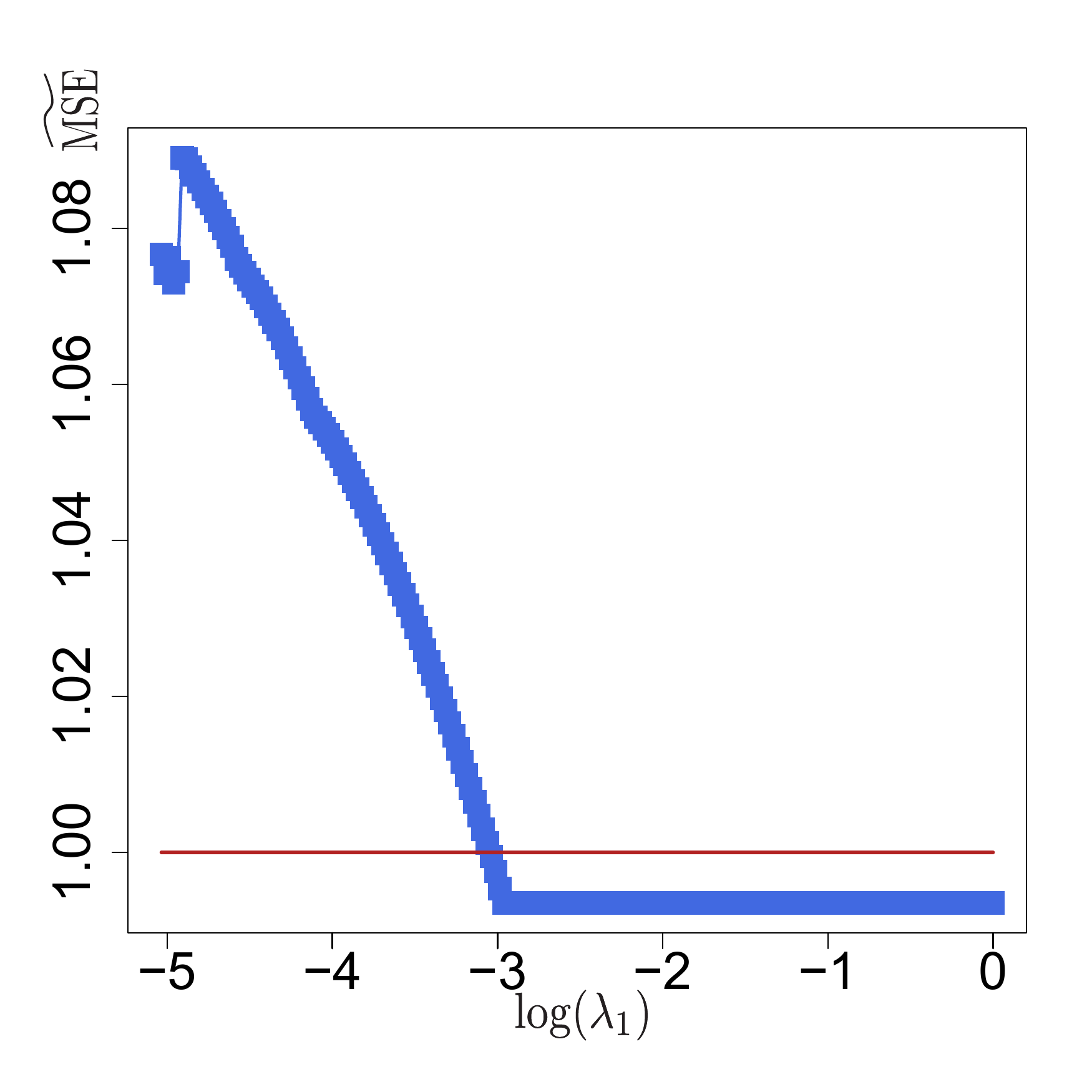}
         \hspace{1.6cm} \tiny{M}
        \end{center}
      \end{minipage}

      \begin{minipage}{0.22\hsize}
        \begin{center}
          \includegraphics[clip, width=3cm]{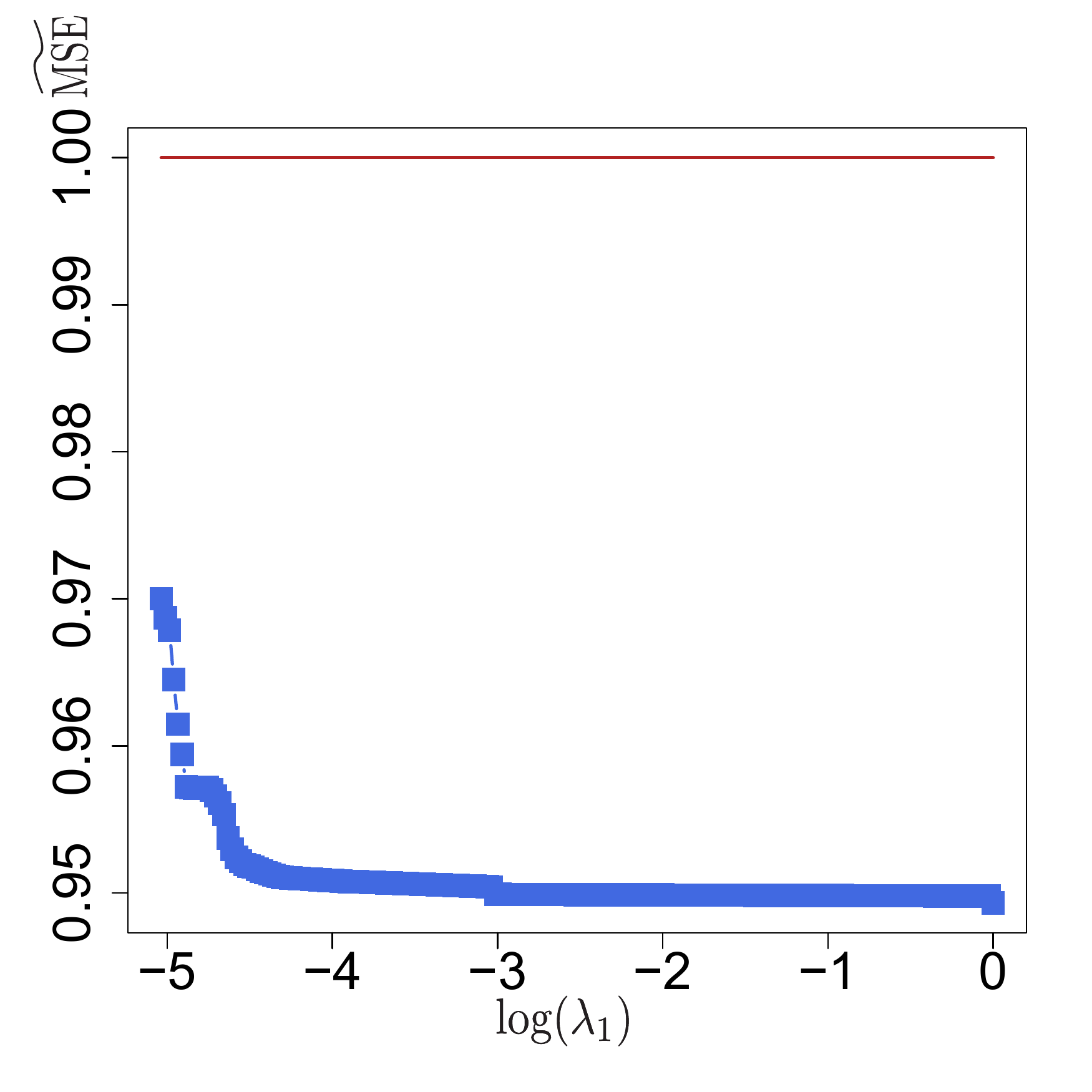}
          \hspace{1.6cm} \tiny{N}
        \end{center}
      \end{minipage}

      \begin{minipage}{0.22\hsize}
        \begin{center}
          \includegraphics[clip, width=3cm]{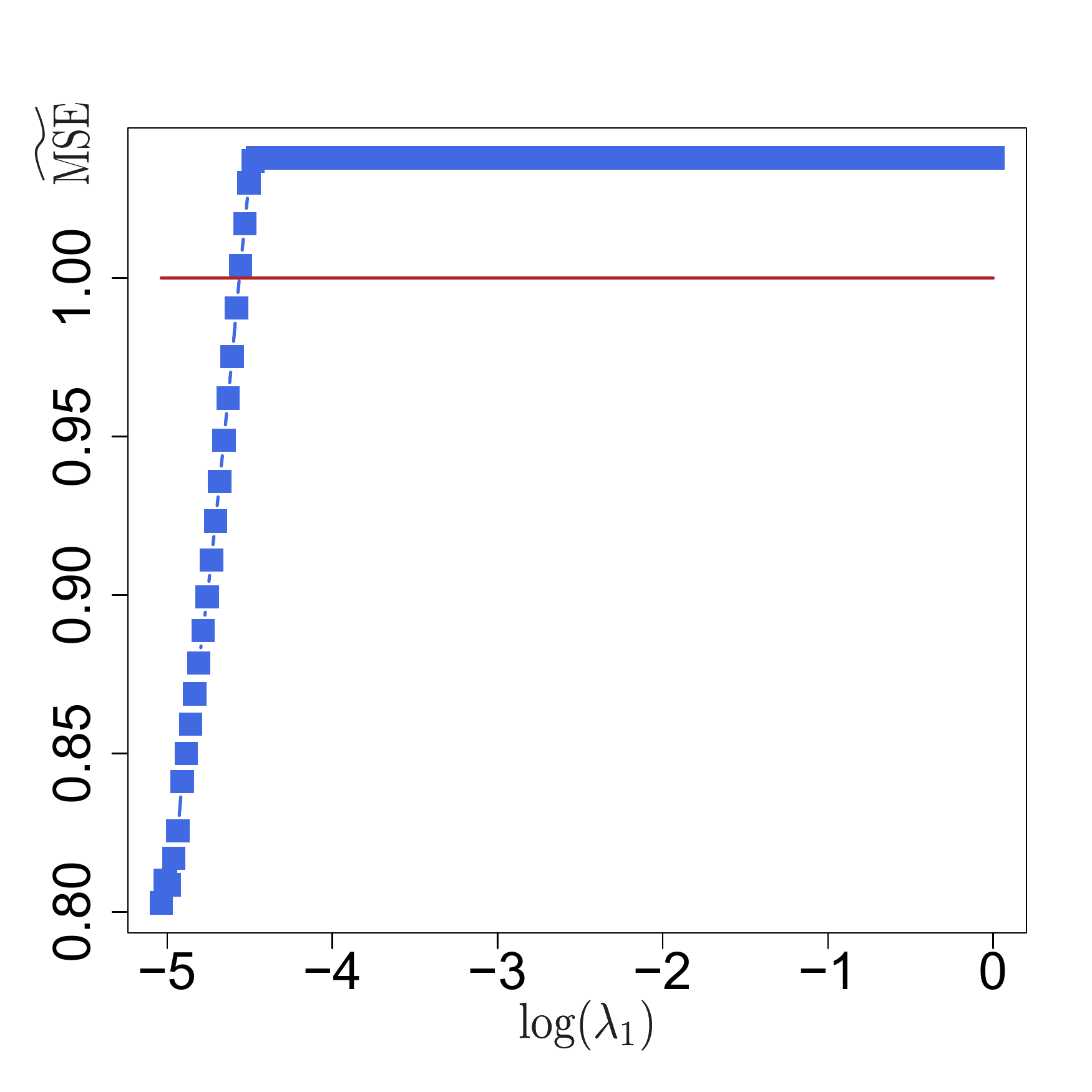}
          \hspace{1.6cm} \tiny{O}
        \end{center}
      \end{minipage} 
      
      \begin{minipage}{0.22\hsize}
        \begin{center}
          \includegraphics[clip, width=3cm]{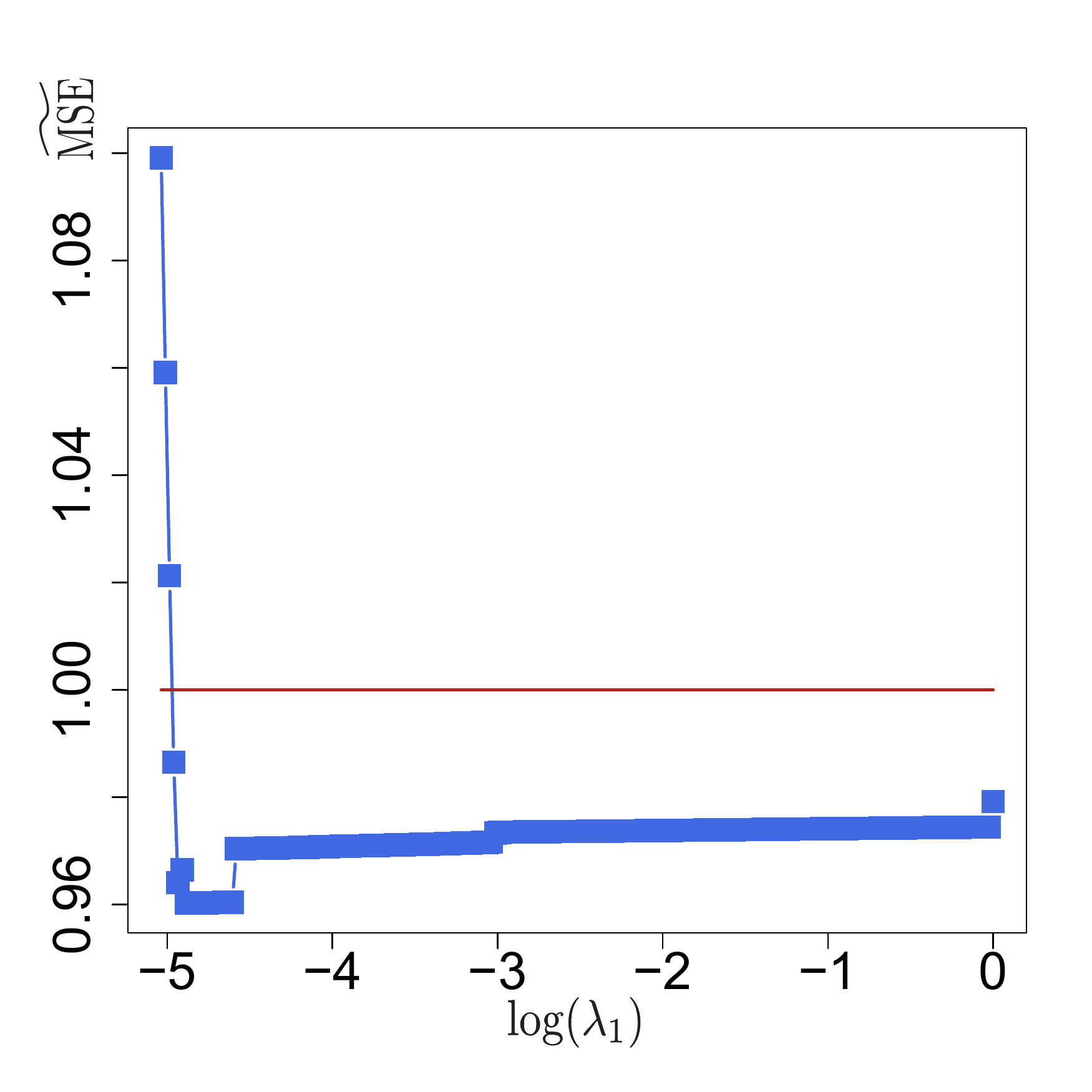}
          \hspace{1.6cm} \tiny{P}
        \end{center}
      \end{minipage}\\
      
      \begin{minipage}{0.22\hsize}
        \begin{center}
          \includegraphics[clip, width=3cm]{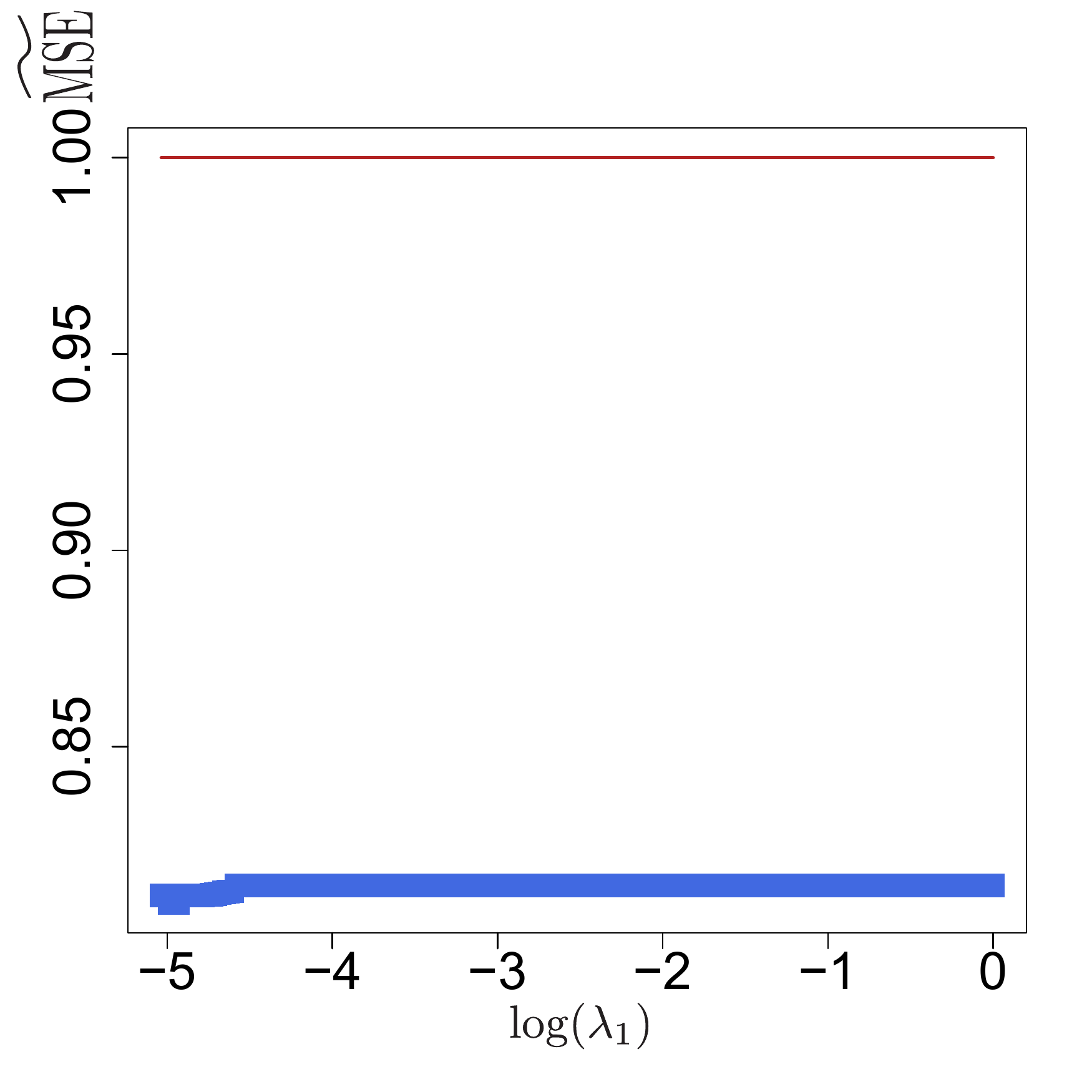}
         \hspace{1.6cm} \tiny{Q}
        \end{center}
      \end{minipage}

      \begin{minipage}{0.22\hsize}
        \begin{center}
          \includegraphics[clip, width=3cm]{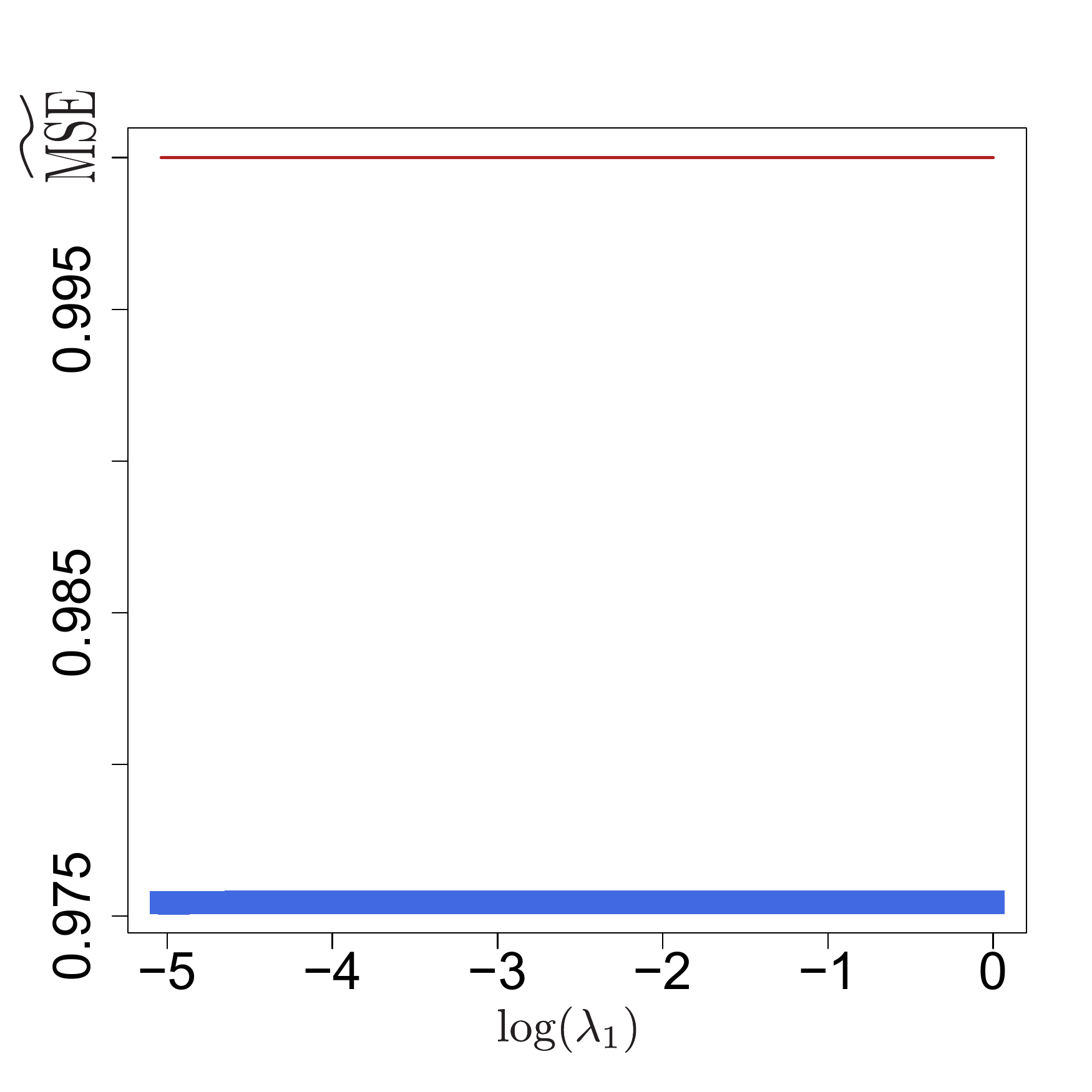}
          \hspace{1.6cm} \tiny{R}
        \end{center}
      \end{minipage}

      \begin{minipage}{0.22\hsize}
        \begin{center}
          \includegraphics[clip, width=3cm]{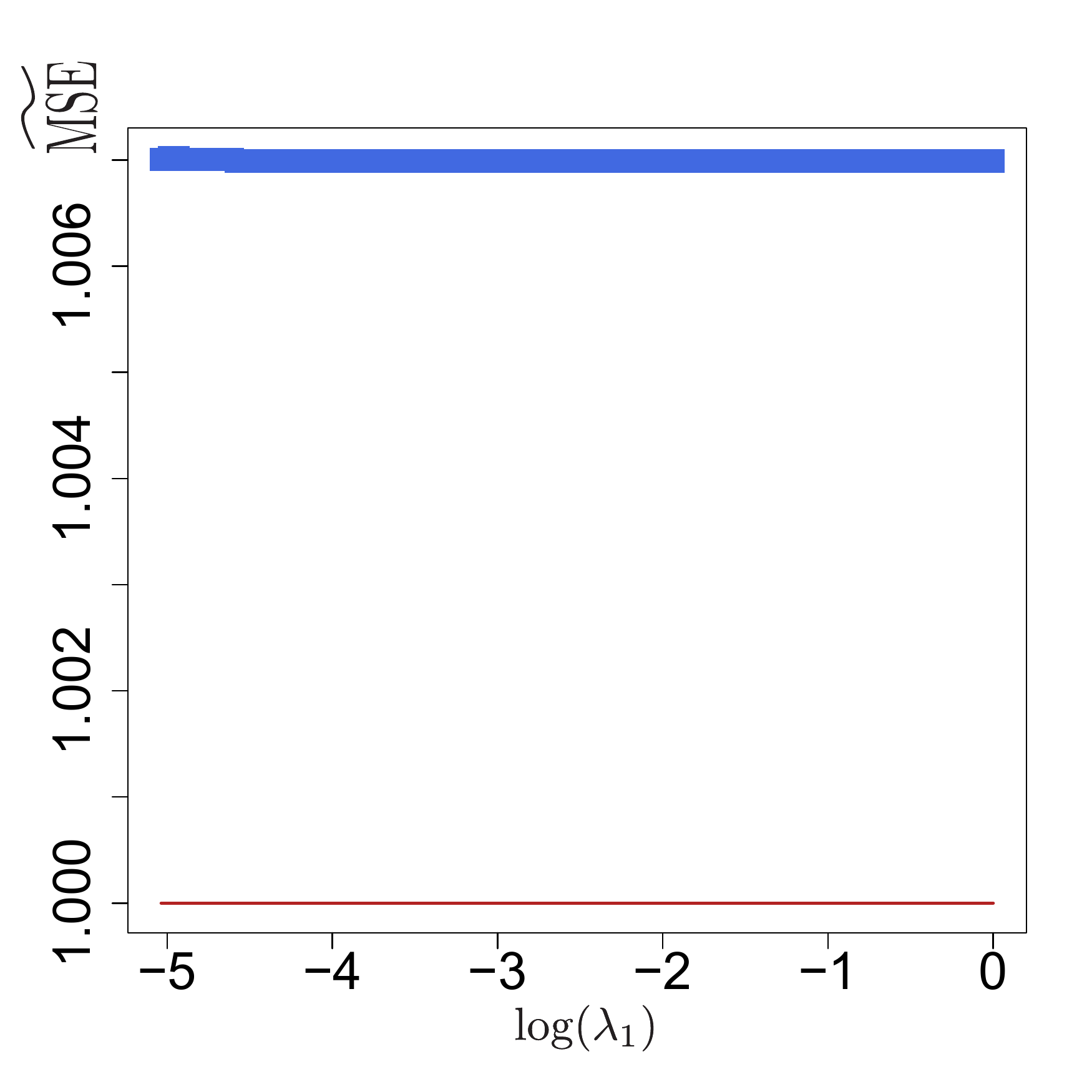}
          \hspace{1.6cm} \tiny{S}
        \end{center}
      \end{minipage} 
      
      \begin{minipage}{0.22\hsize}
        \begin{center}
          \includegraphics[clip, width=3cm]{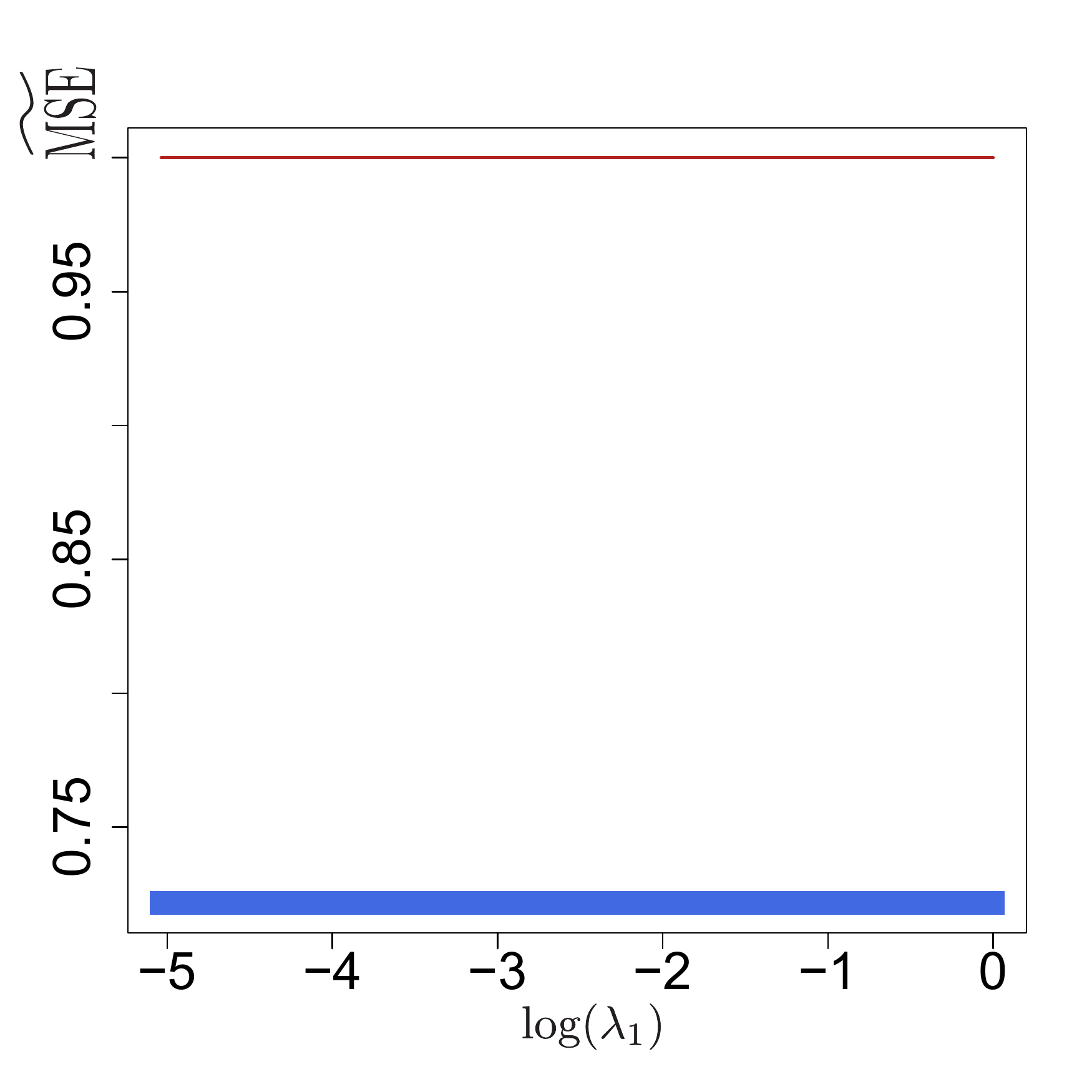}
          \hspace{1.6cm} \tiny{T}
        \end{center}
      \end{minipage}\\
      
      \begin{minipage}{0.22\hsize}
        \begin{center}
          \includegraphics[clip, width=3cm]{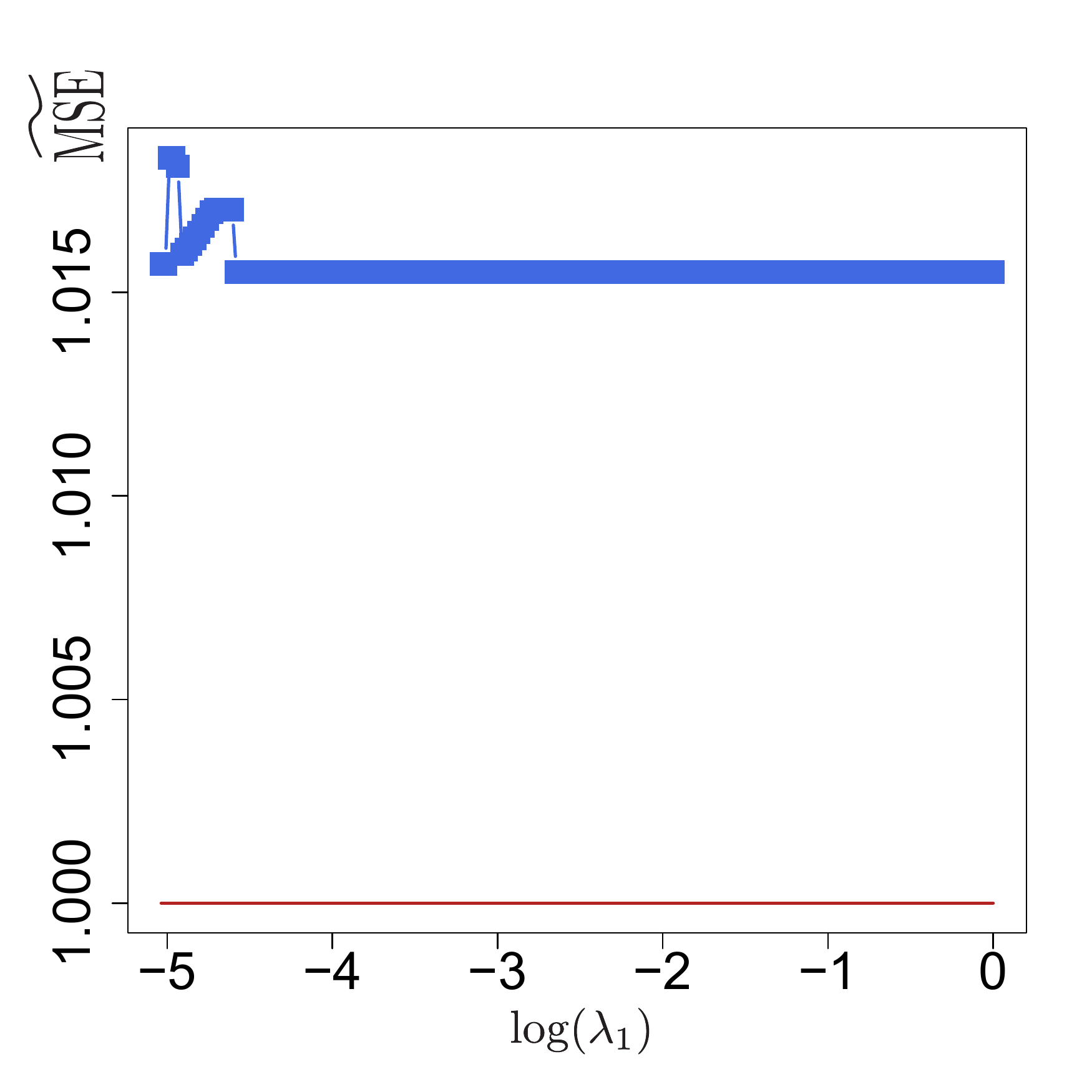}
        \hspace{1.6cm} \tiny{U}
        \end{center}
      \end{minipage}

      \begin{minipage}{0.22\hsize}
        \begin{center}
          \includegraphics[clip, width=3cm]{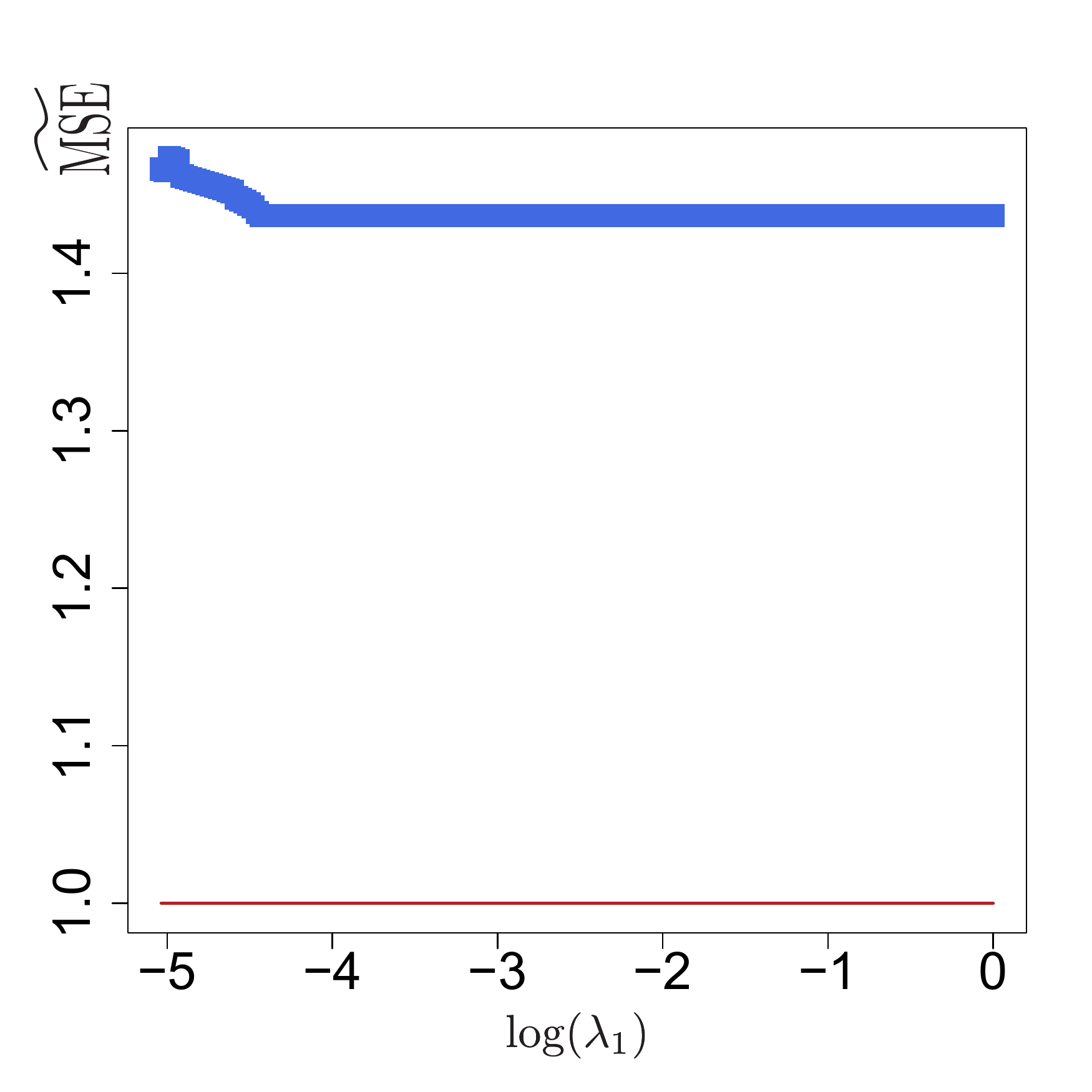}
          \hspace{1.6cm} \tiny{V}
        \end{center}
      \end{minipage}
      
    \end{tabular}
    \caption{$\wtil{\mse}^{\smrm}$ (blue) and $\wtil{\mse}^{\lasso}$ (red) for each mechanical characters. 
    The $x$ axis represents $\log(\lambda_1)$ and 
    the $y$ axis $\wtil{\mse}$.}
    \label{fig:char_mse}
  \end{center}
\end{figure}
\endgroup

\end{document}